%% file: causal_asymptotic_expansion-general_v40-20230408-b+.tex
\documentclass[a4paper, 12pt]{article}
\usepackage{dsfont}
\usepackage{amssymb}
\usepackage{latexsym}
\usepackage{amsmath}
\usepackage{xcolor}
\usepackage{comment}
\usepackage{amsthm}
\usepackage[dvips]{graphicx}
\usepackage{layout} 
\usepackage{ulem}
\usepackage{enumerate}
\usepackage{bm}
\usepackage{bbm} 
\usepackage[bbgreekl]{mathbbol} 
\usepackage{authblk}
\usepackage{setspace} 
\usepackage{here}
\newif\ifrs
\rstrue
\ifrs \usepackage{mathrsfs} \fi  
\newif\ifcol
\coltrue 

\newtheorem{theorem}{Theorem}[section]

\newtheorem{lemma}[theorem]{Lemma}

\newtheorem{proposition}[theorem]{Proposition}

\newtheorem{remark}[theorem]{Remark}
\newtheorem{example}[theorem]{Example}
\numberwithin{equation}{section}
\newtheorem{theorem*}{Theorem}
\newtheorem{ass*}[theorem*]{Assumption}
\newtheorem{note*}[theorem*]{Note}
\newtheorem{lemma*}[theorem*]{Lemma}
\newtheorem{definition*}[theorem*]{Definition}
\newtheorem{proposition*}[theorem*]{Proposition}
\newtheorem{corollary*}[theorem*]{Corollary}
\newtheorem{remark*}[theorem*]{Remark}
\newtheorem{example*}[theorem*]{Example}
\numberwithin{equation}{section}
\newif\ifcol
\colfalse
\ifcol
\newcommand{\colorr}{\color{black}}

\newcommand{\colorb}{\color[rgb]{0,0,0.8}}

\newcommand{\colorn}{\color[rgb]{1,1,1}}

\newcommand{\coloroy}{\color[rgb]{1,0.95,0}}

\else
\newcommand{\colorb}{\color{black}}

\newcommand{\colorr}{\color{black}}

\newcommand{\colorn}{\color{black}}

\newcommand{\coloroy}{\color{black}}
\fi
\newif\ifcol
\colfalse
\ifcol
\newcommand{\sred}{\color[rgb]{0.8,0,0}}

\newcommand{\sblue}{\color[rgb]{0,0,0.8}}
\else
\newcommand{\sred}{\color{black}}

\newcommand{\sblue}{\color{black}}
\fi
\newif\ifcol
\colfalse
\ifcol
\newcommand{\tred}{\color[rgb]{0.8,0,0}}

\newcommand{\colory}{\color{yellow}}
\else
\newcommand{\tred}{\color{black}}
\newcommand{\colory}{\color{yellow}}
\fi
\newif\ifcol
\colfalse
\ifcol
\newcommand{\fred}{\color[rgb]{0.8,0,0}}

\newcommand{\fblue}{\color[rgb]{0,0,0.8}}
\renewcommand{\colory}{\color{yellow}}
\else
\newcommand{\fred}{\color{black}}
\newcommand{\fblue}{\color{black}}
\renewcommand{\colory}{\color{yellow}}
\fi
\newif\ifcol
\colfalse
\ifcol
\newcommand{\vred}{\color[rgb]{0.8,0,0}}

\renewcommand{\colory}{\color{yellow}}
\else
\newcommand{\vred}{\color{black}}
\renewcommand{\colory}{\color{yellow}}
\fi
\def\indep{\perp\!\!\perp}
\excludecomment{en-text}
\includecomment{jp-text}
\includecomment{comment}
\input nakamacro030101+.tex

\def\mfkT{\mathfrak T}
\def\mfkU{\mathfrak U}

\def\mfkC{{\mathfrak C}}

\def\wh{\widehat}
\def\wt{\widetilde}
\def\ol{\overline}
\def\ul{\underline}
\renewcommand{\koko}{{\color{yellow}koko}}
\begin{document}

\title{
Asymptotic expansion for batched bandits
\footnote{
This work was in part supported by 
Japan Science and Technology Agency CREST 
JPMJCR2115; 
Japan Society for the Promotion of Science Grants-in-Aid for Scientific Research 
(Scientific Research);  
and by a Cooperative Research Program of the Institute of Statistical Mathematics. 
}
}
\author[1]{Yechan Park}
\author[2,3,4]{Nakahiro Yoshida}
\affil[1]{The University of Tokyo, Faculty of Economics
\footnote{Faculty of Economics, University of Tokyo: 7-3-1 Hongo, Bunkyo-ku, Tokyo 113-0033, Japan. e-mail: yechanparkjp@g.ecc.u-tokyo.ac.jp}
        }
\affil[2]{The University of Tokyo, Graduate School of Mathematical Sciences
\footnote{Graduate School of Mathematical Sciences, University of Tokyo: 3-8-1 Komaba, Meguro-ku, Tokyo 153-8914, Japan. e-mail: nakahiro@ms.u-tokyo.ac.jp}
        }
\affil[3]{Japan Science and Technology Agency CREST
        }
\affil[4]{Institute of Statistical Mathematics
        }        
\maketitle
\ \\
{\it Summary} \ 
In bandit algorithms, the randomly time-varying adaptive experimental design makes it difficult to apply traditional limit theorems to off-policy evaluation of the treatment effect. 
Moreover, the normal approximation by the central limit theorem becomes unsatisfactory for lack of information due to the small sample size of the inferior arm. 
To resolve this issue, we introduce a backwards asymptotic expansion method and prove the validity of this scheme 
based on the partial mixing, that was originally introduced for the expansion of the distribution of a functional of a jump-diffusion process in a random environment. 
The theory is generalized in this paper to incorporate the backward propagation of random functions in the bandit algorithm. 
Besides the analytical validation, 
the simulation studies also support the new method. 
Our formulation is general and applicable to nonlinearly parametrized differentiable statistical models having an adaptive design. 

\section{Introduction}
The bandit algorithm is a typical adaptive experimental design. 
Free from the classical deterministic designs, 
it enables us to treat time-dependent stochastic designs, that are attracting attentions from various areas such as clinical trials, advertisement, online education and policy making among many others. 
In this paper, we consider inference for batched bandit algorithms, in particular taking asymptotic theoretical approach. 

We will model a population consisting $n_s$ individuals at Stage $s\in\bbS=\{1,...,S\}$. 
The $i_s$-th individual at Stage $s$ is  
identified with an element $j=(s,i_s)$ 
of the set $\bbJ^n=\{{\fred(s,i_s)};\>i_s\in\bbI_s^n,\>s\in\bbS\}$, where 
$\bbI_s^n=\{1,...,n_s\}$ and the numbers $(n_s)_{s\in\bbS}$ depend on $n\in\bbN$, 
that is a parameter driving the asymptotic theory we develop. 
The subgroup at Stage $s\in\bbS$ is denoted by 
$\bbJ^n_s=\{(s,i_s);\>i_s{\fred\in\bbI_s^n}\}$. 
Let $s(j)=s$ for $j\in\bbJ^n_s$. 
Given a probability space $(\Omega,\calf,P)$, 
the action (treatment) for the individual $j\in\bbJ_s^n$ is expressed by 
a $\ol{k}_s$-dimensional random column vector $A_j=(A_{j,k_s})_{k_s\in\calk_s}$, 
$\calk_s=\{1,...,\ol{k}_s\}$, 
such that each entry $A_{j,k_s}$ takes values in $\{1,0\}$ and $\sum_{k_s\in\calk_s} A_{j,k_s}=1$. 
In the study of the batched bandits, 
the reward (effect) of the action $A_j$ is denoted by $R_j$, 
and it is assumed to be written as 
\bea\label{0311180317}
R_j
&=&
A_j^\star\beta_{s(j)}+{\sblue\dot{\ep}_j}
\quad
(j\in\bbJ^n)
\eea
where $\beta_s$ is a $\ol{k}_s$-dimensional deterministic column vector 
expressing the effect of the actions, and $\dot{\ep}_j$ is an random variable of the error. 
The star $\star$ denotes the matrix transpose. 
%

Problematic in applications of classical inferential theories is that the experimental design is randomly time-varying in analysis of bandit algorithms. 
It is then quite easy to fall in a pitfall by using ad hoc estimators. 
As Hadad et al. \cite{hadad2020} commented it with a simple example, 
a seemingly natural estimator of the treatment effect can lead to biased estimation, due to randomness of the weighs. 
This issue is also the case in the batched bandit to be treated  in this paper.
For the batched bandit model (\ref{0311180317}), 
Zhang et al. \cite{zhang2020inference} proved that asymptotic normality can break with standard estimators such as the ordinary least squares estimator (OLS). 
There, conditional asymptotic normality occurs at each stage thanks to the divergence of conditional information, however, 
the standard estimator integrating these normalities results in a mixture of normal distributions not straightforwardly tractable. 
Zhang et al. \cite{zhang2020inference} proposed a batch wise studentized statistic (batched OLS, BOLS)
and resolved the problem by proving its asymptotic normality. 

The central limit theorem offers a universal method in asymptotic analysis thanks to the invariance principle, that is, 
the asymptotic variance of the variable specifies the approximation. 
On the other hand, it is well known that the precision of normal approximation is not always satisfactory especially for small samples. 
Small sample is a common issue in clinical trials whatever the experimental design, but even when a middle sample size can be assumed, 
in the multi-stages batched bandit, imbalance between assignments to different arms can cause few observations to some arm. 
The difference between the p-values $0.04$ and $0.06$ is serious in clinical trials crossing the valley of death. 
In this paper, we adopt an asymptotic expansion method and try to achieves more precise approximation to the distribution of the test statistic. 
The asymptotic expansion is a natural higher-order extension of the invariance principle since it uses first several cumulants of the data, 
that are nonparametrically estimable from the data. 
Robustness is important in applications. 
It is then a big advantage that the Edgeworth expansion does not need precise information about the distribution of the error terms, 
compared with other strongly model-dependent methods such as the saddle point approximation and Monte Carlo methods. 
The numerical examples in Section \ref{0411080600} show how strongly the distribution of the test statistic is affected by the distribution of the noise, 
as well as how much the approximation is improved by the asymptotic expansion. 

In the theory of asymptotic expansion, 
various methodologies have been developed for different dependencies: Bhattacharya and Rao for independent observations \cite{bhattacharya2010normal}, 
G\"otze and Hipp \cite{GotzeHipp1983, GotzeHipp1994} and Kusuoka and Yoshida \cite{KusuokaYoshida2000} for mixing processes, 
and Mykland \cite{Mykland1992, Mykland1993} and Yoshida \cite{Yoshida1997, yoshida2001malliavin, yoshida2016asymptotic} for martingales as well. 
Even if the treatment effects are independent conditionally on them, the assignments are depending on the history, 
as a result, the characteristics specifying the asymptotic expansion become random. 
Thus, the resulting limit distribution in general becomes a complicated mixture of conditionally Gaussian distributions. 
Yoshida \cite{yoshida2004partial} proposed 
an asymptotic expansion scheme based on partial mixing and applied it to 
higher-order approximation of the distribution of an additive functional of a partially mixing $\ep$-Markov process such as 
a jump-diffusion process 
in the random environment. 
Since the outcomes at the present stage determine the environment of the next stage in the batched bandit, 
each stage has the structure of the partial mixing. 
In this paper, 
with this formulation, we introduce a backwards asymptotic expansion formula and assess the backward propagation of errors. 
As the martingale central limit theorem was used in Zhang et al. \cite{zhang2020inference}, 
it is possible to apply the martingale expansion mentioned above. 
However, by nature of the model we consider, the approach by the partial mixing is natural and more effective since it gives any order of expansion 
without specifying the higher-order structure of the associated martingale.
{\fblue A sophisticated random weighting like the batched OLS of Zhang et al. \cite{zhang2020inference} gives asymptotic normality of the estimator. 
However, as they showed asymptotic non-normality of other estimators, the batched bandit involves a structure of the so-called non-ergodic statistics 
(cf. Basawa and Scott \cite{BasawaScott1983}) 
in that the random design caused by the outcomes in the previous layer produces a random mixture of normal distributions in the present layer. 
Modern typical examples of limit theorems stem from the inference for volatility of stochastic processes by high frequency data under finite time horizon. 
There, infinitely many stages appear along the time axis and infinite number of CLTs each of which occurs in an infinitesimal time interval are collected to make a random mixture of normal distributions. Such limit theorems are in the same direction as those of this paper, conceptually. 
Recently theories of asymptotic expansion have been developed and applied: 
Yoshida \cite{Yoshida2013ME,yoshida2020asymptotic}\footnote{\cite{Yoshida2013ME} updated by arXiv:1210.3680v3}, 
Podolskij and Yoshida \cite{podolskij2016edgeworth}, 
Podolskij et al. \cite{podolskij2017edgeworth,podolskij2018edgeworth}, 
Nualart and Yoshida \cite{nualart2019asymptotic}, 
Yamagishi and Yoshida \cite{yamagishi2022order}. 
}

Before we specifically get into constructing the theory, we would like to map our work onto the general literature of statistical inference for adaptively collected data. Our method here fits in the class of estimation methods using asymptotic approximations.  In classical literature of off-policy evaluation, 
the commonly used direct method (DM), which uses a regression based estimator to  predict outcomes based on historical data, are often implemented, leading to biased results, as illustrated in, for example, Villar et al. \cite{villar2015multi}, Nie et al. \cite{nie2018adaptively}, 
and Shin et al.\cite{shin2019sample, shin2020conditional}. 
As an unbiased estimator, 
the inverse probability weighted (IPW) estimator that weights each arm inversely with the probability of assignment to that specific arm is unbiased (cf. Horvitz \cite{horvitz1952generalization}), but often leads to high variance, eschewing asymptotic normality. 
A combination of the previous two methods is the augmented inverse probability weighted (AIPW) estimator proposed by Van der Laan and Lendle \cite{van2014online}, which in addition to the IPW term, adds a term that is an inverse weighted error of the prediction based on the DM. 
It is unbiased and also reduces the variance relative to the IPW estimator. 
When the assignment probability is estimated, the analogue estimator of AIPW is termed the doubly robust (DR) estimator (Robins et al. \cite{robins1994estimation}, Dudik et al. \cite{dudik2011doubly})     

Despite its improvement over the IPW estimator, in the case of bandit algorithms which stress aggressive learning in order to minimize regret, the inferior arm's assignment probability rapidly decreases, the variance of AIPW estimators still remains too high (Hadad et al. \cite{hadad2020}).
Given such difficulties, in recent years, there has been research to use more variance stabilizing weights as Luedtke and Van der Laan \cite{luedtke2016statistical,luedtke2018parametric} proposed an inverse standard deviation weighted average. 
Recently, Hadad et al. \cite{hadad2020} and Zhan et al. \cite{ Zhan2021off} extended their work, 
proposing an inverse variance weighted estimator in order to minimize the variance of the estimator.  A slightly different adaptive weighting scheme 
has been proposed by using debiasing techniques that are common in high dimensional statistics. 
Deshpande et al. \cite{deshpande2018} proposed a W-decorrelated estimator using the serial correlation between arm estimates to construct adaptive weights producing asymptotically normal guarantees, with recent extensions by Khamaru et al. \cite{Khamaru2022near} sharpening the bounds.

An alternative approach for constructing confidence intervals is the one based on using concentration inequalities to construct high probability confidence intervals often based on martingale structure (Ramdas et al. \cite{ramdas2020admissible}), 
with classical results like the mixture-martingale (Robbins \cite{robbins1970statistical}),  and recent extensions like Howard et al. \cite{Howard2021} and 
Waudby-Smith and Ramdas \cite{waudby2020estimating} that build on those past research. 
Although these methods often provide anytime-valid-type I error being controlled at a user-specified level 
and can be used for random stopping time without placing assumptions on the assignment process, they require assumptions on the potential outcomes of the rewards, like support or an upper bound on their variance, 
and  they may be conservative. Past simulation results have shown how normal approximation based methods 
while having weaker theoretical guarantees does a relatively good job in balancing the type I error and power, both of which are essential for hypothesis testing in practice (Zhang et al. \cite{zhang2020inference}).  
Nevertheless, by observing how each approach places distinct assumptions which produce different guarantees, we think that they are mostly complementary to the asymptotic approximation based methods. 

The organization of this paper is as follows. 
In Section \ref{0411101358}, apart from the specific model (\ref{0311180317}), we reformulate the problem in a more general setting, 
and consider a backward recursion formula to compute a target expectation. 
Our formulation is general and applicable to nonlinearly parametrized differentiable statistical models having an adaptive design. 
To implement this formula,  in Section \ref{0404190428}, we introduce a backward approximation scheme 
and evaluate the backward propagation of errors of the sequence of approximations. 
The proposed backward scheme is defined in Section \ref{0411110511} by Formula (\ref{0311211418b}) 
involving a generic signed measure $\Psi^n_{s,\sfp,{\bf w}^n_s}$. 
The sections thereafter are devoted to deriving a conditional asymptotic expansion as $\Psi^n_{s,\sfp,{\bf w}^n_s}$ and estimation of the error. 
To accommodate the adaptive designs, Section \ref{03112240118} reconstructs the theory of asymptotic expansion under partial mixing. 
The main analytic results of this paper are given in Sections \ref{0311240905} and \ref{0411110433}. 
The backward approximation scheme (\ref{0311211418b}) is combined with the asymptotic expansion of Section \ref{03112240118}, 
and applied to a sequentially partially mixing process in Section \ref{0311240905}. 
%
The asymptotic expansion formulas are more concretely embodied in Section \ref{0411110433} 
for linear and nonlinear statistics of hierarchically conditionally i.i.d. sequences. 
In particular, Section \ref{0404200515} presents an expansion formula for nonlinear functionals, which is applied to Section \ref{0411101429} 
treating the batched OLS estimator for the reward model (\ref{0311180317}) when the variance of the error terms is unknown. 
Section \ref{0411110634} collects proofs of the results in Section \ref{03112240118}.

\section{A model of bandit algorithms and a backward recursion formula}\label{0411101358}
\subsection{A model of bandit algorithms}\label{0404101857}
{\fblue Let $\A^n_{s}$ be an random element for $s\in\bbS$ taking values in a measurable space ${\mathfrak A}^n_s$.} 
Let ${\bm \ep}^n_{s}=(\ep_j;\>j\in\bbJ_s^n)$ for $s\in\bbS$, where 
{\sblue $\ep_j$ is an $\sfr_s$-dimensional random vector. 
} 
For $j\in\bbJ^n_s$, let 
$W^n_j$ be a {\sblue $\sfd_s\times\sfr_s$ random matrix} 
measurable with respect to $\sigma[\A^n_{s(j)}]$.
Write $\W^n_s=(W^n_j)_{j\in\bbJ^n_s}$. 
{\fred Apart from the reward model (\ref{0311180317}), more generally we}
consider a weighted sum 
\bea\label{0312071101}
\bbZ^n_s &=& \sum_{j\in\bbJ^n_s}W^n_j\ep_j. 
\eea
%
{\fred For example, $\ep_j=(\dot{\ep}_j,\dot{\ep}_j^2-\sigma^2_{s(j)})^\star$ for the model (\ref{0311180317}) in Section \ref{0404200515}, 
though we do not assume (\ref{0311180317}) 
nor use $R_j$ hereafter 
unless otherwise mentioned.}
{\fblue Moreover, the variable $\A^n_s$ is abstract and no longer a variable describing the assignments $(A_j;\>j\in\bbJ_s^n)$ in State $s$ 
as in the model (\ref{0311180317}). 
See Remark \ref{0412010652}.}
\begin{en-text}
$\bbZ_n=\sum_{t\in\bbT}\bbZ^n_s$, i.e., 
\bea
\bbZ_n &=& \sum_{j\in\bbJ^n}W^n_j\ep_j
\eea
\end{en-text}
{\fred 
Let $\call_s$ be a measurable space for $s\in\bbS$. 
We consider a $\sigma[\A^n_s]$-measurable random map $L^n_s:\Omega\to\call_s$ for every $(n,s)\in\bbN\times\bbS$. 
We write $\ul{L}_s^n=(L_1^n,...,L_s^n)$ for $(L_s)_{s\in\bbS}$. 
This operation by the underline $\ul{\>\cdot\>}_s$ will apply to other vectors.
In the batched bandits, the strategy (the distribute of $\A^n_s$) is determined 
by information up to Stage $s-1$, e.g.,
the average effects of actions at Stage $s-1$. 
The variables $\ul{L}^n_{s-1}$ and $\ul{Z}^n_{s-1}$ will be used for making a criterion for selection of a treatment at Stage $s$, as follows. 
A distribution labeled by $c_s$ is chosen from a measurable set $\mfkC_s$ 
for each $s\in\bbT$ following the probability distribution $q\big((\ul{L}^n_{s-1},\ul{\bbZ}^n_{s-1}),dc_s\big)$ on $\mfkC_s$.
This formulation is natural since it is common in statistical decision theory to consider randomized decision rules. 
Note that $q(\cdot,\cdot)$ depends on $s$. 
We assume that $\#\mfkC_1=1$ and $q(\emptyset,dc_1)$ is the Dirac measure on a singleton $c_1\in\mfkC_1$. 
}
\begin{en-text}
, and that 
the choice $c_s$ is equivalent to 
\beas 
({\sred\ul{L}^n_{s-1},\ul{\bbZ}^n_{s-1}})\in B(c_s)
\eeas
for some measurable set $B(c_s)$ in $\big(\Pi_{s'=1}^s\call_{s'}\big)\times\big(\Pi_{s'=1}^s\bbR^{\sfd_{s'}}\big)$. 
Since some $c_s$ is chosen at Stage $s$, 
\bea\label{0311190741} 
\sum_{c_s\in\mfkC_s}1_{\big\{({\sred\ul{L}^n_{s-1},\ul{\bbZ}^n_{s-1}})\in B(c_s)\big\}}
&=& 
1\quad a.s.
\eea
%
%
From (\ref{0311190741}), 
\bea\label{0402090414}
1_{\big\{({\sred\ul{L}^n_{s-1},\ul{\bbZ}^n_{s-1}})\in B(k)\big\}}
1_{\big\{({\sred\ul{L}^n_{s-1},\ul{\bbZ}^n_{s-1}})\in B(k')\big\}}
&=&
1_{\{k=k'\}}
1_{\big\{({\sred\ul{L}^n_{s-1},\ul{\bbZ}^n_{s-1}})\in B(k)\big\}}
\quad a.s.
\eea
for $k,k'\in\mfkC_s$. 
\end{en-text}
\begin{en-text}
We assume $\#\mfkC_1=1$ and set 
$\big\{\omega\in\Omega;\>({\sred\ul{L}^n_{0},\ul{\bbZ}^n_{0}})\in B(c_1)\big\}=\Omega$ 
for convenience. 
In particular, (\ref{0311190741}) is satisfied for all $s\in\bbS$. 
Furthermore, we set 
$B(c_{S+1})$ as the whole space {\sred and $\#\mfkC_{S+1}=1$,} therefore 
$\big\{\omega\in\Omega;\>(L^n_{S},\bbZ^n_{S})\in B(c_{S+1})\big\}=\Omega$. 
\end{en-text}
{\tred 
In Section \ref{0404200515}, 
we will treat a linear combination of some basic random variables associated to $\bbZ_s^n$, with 
$\sigma[{\bf A}^n_s]$-measurable random weights 
incorporated into the variable $L^n_s$; see (\ref{0404200852}). 
So, maybe having two components, $L^n_s$ plays two roles: 
defining assignment mechanism and integrating basic statistics. 
}
\begin{en-text}
{\sred 
For each $(n,t)\in\bbN\times\bbT$, we consider a $\sigma[\A^n_{t}]$-measurable random variable $V^n_s:\Omega\to\calv_t$ 
for a measurable space $\calv_t$. 
}
\end{en-text}
The history is recorded by the $\sigma$-fields 
$\calh^n_s=\sigma[\A^n_{s'},{\bm\ep}_{s'}\>s'\leq s]$ for $s\in\{0\}\cup\bbS$ 
with the trivial $\sigma$-field $\calh^n_0$. 
An often used assumption is the independency between 
${\bm\ep}^n_s$ and $\calh^n_{s-1}\vee\sigma[\A^n_{s}]$, i.e., 
\bea\label{0311190625}
{\bm\ep}^n_s\indep \big(\calh^n_{s-1}\vee\sigma[\A^n_{s}]\big)
\quad(s\in\bbS).
\eea
Condition (\ref{0311190625}) is standard one but we can relax this condition for our use. 
Let 
\beas 
\calg^n_s=\sigma\big[L^n_{s'},\bbZ^n_{s'};\>s'\leq s\big]
\eeas
for $s\in\bbS$, and $\calg^n_0$ the trivial $\sigma$-field. 
Instead of (\ref{0311190625}), we assume the condition 
\bea\label{0311191900}
{\bm\ep}^n_s\indep \big(\calg^n_{s-1}\vee\sigma[\A^n_{s}]\big)
\quad(s\in\bbS)
\eea
unless otherwise stated. 
The conditional expectation $E[\>\cdot\>|\calg^n_{s-1}]$ is denoted by $E_{s-1}[\>\cdot\>]$, 
and the conditional expectation $E\big[\>\cdot\>|\calg^n_{s-1}\vee\sigma[\A^n_s]\big]$ 
by $E_{s-1,\A^n_s}[\>\cdot\>]$. 

For ${\bf w}^n_s=(w^n_j)_{j\in\bbJ^n_s}\in\bbR^{\sfd_s \sfr_sn_s}$, let  
\bea\label{0311191302} 
P^n_s({\bf w}^n_s,dz_s)
&=& 
P^{\sum_{j\in\bbJ^n_s}w^n_j\ep_j}(dz_s), 
\eea
the distribution of $\sum_{j\in\bbJ^n_s}w^n_j\ep_j$. 
In other words, $P^n_s(\W^n_s,dz_s)$ is a regular conditional probability of $\bbZ^n_s$ 
given $\W^n_s$ becuase of (\ref{0311191900}). 

{\fred 
We assume 
\bea\label{0311200515}
P^{\A^n_s}\big(da_s|\calg^n_{s-1}\big)
&=& 
\int_{\mfkC_s}q\big((\underline{L}^n_{s-1},\underline{\bbZ}^n_{s-1}),dc_s)
\eta_{c_s}(da_s)
\eea
for some probability distributions $\eta_{c_s}$ ($c_s\in\mfkC_s$) 
for a regular conditional distribution $P^{\A^n_s}\big(da_s|\calg^n_{s-1}\big)$ 
of $\A^n_s$ given $\calg^n_{s-1}$; we implicitly assume that ${\mathfrak A}^n_s$ is standard. 
Since $L^n_s$ and $\W^n_s$ are $\sigma[\A^n_s]$-measurable, 
the assumption (\ref{0311200515}) yields a representation 
of a regular conditional distribution of $(L^n_s,\W^n_s)$ as 
\bea\label{0311200519}
P^{(L^n_s,\W^n_s)}(dl_s,d{\bf w}^n_s|\calg^n_{s-1})
&=& 
\int_{\mfkC_s}q\big((\underline{L}^n_{s-1},\underline{\bbZ}^n_{s-1}),dc_s)
\nu^n_{c_s}(dl_s,d{\bf w}^n_s)
\eea
for the probability distributions $\nu^n_{c_s}=\eta_{c_s}^{(L^n_s,\W^n_s)}$, the induced measure from $\eta_{c_s}$ by $(L^n_s,\W^n_s)$. 
}

{\fred Let $\call=\Pi_{s\in\bbS}\call_s$ and 
$\sfd=\sum_{s\in\bbS}\sfd_s$.}
{\tred 
Consider an $\sfm_s$-dimensional random variable
\bea\label{0404200532}
\bbY^n_s
&=& 
Y^n_s(L^n_s,\bbZ^n_s)
\eea
for a measurable map $Y^n_s:\call_s\times\bbR^{\sfd_s}\to\bbR^{\sfm_s}$. 
{\fred 
Thanks to the generality of the functional $\bbY^n_s$ of (\ref{0404200532}), 
our scheme is applicable to nonlinearly parametrized differentiable statistical models having an adaptive design.}
Equation (\ref{0404231626}) of Section \ref{0404200515} gives an example of the function $Y^n_s$ as the so-called 
Bhattacharya-Ghosh map, that was used in Bhattacharya and Ghosh \cite{BhattacharyaGhosh1978} 
to validate the formal Edgeworth expansion for various statistics. 
}
Given a measurable function 
\bea\label{0404242304}
f:{\sred\call\times}{\tred\bbR^{\sfm}}\to\bbR, 
\eea
{\tred $\sfm=\sum_{s\in\bbS}\sfm_s$,} 
the aim of this article is to compute the value of the expectation 
\bea\label{0311191157}
\ol{{\bf E}}_n &=& 
E\big[f\big({\sred\ul{L}_S^n,{\tred\ul{\bbY}^n_S}}\big)
\big], 
\eea
suppose that $f\big({\sred\ul{L}_S^n,{\tred\ul{\bbY}^n_S}}\big)$ is integrable. 
{\fblue 
\begin{remark}\label{0412010652}\rm
In the abstract setting of this section, the variable $\A^n_{s}$ is just playing a role for specifying the measurability of other quantitative variables, 
through the $\sigma$-field generated by it. 
Thus, it is possible to work with an abstract $\sigma$-field $\cald^n_s$ for $\sigma[\A^n_{s}]$ without introducing the variable $\A^n_{s}$. 
This aspect is important because, according to it, our model is already possible to incorporate time-dependent covariates and also the history of variables. 
The variable $\A^n_{s}$ has the ability to represent more information than the assignments. 
\end{remark}

An example of the measure $q$ is the $\ep$-Greedy algorithm having $\mfkC_s=\{c_s^{(1)},c_s^{(2)}\}$ with
\bea\label{0412030158}
q\big((\ul{l}_{s-1},\ul{z}_{s-1}),dc_{s}\big)
&=& 
1_{\{{\tt h}_{s-1}(\ul{l}_{s-1},\ul{z}_{s-1})\geq0\}}\delta_{c_s^{(1)}}(dc_s)
+1_{\{{\tt h}_{s-1}(\ul{l}_{s-1},\ul{z}_{s-1})<0\}}\delta_{c_s^{(2)}}(dc_s)
\eea
for some function ${\tt h}_{s-1}$. 
Multi-armed bandits can be treated similarly. 
On the other hand, the Thompson sampling is realized as 
\bea\label{0408251130}
q\big((\ul{l}_{s-1},\ul{z}_{s-1}),dc_{s}\big)
&=& 
1_{\{{\tt h}_{s-1}(\ul{l}_{s-1},\ul{z}_{s-1})\leq a_{s-1}^{(1)}\}}\delta_{c_s^{(1)}}(dc_s)
+1_{\{{\tt h}_{s-1}(\ul{l}_{s-1},\ul{z}_{s-1})>a_{s-1}^{(2)}\}}\delta_{c_s^{(2)}}(dc_s)
\nn\\&&
+1_{\{a_{s-1}^{(1)}<{\tt h}_{s-1}(\ul{l}_{s-1},\ul{z}_{s-1})\leq a_{s-1}^{(2)}\}}\delta_{C_s(\ul{l}_{s-1},\ul{z}_{s-1})}(dc_s)
\eea
with some constants $a_{s-1}^{(1)},a_{s-1}^{(2)}$ and some $\mfkC_s$-valued function $C_s$ of $(\ul{l}_{s-1},\ul{z}_{s-1})$. 
Obviously, the right-hand side of (\ref{0408251130}) can be written as 
$\delta_{C_s'(\ul{l}_{s-1},\ul{z}_{s-1})}(dc_s)$ with a $\mfkC_s$-valued function $C_s'$ of $(\ul{l}_{s-1},\ul{z}_{s-1})$ defined by 
\beas 
C_s'(\ul{l}_{s-1},\ul{z}_{s-1})
&=&
\left\{\begin{array}{cl}
c_s^{(1)}&({\tt h}_{s-1}(\ul{l}_{s-1},\ul{z}_{s-1})\leq a_1)\y
C_s(\ul{l}_{s-1},\ul{z}_{s-1})&(a_{s-1}^{(1)}<{\tt h}_{s-1}(\ul{l}_{s-1},\ul{z}_{s-1})\leq a_{s-1}^{(2)})\y
c_s^{(2)}&({\tt h}_{s-1}(\ul{l}_{s-1},\ul{z}_{s-1})>a_s^{(2)}). 
\end{array}\right.
\eeas
}
\subsection{A backward recursion formula}
{\fblue
We propose a backward recursion scheme to calculate (\ref{0311191157}). 
A basic idea is to use the backward shift operator $\B_{s-1}^n$ for a measurable function $g:\prod_{s'=1}^s\call_{s'}\times\prod_{s'=1}^s\bbR^{\sfd_{s'}}$ 
for $s\in\bbS$ and $n\in\bbN$, defined by 
\bea\label{0311251723} &&
(\B_{s-1}^ng)(\ul{l}_{s-1},\ul{z}_{s-1})
\nn\\&=&
E\bigg[ \int_{\mfkC_{s}}\int_{\call_s\times{\fblue\bbR^{\sfd_s{\sfr_sn_s}}}}
g\big(\ul{l}_{s-1},l_s,\ul{z}_{s-1},\sum_{j\in\bbJ^n_s}w^n_j\ep_j\big)
1_{\{l_s\in\Lambda^n_{s}\}}
\nu^n_{c_s}(dl_s,d{\bf w}^n_s)
\nn\\&&\hspace{250pt}\times
q\big((\ul{l}_{s-1},\ul{z}_{s-1}),dc_{s}\big)
\bigg]
\nn\\&=&
\int_{\mfkC_{s}}
\int_{\call_s{\fblue\times\bbR^{\sfd_s{\sfr_sn_s}}}\times\bbR^{\sfd_s}} 
g\big(\ul{l}_{s-1},l_s,\ul{z}_{s-1},z_s\big)
P^n_s({\bf w}^n_s,dz_s)
1_{\{l_s\in\Lambda^n_{s}\}}
\nu^n_{c_s}(dl_s,d{\bf w}^n_s)
\nn\\&&\hspace{250pt}\times
q\big((\ul{l}_{s-1},\ul{z}_{s-1}),dc_{s}\big)
\eea
if the integral exists, 
where $\Lambda^n_s$ is a measurable set in $\call_s$. 
The operator $\B_{s-1}^n$ maps a function of $(\ul{l}_s,\ul{z}_s)$ to a function of $(\ul{l}_{s-1},\ul{z}_{s-1})$. 
}

{\fblue
We have the following simple formula. 
\begin{lemma}\label{0311190724} Assume (\ref{0311191900}). Let $s\in\bbS$ and $n\in\bbN$. 
Let $g:\prod_{s'=1}^s\call_{s'}\times\prod_{s'=1}^s\bbR^{\sfd_{s'}}\to\bbR$ be a measurable function. 
Then
\bea\label{0311180730} 
E_{s-1}\bigg[g\big({\sred\ul{L}_s^n,\>}\ul{\bbZ}^n_s\big)
\prod_{s'=1}^s1_{\{L^n_{s'}\in\Lambda^n_{s'}\}}
\bigg]
&=&
(\B_{s-1}^ng)({\sred\ul{L}_{s-1}^n,\>}\ul{\bbZ}^n_{s-1})
\prod_{s'=1}^{s-1}1_{\{L^n_{s'}\in\Lambda^n_{s'}\}}
\quad a.s.,
\eea
suppose that $g\big({\sred\ul{L}_s^n,\>}\ul{\bbZ}^n_s\big)\prod_{s'=1}^s1_{\{L^n_{s'}\in\Lambda^n_{s'}\}}$ is integrable. 
In particular, the random variable on the right-hand side of (\ref{0311180730}) is integrable. 
Here the product $\prod_{s'=1}^0$ reads $1$. 
\end{lemma}
\proof 
Suppose that $g$ is bounded. 
Then we have 
\bea\label{0311180727} &&
E_{s-1}\bigg[g\big({\sred\ul{L}_s^n,\>}\ul{\bbZ}^n_s\big)
1_{\{L^n_s\in\Lambda^n_{s}\}}\bigg]
\nn\\&=&
E_{s-1}\bigg[E_{s-1,\A^n_s}\bigg[
g\big({\sred\ul{L}_s^n,\>}\ul{\bbZ}^n_{s-1},\sum_{j\in\bbJ^n_s}W^n_j\ep_j\big)
\bigg]
1_{\{L^n_s\in\Lambda^n_{s}\}}\bigg]
\nn\\&=&
E_{s-1}\bigg[E\bigg[
g\big({\sred\ul{l}_s,\>}\ul{z}_{s-1},\sum_{j\in\bbJ^n_s}w^n_j\ep_j\big)
\bigg]\bigg|_{
\ul{z}_{s-1}=\ul{\bbZ}^n_{s-1}
\atop {{\bf w}^n_s=\W^n_s
\atop {\sred\ul{l}_s=\ul{L}^n_s}
}}
1_{\{L^n_s\in\Lambda^n_{s}\}}\bigg]
\quad(\because(\ref{0311191900}))
\nn\\&=&
\int E\bigg[
g\big({\sred\ul{l}_{s-1},l_s,\>}\ul{z}_{s-1},\sum_{j\in\bbJ^n_s}w^n_j\ep_j\big)
\bigg]
1_{\{l_s\in\Lambda^n_{s}\}}
\nn\\&&\hspace{100pt}\times
\nu^n_{c_s}(dl_s,d{\bf w}^n_s)
q\big((\ul{l}_{s-1},\ul{z}_{s-1}),dc_s\big)
\bigg|_{
\ul{z}^n_{s-1}=\ul{\bbZ}^n_{s-1}
\atop {\sred \ul{l}_{s-1}=\ul{L}_{s-1}^n}
}
\nn\\&=&
(\B^n_{s-1}g)(\ul{l}_{s-1},\ul{z}_{s-1})\big|_{
\ul{z}^n_{s-1}=\ul{\bbZ}^n_{s-1}
\atop {\sred \ul{l}_{s-1}=\ul{L}_{s-1}^n}}
\eea
by (\ref{0311200519}). 
Therefore we obtain (\ref{0311180730}). 
For a general $g$, we may assume that $g$ is nonnegative. 
For $K>0$, the equality (\ref{0311180730}) is valid for $g\wedge K$ in place of $g$. 
When $K\up\infty$, the left-hand side of (\ref{0311180730}) for $g\wedge K$ converges to that for $g$ a.s. (and in $L^1$). 
The right-hand side of (\ref{0311180730}) for $\B^n_{s-1}(g\wedge K)$ converges to that for $\B^n_{s-1}g$, which is by the point-wise convergence 
of $\B^n_{s-1}(g\wedge K)\to\B^n_{s-1}g$. 
\begin{en-text}
\bea
&=&
E_{s-1}\bigg[E\bigg[
f^n_{s,c}\big(\ul{l}_s,\ul{z}^n_{s-1},\sum_{j\in\bbJ^n_s}w^n_j\ep_j\big)
1_{\big\{(l_{s},\sum_{j\in\bbJ^n_s}w^n_j\ep_j)\in B(c_{s+1})\big\}}
\bigg]\bigg|_{
\ul{z}^n_{s-1}=\ul{\bbZ}^n_{s-1}}
1_{\{l_s\in\Lambda^n_{s,c}\}}
\nn\\&&\hspace{100pt}\times
\nu^n_{c_s}(dl_s,d{\bf w}^n_s)1_{\big\{(L^n_{s-1},\bbZ^n_{s-1})\in B(c_s)\big\}}
\bigg]
\nn\\&=&\koko
E_{s-1}\bigg[
\int 
f^n_{s,c}\big(\ul{l}_s,\ul{z}^n_{s-1},z_s\big)P^n_s({\bf w}^n_s,dz_s)
\bigg]\bigg|_{
l_s=L^n_s,\>
\ul{z}^n_{s-1}=\ul{\bbZ}^n_{s-1}
\atop {\bf w}^n_s=\W^n_s}
1_{\{L^n_s\in\Lambda^n_{s,c}\}}\bigg]
\nn\\&=&
E_{s-1}\bigg[
\int 
f^n_{s,c}\big(\ul{L}^n_s,\ul{\bbZ}^n_{s-1},z_s\big)P^n_s(\W^n_s,dz_s)
1_{\{L^n_s\in\Lambda^n_{s,c}\}}\bigg]
\eea
\end{en-text}
\begin{en-text}
With (\ref{0311180727}), we obtain 
\beas 
&&
E_{s-1}\bigg[f^n_{s,c}\big({\sred\ul{L}_s^n,\>}\ul{\bbZ}^n_s\big)
\prod_{s'=1}^{s}1_{\big\{({\sred\ul{L}^n_{s'},\ul{\bbZ}^n_{s'}})\in B(c_{s'+1})\big\}}
\prod_{s'=1}^{s}1_{\{L^n_{s'}\in\Lambda^n_{s',c}\}}
\bigg]
\nn\\&=&
E_{s-1}\bigg[f^n_{s,c}\big({\sred\ul{L}_s^n,\>}\ul{\bbZ}^n_{s}\big)
1_{\big\{({\sred\ul{L}^n_{s},\ul{\bbZ}^n_{s}})\in B(c_{s+1})\big\}}
1_{\{L^n_s\in\Lambda^n_{s,c}\}}
\bigg]
\prod_{s'=1}^{s-1}1_{\big\{({\sred\ul{L}^n_{s'},\ul{\bbZ}^n_{s'}})\in B(c_{s'+1})\big\}}
\prod_{s=1}^{s-1}1_{\{L^n_{s'}\in\Lambda^n_{s',c}\}}
\nn\\&=&
\int E\bigg[
f^n_{s,c}\big({\sred\ul{l}_{s-1},l_s,\>}\ul{z}_{s-1},\sum_{j\in\bbJ^n_s}w^n_j\ep_j\big)
1_{\big\{({\sred\ul{l}_{s-1},\>}l_{s},\sum_{j\in\bbJ^n_s}w^n_j\ep_j)\in B(c_{s+1})\big\}}
\bigg]\bigg|_{
\ul{z}_{s-1}=\ul{\bbZ}^n_{s-1}
\atop{\sred \ul{l}_{s-1}=\ul{L}_{s-1}^n}
}
1_{\{l_s\in\Lambda^n_{s,c}\}}
\nn\\&&\hspace{100pt}\times
\sum_{k\in\mfkC_s}\nu^n_{k}(dl_s,d{\bf w}^n_s)
1_{\big\{({\sred\ul{L}^n_{s-1},\ul{\bbZ}^n_{s-1}})\in B(k)\big\}}
\prod_{s'=1}^{s-1}1_{\big\{({\sred\ul{L}^n_{s'},\ul{\bbZ}^n_{s'}})\in B(c_{s'+1})\big\}}
\prod_{s'=1}^{s-1}1_{\{L^n_{s'}\in\Lambda^n_{s',c}\}}.
\eeas
{\sred Due to (\ref{0402090414}), the last expression is equal to}
\beas
&&
\int E\bigg[
f^n_{s,c}\big({\sred\ul{l}_{s-1},l_s,\>}\ul{z}_{s-1},\sum_{j\in\bbJ^n_s}w^n_j\ep_j\big)
1_{\big\{({\sred\ul{l}_{s-1},\>}l_{s},\sum_{j\in\bbJ^n_s}w^n_j\ep_j)\in B(c_{s+1})\big\}}
\bigg]\bigg|_{
\ul{z}_{s-1}=\ul{\bbZ}^n_{s-1}
\atop{\sred \ul{l}_{s-1}=\ul{L}_{s-1}^n}
}
1_{\{l_s\in\Lambda^n_{s,c}\}}
\nn\\&&\hspace{50pt}\times
\nu^n_{c_s}(dl_s,d{\bf w}^n_s)
1_{\big\{({\sred\ul{L}^n_{s-1},\ul{\bbZ}^n_{s-1}})\in B(c_s)\big\}}
\prod_{s'=1}^{s-1}1_{\big\{({\sred\ul{L}^n_{s'},\ul{\bbZ}^n_{s'}})\in B(c_{s'+1})\big\}}
\prod_{s'=1}^{s-1}1_{\{L^n_{s'}\in\Lambda^n_{s',c}\}}
\nn\\&=&
\int E\bigg[
f^n_{s,c}\big({\sred\ul{l}_{s-1},l_s,\>}\ul{z}_{s-1},\sum_{j\in\bbJ^n_s}w^n_j\ep_j\big)
1_{\big\{({\sred\ul{l}_{s-1},\>}l_{s},\sum_{j\in\bbJ^n_s}w^n_j\ep_j)\in B(c_{s+1})\big\}}
\bigg]
1_{\{l_s\in\Lambda^n_{s,c}\}}
\nu^n_{c_s}(dl_s,d{\bf w}^n_s)
\bigg|_{
\ul{z}_{s-1}=\ul{\bbZ}^n_{s-1}
\atop{\sred \ul{l}_{s-1}=\ul{L}_{s-1}^n}}
\nn\\&&\hspace{50pt}\times
\prod_{s'=1}^{s-1}1_{\big\{({\sred\ul{L}^n_{s'},\ul{\bbZ}^n_{s'}})\in B(c_{s'+1})\big\}}
\prod_{s'=1}^{s-1}1_{\{L^n_{s'}\in\Lambda^n_{s',c}\}}
\nn\\&=&
f^n_{s-1,c}({\sred\ul{L}_{s-1}^n,\>}\ul{\bbZ}^n_{s-1})
\prod_{s'=1}^{s-1}1_{\big\{({\sred\ul{L}^n_{s'},\ul{\bbZ}^n_{s'}})\in B(c_{s'+1})\big\}}
\prod_{s'=1}^{s-1}1_{\{L^n_{s'}\in\Lambda^n_{s',c}\}}
\eeas
\end{en-text}
\begin{en-text}
\beas
\nn\\&=&
E_{t-1}\bigg[
\int 
f^n_{t,c}\big(\ul{L}^n_s,\ul{\bbZ}^n_{t-1},z_t\big)1_{\{L^n_s\in\Lambda^n_{t,c}\}}P^n_s(\W^n_s,dz_t)
\bigg]
\prod_{s=1}^{t-1}1_{\big\{h_{s}(L^n_{s},\bbZ^n_{s})\in B(c_{s+1})\big\}}
\prod_{s=1}^{t-1}1_{\{L^n_s\in\Lambda^n_{s,c}\}}
\nn\\&=&
\int 
E_{t-1}\big[
f^n_{t,c}\big(\ul{L}^n_s,\ul{\bbZ}^n_{t-1},z_t\big)1_{\{L^n_s\in\Lambda^n_{t,c}\}}P^n_s(\W^n_s,dz_t)
\big]
\prod_{s=1}^{t-1}1_{\big\{h_{s}(L^n_{s},\bbZ^n_{s})\in B(c_{s+1})\big\}}
\prod_{s=1}^{t-1}1_{\{L^n_s\in\Lambda^n_{s,c}\}}
\nn\\&=&
f^n_{t-1,c}(\ul{L}_{t-1},\ul{\bbZ}^n_{t-1})
\prod_{s=1}^{t-1}1_{\big\{h_{s}(L^n_{s},\bbZ^n_{s})\in B(c_{s+1})\big\}}
\prod_{s=1}^{t-1}1_{\{L^n_s\in\Lambda^n_{s,c}\}}
\eeas
for $s\in\bbS$.
\end{en-text}
\qed\halflineskip
}
\begin{en-text}
Due to (\ref{0311190741}), 
\bea\label{0404200536}
\ol{{\bf E}}_n &=& \sum_{c\in\mfkC} \ol{{\bf E}}_{n,c},
\eea
where 
\bea\label{0311191123} 
\ol{{\bf E}}_{n,c} &=& 
E\bigg[f\big({\sred\ul{L}_S^n,{\tred\ul{\bbY}^n_S}}\big)
\prod_{s=1}^S1_{\big\{{\sred(\underline{L}^n_{s},\underline{\bbZ}^n_{s}})\in B(c_{s+1})\big\}}
\bigg]
\eea
and $c=(c_s)_{s\in\bbS}\in\mfkC=\prod_{s\in\bbS}\mfkC_s$. 
For $\ol{{\bf E}}_n$, it is sufficient to compute 
the value of $\ol{{\bf E}}_{n,c}$ for each $c\in\mfkC$. 
\end{en-text}
We will consider a modification of $\ol{{\bf E}}_{n}$ 
by a truncation $1_{\{L^n_s\in\Lambda^n_{s}\}}$ with a measurable set $\Lambda^n_{s}$ of $\call_s$ at each stage $s$. 
Let 
{\fred 
\bea\label{0311191836} 
{\bf E}_{n} &=& 
E\bigg[f\big({\sred\ul{L}_S^n,{\tred\ul{\bbY}^n_S}}\big)
\prod_{s=1}^S1_{\{L^n_s\in\Lambda^n_{s}\}}\bigg]. 
\eea
The choice of $\Lambda^n_{s}$ is important so that it will determine the accuracy of the approximation ${\bf E}_{n}$ 
to $\ol{{\bf E}}_{n}$.
{\fblue 
Since $L^n_s$ is only assumed to be $\sigma[\A^n_s]$-measurable, the indicator function $1_{\{L^n_s\in\Lambda^n_{s}\}}$ can select 
a good event depending on the assignment, and more generally on the covariates if $\sigma[\A^n_s]$ include their information.}

{\fred
Let 
{\fred 
\bea\label{0404100341}
f^n_{S}(\ul{l}_S,\ul{z}_S)
&=& 
f\big(\ul{l}_S,(Y^n_s(l_s,z_s))_{s\leq S}\big),
\eea
}
and under the integrability condition, define 
a function $f^n_{s-1}$ 
by 
{\fblue $f^n_{s-1}(\ul{l}_{s-1},\ul{z}_{s-1})=\B_{s-1}^nf^n_s$, i.e.,}
\bea\label{0311192029} &&
f^n_{s-1}(\ul{l}_{s-1},\ul{z}_{s-1})
\nn\\&=&
\int_{\mfkC_{s}}
\int_{\call_s{\fblue\times\bbR^{\sfd_s{\sfr_sn_s}}}\times\bbR^{\sfd_s}} 
f^n_{s}\big(\ul{l}_{s-1},l_s,\ul{z}_{s-1},z_s\big)
P^n_s({\bf w}^n_s,dz_s)
1_{\{l_s\in\Lambda^n_{s}\}}
\nu^n_{c_s}(dl_s,d{\bf w}^n_s)
\nn\\&&\hspace{250pt}\times
q\big((\ul{l}_{s-1},\ul{z}_{s-1}),dc_{s}\big)
\nn\\&&
\eea
for $s\in\bbS$.
}
In particular, the function $f^n_{0}$ is a constant given by 
\bea\label{0311191133} 
f^n_{0} 
&=& 
\int 
f^n_{1}\big({\sred l_1,\>}z_1\big)
P^n_1({\bf w}^n_1,dz_1)
1_{\{l_1\in\Lambda^n_{1}\}}
\nu^n_{c_1}(dl_1,d{\bf w}^n_1).
\eea
}

Lemma \ref{0311190724} gives a backwards approximation to $\ol{{\bf E}}_{n}$ 
of (\ref{0311191157}). 
\begin{theorem}\label{0311191131}
Suppose that $f\big({\sred\ul{L}_S^n,\ul{\bbZ}^n_S}\big)$ is integrable. Then 
\bd
\im[(a)] 
$
{\bf E}_{n} \yeq f^n_{0}$, 
where $f^n_{0}$ is given by (\ref{0311191133}) and 
the functions $f^n_{s}$ for {\fred$s\in\bbS$} are recursively given by {\fred (\ref{0404100341}) and}
(\ref{0311192029}). 
\im[(b)] 
$\ds
\ol{{\bf E}}_n \yeq 
E\big[f\big({\sred\ul{L}_S^n,{\tred\ul{\bbY}^n_S}}\big)
\big]
\yeq
{\bf E}_{n}+\rho_n
$ 
with 
\beas 
|\rho_n|
&\leq& 
E\bigg[\big|f\big({\sred\ul{L}_S^n,\ul{\bbZ}^n_S}\big)\big|
\sum_{s\in\bbS}1_{\{L^n_s\not\in\Lambda^n_{s}\}}\bigg]. 
\eeas
\ed
\end{theorem}
\begin{en-text}
{\sred 
\begin{remark}\rm 
It was possible to formally refrain from introducing the variables $V_t^n$ by 
incorporating $V_t^n$ into $L_t^n$. 
However, we kept $V_t^n$ because they have a different role from $L_t^n$ 
in the causal inference. 
\end{remark}
}
\end{en-text}

{\tred 
\section{
A backward approximation scheme and the backward propagation of errors
}\label{0404190428}
\subsection{Backward approximation}\label{0411110511}
Theorem \ref{0311191131} gave an approximation to the expected value 
$E\big[f\big(\ul{L}_S^n,\ul{\bbY}^n_S\big)\big]$ 
with a backwards formula for ${\bf E}_{n}$. 
The value of ${\bf E}_{n}$ is inductively 
computable by the formulas {\fred(\ref{0404100341})-(\ref{0311191133})} 
if we know 
the distributions $P^n_s({\bf w}^n_s,dz_s)$ $(s\in\bbS)$ completely. 
However, they rarely have a closed form, and hence we need some approximation 
to $P^n_s({\bf w}^n_s,dz_s)$ by a signed measure $\Psi^n_{s,\sfp,{\bf w}^n_s}$ on $\bbR^{\sfd_s}$ 
depending on ${\bf w}^n_s$. Asymptotic expansion method is promising in this situation. 
In fact, for batched bandits, in Section \ref{0311240905}, 
we will apply 
an expansion scheme under partial mixing (Yoshida \cite{yoshida2004partial}) 
to derive asymptotic expansion of ${\bf E}_{n}$. 
{\fred The notation $\Psi^n_{s,\sfp,{\bf w}^n_s}$ is taking in advance the index $\sfp$ determining the order of the asymptotic expansion.}
Before carrying out this plot, we shall prepare a framework of our method in {\fred Section \ref{0404190428}}.
}

Recall the formula (\ref{0311192029}): 
{\fred 
\beas
&&
f^n_{s-1}({\sred\ul{l}_{s-1},\>}\ul{z}_{s-1})
\nn\\&=&
\begin{en-text}
\int E\bigg[
f^n_{s,c}\big({\sred\ul{l}_{s-1},l_s,\>}\ul{z}_{s-1},\sum_{j\in\bbJ^n_s}w^n_j\ep_j\big)
1_{\big\{({\sred\ul{l}_{s-1},\>}l_{s},\sum_{j\in\bbJ^n_s}w^n_j\ep_j)\in B(c_{s+1})\big\}}
\bigg]
1_{\{l_s\in\Lambda^n_{s,c}\}}
\nu^n_{c_s}(dl_s,d{\bf w}^n_s)
\nn\\&=&
\end{en-text}
\int_{\mfkC_{s}}
\int_{\call_s{\fblue\times\bbR^{\sfd_s{\sfr_sn_s}}}\times\bbR^{\sfd_s}} 
f^n_{s}\big(\ul{l}_{s-1},l_s,\ul{z}_{s-1},z_s\big)
P^n_s({\bf w}^n_s,dz_s)
1_{\{l_s\in\Lambda^n_{s}\}}
\nu^n_{c_s}(dl_s,d{\bf w}^n_s)
\nn\\&&\hspace{250pt}\times
q\big((\ul{l}_{s-1},\ul{z}_{s-1}),dc_{s}\big)
\eeas
}
for $s\in\bbS$. 
We 
define $\wh{f}^n_{s}({\sred\ul{l}_{s},\>}\ul{z}_{s})$ by 
$\wh{f}^n_{S}={\tred f^n_{S}}$ {\tred for $f^n_{S}$ of (\ref{0404100341})} and 
{\fred 
\bea\label{0311211418b} 
\wh{f}^n_{s-1}({\sred\ul{l}_{s-1},\>}\ul{z}_{s-1})
&=&
\int_{\mfkC_{s}}
\int_{\call_s{\fblue\times\bbR^{\sfd_s{\sfr_sn_s}}}\times\bbR^{\sfd_s}} 
\wh{f}^n_{s}\big(\ul{l}_{s-1},l_s,\ul{z}_{s-1},z_s\big)
\Psi^n_{s,\sfp,{\bf w}^n_s}(dz_s)
1_{\{l_s\in\Lambda^n_{s}\}}
\nu^n_{c_s}(dl_s,d{\bf w}^n_s)
\nn\\&&\hspace{200pt}\times
q\big((\ul{l}_{s-1},\ul{z}_{s-1}),dc_{s}\big)
\eea
}
for $s\in\bbS$, 
if the integral (\ref{0311211418b}) exists. 
\begin{en-text}
{\tred Denote by ${\tred\ul{\cale}_t(M,\gamma)}$ the set of measurable functions 
$\sfg:\bbR^{\sum_{s=1}^t\sfd_s}\to\bbR$ such that 
\beas 
\big|\sfg(\ul{z}_t\big)\big| &\leq& M(1+|\ul{z}_t|^\gamma)
\qquad\big(\ul{z}_t\in\bbR^{\sum_{s=1}^t\sfd_s}\big). 
\eeas
Suppose that 
\beas
\big\{f^n_{S,c};\>\ul{l}_S\in\call\big\}&\subset&\ul{\cale}_S(M_S,\gamma)
\eeas
for some $M_S,\gamma>0$. 
}
\end{en-text}
\begin{en-text}
{\sred Denote by $\cale_s(M,\gamma)$ the set of measurable functions 
$f:\big(\Pi_{s'=1}^t\call_{s'}\big)\times\bbR^{\sum_{s'=1}^t\sfd_{s'}}\to\bbR$ such that 
\beas 
\big|f(\ul{l}_t,\ul{z}_t\big)\big| &\leq& M(1+|\ul{z}_t|^\gamma)
\qquad\big(\ul{l}_t\in\Pi_{s=1}^t\call_s, \>\ul{z}_t\in\bbR^{\sum_{s=1}^t\sfd_s}\big). 
\eeas
}
\end{en-text}

We will use a conditional asymptotic expansion as $\Psi^n_{s,\sfp,{\bf w}^n_s}$ and estimate the error. 
This plot is carried out in Sections \ref{0311240905} and \ref{0411110433}, 
after reconstruction of the theory of asymptotic expansion under partial mixing in 
Section \ref{03112240118} to accommodate the adaptive designs. 
We will combine the backward approximation scheme (\ref{0311211418b}) with the asymptotic expansion of Section \ref{03112240118}, 
and apply this result to a sequentially partial mixing process in Section \ref{0311240905}. 
Section \ref{0411110433} treats linear and nonlinear statistics of hierarchically conditionally i.i.d. sequences.

\subsection{Backward propagation of errors}\label{0411110512}
{\tred 
Some condition for integrability of the integrand is necessary to validate 
Formula (\ref{0311211418b}) 
at Stage $s$. 
Moreover, 
we will need to restrict the class of functions $f$ since 
the error bound of the asymptotic expansion depends on a measure-theoretic modulus of continuity of 
the function $f$. 
\begin{en-text}
(This is clear if one considers approximation of the distribution function of a binomial distribution 
by an Edgeworth expansion. The error cannot be smaller than the order of $n^{-1/2}$ 
with any continuous distribution function for approximation.)
\end{en-text}
{\tred Use of such a measure-theoretic modulus of continuity of the function $f$ 
is inevitable and standard in the usual theory of asymptotic expansion. 
The reader is referred to Bhattacharya and Rao \cite{bhattacharya2010normal} 
for a construction of the theory of asymptotic expansion for independent variables.}

{\fblue 
Let $r_0=r_0(\ul{z}_0)=1$ and 
$r_s(\ul{z}_s)$ a measurable function such that $0<r_s(\ul{z}_s)\leq1$ for all $\ul{z}_s$, for each $s\in\bbS$.
Two examples of $r_s(\ul{z}_s)$ are the constant $1$ and the function 
$\prod_{s'=1}^{{\fred s}}(1+|z_{s'}|)^{-\gamma}$.
However, we treat a generic function $r_s(\ul{z}_s)$.
Define $\wh{g}^n_s=(\wh{g}_f)^n_s$ by 
\beas 
\wh{g}^n_s(\ul{l}_{s},\ul{z}_{s}\big)\yeq (\wh{g}_f)^n_s(\ul{l}_{s},\ul{z}_{s}\big)
&=& 
\Pi_{s'=1}^{s}1_{\{l_{s'}\in\Lambda^n_{s'}\}}r_{s-1}(\ul{z}_{s-1})
\wh{f}^n_{s}\big(\ul{l}_{s-1},l_s,\ul{z}_{s-1},z_s\big). 
\eeas
In particular, $\wh{g}^n_S(\ul{l}_{S},\ul{z}_{S}\big)=\Pi_{s'=1}^{S}1_{\{l_{s'}\in\Lambda^n_{s'}\}}r_{S-1}(\ul{z}_{S-1})f^n_{s}\big(\ul{l}_{S},\ul{z}_{S}\big)$. 
Let 
\beas 
R^n_{s-1} 
&=& 
r_{s-1}\big(
\ul{\bbZ}^n_{s-1}\big)^{-1}\prod_{s'=1}^{s-1}1_{\{L^n_{s'}\in\Lambda^n_{s'}\}}
\eeas
for $s\in\bbS$. In particular, $R^n_0=1$.

For the meantime, suppose that 
\bea\label{0411261109}
E\bigg[\big|f^n_S\big({\sred\ul{L}_S^n,\>}\ul{\bbZ}^n_S\big)\big|\prod_{s=1}^S1_{\{L^n_{s}\in\Lambda^n_{s}\}}\bigg] &<& \infty
\eea
and 
\bea\label{0411261111}
E\bigg[
|\Psi^n_{s,\sfp,{\bf W}^n_s}|\big[\big|\wh{f}^n_s\big(\ul{L}^n_{s-1},L^n_s,\ul{\bbZ}^n_{s-1},\cdot\big)\big|\big]
\prod_{s'=1}^s1_{\{L_{s'}^n\in\Lambda^n_{s'}\}}\bigg] &<& \infty
\qquad(s\in\bbS), 
\eea
where $|\Psi^n_{s,\sfp,{\bf w}^n_s}|$ is the variation of $\Psi^n_{s,\sfp,{\bf w}^n_s}$. 
The condition (\ref{0411261111}) is equivalent to 
\bea\label{0411261111b}
E\bigg[R^n_{s-1} |\Psi^n_{s,\sfp,{\bf W}^n_s}|\big[\big|(\wh{g}_f)^n_s\big(\ul{L}^n_{s-1},L^n_s,\ul{\bbZ}^n_{s-1},\cdot\big)\big|\big]\bigg] &<& \infty
\qquad(s\in\bbS). 
\eea

As verified backwardly, under (\ref{0411261111}), $\wh{f}^n_{s-1}\big(\ul{L}_{s-1}^n,\ul{\bbZ}^n_{s-1}\big)$ 
is well defined a.s. on the event $\big\{\prod_{s'=1}^{s-1}1_{\{L^n_{s'}\in\Lambda^n_{s'}\}}=1\big\}$. 
Moreover, as already known from Lemma \ref{0311190724}, $f^n_{s-1}\big(\ul{L}_{s-1}^n,\ul{\bbZ}^n_{s-1}\big)$ 
is also well defined a.s. on the event $\big\{\prod_{s'=1}^{s-1}1_{\{L^n_{s'}\in\Lambda^n_{s'}\}}=1\big\}$. 
}

{\fblue 
Let 
\bea\label{0411261153}
\bbD^n_s(f)
&=&
\bigg|E\bigg[\wh{f}^n_{s}\big(\ul{L}^n_{s},\ul{\bbZ}^n_{s}\big)\prod_{s'=1}^{s}1_{\{L^n_{s'}\in\Lambda^n_{s'}\}}\bigg]
-E\bigg[\wh{f}^n_{s-1}\big(\ul{L}^n_{s-1},\ul{\bbZ}^n_{s-1}\big)\prod_{s'=1}^{s-1}1_{\{L^n_{s'}\in\Lambda^n_{s'}\}}\bigg]\bigg|. 
\eea
We then have
\beas 
\bbD^n_s(f)
&=&
\bigg|E\bigg[r_{s-1}\big(\ul{L}^n_{s-1},\ul{\bbZ}^n_{s-1}\big)^{-1}
E_{s-1}\bigg[
r_{s-1}\big(\ul{L}^n_{s-1},\ul{\bbZ}^n_{s-1}\big)\wh{f}^n_{s}\big(\ul{L}^n_{s},\ul{\bbZ}^n_{s}\big)\prod_{s'=1}^{s}1_{\{L^n_{s'}\in\Lambda^n_{s'}\}}\bigg]\bigg]
\nn\\&&
-E\bigg[\wh{f}^n_{s-1}\big(\ul{L}^n_{s-1},\ul{\bbZ}^n_{s-1}\big)\prod_{s'=1}^{s-1}1_{\{L^n_{s'}\in\Lambda^n_{s'}\}}\bigg]\bigg|
\nn\\&=&
\bigg|E\bigg[r_{s-1}\big(\ul{L}^n_{s-1},\ul{\bbZ}^n_{s-1}\big)^{-1}\prod_{s'=1}^{s-1}1_{\{L^n_{s'}\in\Lambda^n_{s'}\}}
\nn\\&&\hspace{30pt}\times
\bigg\{\big(\B^n_{s-1}(\Pi_{s'=1}^{s-1}1_{\{\cdot\in\Lambda^n_{s'}\}}r_{s-1}\wh{f}^n_s)\big)\big(\ul{L}^n_{s-1},\ul{\bbZ}^n_{s-1}\big)
\nn\\&&\hspace{60pt}
-\big(\Pi_{s'=1}^{s-1}1_{\{\cdot\in\Lambda^n_{s'}\}}r_{s-1}\wh{f}^n_{s-1}\big)\big(\ul{L}^n_{s-1},\ul{\bbZ}^n_{s-1}\big)\bigg\}\bigg]\bigg|
\eeas
by (\ref{0311180730}) of Lemma \ref{0311190724}. 
Consequently, 
\beas
\bbD^n_s(f)
&=&
\bigg|E\bigg[R^n_{s-1}
\triangle^n_{s}(f)\bigg]\bigg|
\eeas
for
\beas 
\triangle^n_{s}(f)
&=& 
\big(\B^n_{s-1}(\Pi_{s'=1}^{s-1}1_{\{\cdot\in\Lambda^n_{s'}\}}r_{s-1}\wh{f}^n_s)\big)\big(\ul{L}^n_{s-1},\ul{\bbZ}^n_{s-1}\big)
-\big(\Pi_{s'=1}^{s-1}1_{\{\cdot\in\Lambda^n_{s'}\}}r_{s-1}\wh{f}^n_{s-1}\big)\big(\ul{L}^n_{s-1},\ul{\bbZ}^n_{s-1}\big). 
\eeas

Obviously, 
\beas 
\bbD^n_s(f)
\yleq
\big\|R^n_{s-1}\triangle^n_{s}(f)\big\|_1
\yleq
(n_s)^{\ep}\big\|\triangle^n_{s}(f)\big\|_1+V^{\ep,n}_{s-1}(f)\qquad(s\in\bbS)
\eeas
for any positive number $\ep$, where 
\bea\label{0411261317}
V^{\ep,n}_{s-1}(f)
&=& 
\big\|1_{\{R^n_{s-1}>(n_s)^{\ep}\}}R^n_{s-1} \triangle^n_{s-1}(f)\big\|_1.
\eea

We have 
\beas 
\triangle^n_{s}(f)
&=& 
\big(\B^n_{s-1}(\Pi_{s'=1}^{s-1}1_{\{\cdot\in\Lambda^n_{s'}\}}r_{s-1}\wh{f}^n_s)\big)\big(\ul{L}^n_{s-1},\ul{\bbZ}^n_{s-1}\big)
-\big(\Pi_{s'=1}^{s-1}1_{\{\cdot\in\Lambda^n_{s'}\}}r_{s-1}\wh{f}^n_{s-1}\big)\big(\ul{L}^n_{s-1},\ul{\bbZ}^n_{s-1}\big)
\nn\\&=&
\int_{\mfkC_{s}}\int_{\call_s{\fblue\times\bbR^{\sfd_s{\sfr_sn_s}}}\times\bbR^{\sfd_s}} 
\Pi_{s'=1}^{s}1_{\{l_{s'}\in\Lambda^n_{s'}\}}r_{s-1}(\ul{z}_{s-1})\wh{f}^n_s\big(\ul{l}_{s-1},l_s,\ul{z}_{s-1},z_s\big)
P^n_s({\bf w}^n_s,dz_s)
\nn\\&&\hspace{150pt}\times
\nu^n_{c_s}(dl_s,d{\bf w}^n_s)q\big((\ul{l}_{s-1},\ul{z}_{s-1}),dc_{s}\big)\bigg|_{\ul{l}_{s-1}=\ul{L}^n_{s-1},\ul{z}_{s-1}=\ul{\bbZ}^n_{s-1}}
\nn\\&&
-\int_{\mfkC_{s}}
\int_{\call_s{\fblue\times\bbR^{\sfd_s{\sfr_sn_s}}}\times\bbR^{\sfd_s}} 
\Pi_{s'=1}^{s}1_{\{l_{s'}\in\Lambda^n_{s'}\}}r_{s-1}(\ul{z}_{s-1})
\wh{f}^n_{s}\big(\ul{l}_{s-1},l_s,\ul{z}_{s-1},z_s\big)\Psi^n_{s,\sfp,{\bf w}^n_s}(dz_s)
\nn\\&&\hspace{150pt}\times
\nu^n_{c_s}(dl_s,d{\bf w}^n_s)q\big((\ul{l}_{s-1},\ul{z}_{s-1}),dc_{s}\big)\bigg|_{\ul{l}_{s-1}=\ul{L}^n_{s-1},\ul{z}_{s-1}=\ul{\bbZ}^n_{s-1}}
\nn\\&=&
E_{s-1}\big[\wh{g}^n_s(\ul{L}^n_{s},\ul{\bbZ}^n_{s}\big)\big]
-E_{s-1}\big[\Psi^n_{s,\sfp,{\bf W}^n_s}\big[\wh{g}^n_s(\ul{L}^n_{s-1},L^n_s,\ul{\bbZ}^n_{s-1},\cdot\big)\big]\big]. 
\eeas
Therefore, 
\beas 
\big\|\triangle^n_{s}(f)\big\|_1
&=&
E\bigg[\bigg|E_{s-1}\bigg[\wh{g}^n_s(\ul{L}^n_{s},\ul{\bbZ}^n_{s}\big)
-\Psi^n_{s,\sfp,{\bf W}^n_s}\big[\wh{g}^n_s(\ul{L}^n_{s-1},L^n_s,\ul{\bbZ}^n_{s-1},\cdot\big)\big]\bigg]\bigg|\bigg]
\nn\\&=&
E\bigg[\bigg|E_{s-1}\bigg[E_{s-1,\A^n_s}\big\{\wh{g}^n_s(\ul{L}^n_{s},\ul{\bbZ}^n_{s}\big)
-\Psi^n_{s,\sfp,{\bf W}^n_s}\big[\wh{g}^n_s(\ul{L}^n_{s-1},L^n_s,\ul{\bbZ}^n_{s-1},\cdot\big)\big]\big\}\bigg]\bigg|\bigg]
\nn\\&\leq&
\Delta^n_s\big((\wh{g}_f)^n_s\big),  
\eeas
where 
\bea\label{0411261150}
\Delta^n_s\big((\wh{g}_f)^n_s\big)
&=&
\bigg\|E_{s-1,\A^n_s}\big[(\wh{g}_f)^n_s(\ul{L}^n_{s},\ul{\bbZ}^n_{s}\big)\big]
-\Psi^n_{s,\sfp,{\bf W}^n_s}\big[(\wh{g}_f)^n_s(\ul{L}^n_{s},\ul{\bbZ}^n_{s-1},\cdot\big)\big]\bigg\|_1.
\eea
Thus, 
\bea\label{0411261151}
\bbD^n_s(f)
&\leq& 
\l\{\begin{array}{ll}
\Delta^n_1((\wh{g}_f)^n_1)&(s=1)
\y
(n_s)^{\ep}\Delta^n_s((\wh{g}_f)^n_s)+V^{\ep,n}_{s-1}(f)&(s\geq2)
\end{array}\r.
\eea
In this way, for 
\beas 
f^n_0\yeq\E_n\yeq 
E\bigg[f^n_{S}\big(\ul{L}^n_{S},\ul{\bbZ}^n_{S}\big)\prod_{s=1}^{S}1_{\{L^n_{s}\in\Lambda^n_{s}\}}\bigg]
\yeq E\bigg[\wh{f}^n_{S}\big(\ul{L}^n_{S},\ul{\bbZ}^n_{S}\big)\prod_{s=1}^{S}1_{\{L^n_{s}\in\Lambda^n_{s}\}}\bigg],
\eeas
we obtain the following result. 
\begin{proposition}\label{0411261159}
Assume (\ref{0311191900}), (\ref{0411261109}) and (\ref{0411261111}). Then both $f^n_0$ and $\wh{f}^n_0$ are well defined, and 
\bea\label{0411261319}
\big|f^n_0-\wh{f}^n_0\big|
&\leq&
\sum_{s=2}^S(n_s)^{\ep}\Delta^n_s\big((\wh{g}_f)^n_s\big)+\sum_{s=2}^SV^{\ep,n}_{s-1}(f)+\Delta^n_1\big((\wh{g}_f)^n_1\big)
\nn\\&\leq&
\sum_{s=1}^S(n_s)^{\ep}\Delta^n_s\big((\wh{g}_f)^n_s\big)+\sum_{s=2}^SV^{\ep,n}_{s-1}(f). 
\eea
\end{proposition}
\halflineskip

In Section \ref{0311240905}, 
we will combine the backward estimate of the approximation error given by Proposition \ref{0411261159} with a conditional type Edgeworth expansion. 
The factor $\Delta^n_s\big((\wh{g}_f)^n_s\big)$ in (\ref{0411261319}) will turn out to be negligible and a small $\ep$ does not matter. 
The term $V^{\ep,n}_{s-1}(f)$, as defined by (\ref{0411261317}), is also controllable if 
boundedness of moments of 
$R^n_{s-1}$ and $\triangle^n_{s-1}(f)$ is available, which is common. 
As in the usual context of the theory of asymptotic expansion, we will treat functions $\wh{f}^n_s(\ul{L}^n_s,\ul{\bbZ}^n_{s-1},z_s)$ of at most polynomial growth in $z_s$. 
However, a careful handling is necessary because they are random functions. 
}

For $(M,\gamma)\in\bbR_+^2$, denote by $\cale_s(M,\gamma)$ the set of measurable functions 
$\sff:\bbR^{\sfd_s}\to\bbR$ such that 
\beas 
\big|\sff(z_s\big)\big| &\leq& M(1+|z_s|)^\gamma
\qquad\big(z_s\in\bbR^{\sfd_s}\big). 
\eeas
Given $\M=(M_s)_{s\in\bbS}\in\bbR_+^S$, 
denote by $\cald(\M,\gamma)$ a family of measurable functions $f:\call\times\bbR^\sfm\to\bbR$ 
satisfying the following condition: 
{\fblue 
\bea\label{0411261348}
(\wh{g}_f)^n_s(\ul{L}^n_s,\ul{\bbZ}^n_{s-1},\cdot)&\subset& 
\cale_s(M_s,\gamma)\qquad a.s. 
\eea
for every $s\in\bbS$ and $n\in\bbN$. 
Then, for $f\in\cald(\M,\gamma)$, 
the functions $\wh{f}^n_{s-1}({\sred\ul{L}^n_{s-1},\>}\ul{\bbZ}^n_{s-1})$ 
are inductively well defined if 
\bea\label{0404111252}
E\big[R^n_{s-1}\> |\Psi^n_{s,\sfp,{\bf W}^n_s}|[1+|z_s|^\gamma]\>\Pi_{s'=1}^{s}1_{\{L^n_{s'}\in\Lambda^n_{s'}\}}\big] &<& \infty
\qquad(s\in\bbS). 
\eea
See (\ref{0411261111b}). 

\begin{theorem}\label{0404231847}
Suppose that the conditions (\ref{0311191900}), (\ref{0411261109}) and (\ref{0404111252}) are satisfied. 
Then both $f^n_0$ and $\wh{f}^n_0$ are well defined, and 
(\ref{0411261319}) holds true 
for all $f\in\cald(\M,\gamma)$, $\M=(M_s)_{s\in\bbS}$. 
\end{theorem}

A natural choice of $r_s(\ul{z}_s)$ may be $\prod_{s'=1}^{{\fred s}}(1+|z_{s'}|)^{-\gamma}$, while we keep a generic $r_s(\ul{z}_s)$ yet. 
For example, Condition (\ref{0404111252}) is satisfied for $\gamma=0$ 
by a suitable choice of $\ul{L}^n_s$ and $\Lambda^n_s$, 
if $\cald({\bf M},\gamma)$ consists of uniformly bounded functions. 
\begin{en-text}
More precisely, under the integrability condition (\ref{0404111252}), 
if {\sred$\big\{f^n_{t,c}(\ul{l}_t,\cdot);\>
\ul{l}_t\in\Pi_{s=1}^t\call_s
\big\}\subset{\tred\ul{\cale}_t(M_t,\gamma)}$ 
for some positive constants $M_t$ and $\gamma$, 
then  $\big\{f^n_{t-1,c}(\ul{l}_{t-1},\cdot);\>\ul{l}_{t-1}\in\Pi_{s=1}^{t-1}\call_s\big\}
\subset{\tred\ul{\cale}_{t-1}(M_{t-1},\gamma)}$\noindent
} for some positive constant $M_{t-1}$. 
\end{en-text}
The class $\cald({\bf M},\gamma)$ will be restricted later
by putting conditions that ensure a measure-theoretic continuity of functions. 
Besides controlling the magnitude of functions, such a restriction is necessary and standard in the theory of asymptotic expansion. 
The fact that the ordinary Edgeworth expansion with a continuous function cannot give any higher-order approximation 
to the distribution function in the lattice case explains this requirement. 
}
{\fblue 
\subsection{A modified backward scheme for the $\ep$-Greedy algorithm}
For the $\ep$-Greedy algorithm like (\ref{0412030158}), the integral with respect to $q\big((\ul{l}_{s-1},\ul{z}_{s-1}),dc_{s}\big)$ 
in the formula (\ref{0311211418b}) becomes the sum $\sum_{c_s\in{\mathfrak C}_s}$, 
and hence the recursion of the formula gives an $(S-1)$-ple sum $\sum_{c_2\in{\mathfrak C}_2}\cdots\sum_{c_S\in{\mathfrak C}_S}$. 
So, it is possible to form a back propagation with a function $\wh{f}^n_{s,c}$ defined similarly to (\ref{0412030158}) but 
along a fixed scenario $c=(c_s)_{s\in\bbS}$. 
Then, $\sum_{c_2\in{\mathfrak C}_2}\cdots\sum_{c_S\in{\mathfrak C}_S}\wh{f}^n_{s,c}$ gives $\wh{f}^n_0$. 
}
\halflineskip

\begin{en-text}
\bea\label{0404061243}
\calm_s(f)&\subset& 
\cale_s(M_s,\gamma)
\eea
for every $s\in\bbS$, where 
{\fred
\beas  
\calm_s(f)
&=&
\bigg\{r_{s-1}(\ul{z}_{s-1})
\big(|f^n_{s}|+|\wh{f}^n_{s}|\big)\big(\ul{l}_{s},\ul{z}_{s-1},\cdot\big)
{\fblue\Pi_{s'=1}^s1_{\{l_{s'}\in\Lambda^n_{s'}\}}}
;\>
\ul{l}_s,\ul{z}_{s-1},n\bigg\}.
\eeas
}
{\fred 
Then, obviously, 
the functions $f^n_{s-1}(\ul{l}_{s-1},\ul{z}_{s-1})$ 
and $\wh{f}^n_{s-1}({\sred\ul{l}_{s-1},\>}\ul{z}_{s-1})$ 
are inductively well defined 
if 
{\fred 
\bea
&&
\hspace{-10pt}
\int 
|z_s|^\gamma
\big\{P^n_s({\bf w}^n_s,dz_s)+|\Psi^n_{s,\sfp,{\bf w}^n_s}|(dz_s)\big\}
1_{\{l_s\in\Lambda^n_{s}\}}
\nu^n_{c_s}(dl_s,d{\bf w}^n_s)
q\big((\ul{l}_{s-1},\ul{z}_{s-1}),dc_{s}\big)
<
\infty
\eea
for every $\ul{l}_{s-1},\ul{z}_{s-1},s,n$. 
{\fblue For example, Condition (\ref{0404111252}) is satisfied for $\gamma=0$ 
by a suitable choice of $\ul{L}^n_s$ and $\Lambda^n_s$, 
if $\cald({\bf M},\gamma)$ consists of uniformly bounded functions.} 
}
The class $\cald({\bf M},\gamma)$ will more precisely be specified later
by putting conditions that ensure a measure-theoretic continuity of functions. 
Besides controlling the magnitude of functions, such a restriction is necessary and standard in the theory of asymptotic expansion. 
The fact that the ordinary Edgeworth expansion with a continuous density cannot give any higher-order approximation in the lattice case 
explains this requirement. 
}

{\fred \koko
Define $\wt{f}^n_{s}({\sred\ul{l}_{s},\>}\ul{z}_{s})$ by 
$\wt{f}^n_{S}={\tred f^n_{S}}$ {\tred for $f^n_{S}$ of (\ref{0404100341})} and 
\beas\label{0311211418} 
\wt{f}^n_{s-1}({\sred\ul{l}_{s-1},\>}\ul{z}_{s-1})
&=&
\int_{\mfkC_{s}}
\int_{\call_s{\fblue\times\bbR^{\sfd_s{\sfr_sn_s}}}\times\bbR^{\sfd_s}} 
f^n_{s}\big(\ul{l}_{s-1},l_s,\ul{z}_{s-1},z_s\big)
\Psi^n_{s,\sfp,{\bf w}^n_s}(dz_s)
1_{\{l_s\in\Lambda^n_{s}\}}
\nu^n_{c_s}(dl_s,d{\bf w}^n_s)
\nn\\&&\hspace{200pt}\times
q\big((\ul{l}_{s-1},\ul{z}_{s-1}),dc_{s}\big)
\eeas}\noindent
for $s\in\bbS$. 
{\fred 
We notice that the functions 
$\wt{f}^n_{s}({\sred\ul{l}_{s},\>}\ul{z}_{s})$ are well defined under Conditions 
(\ref{0404061243}) and (\ref{0404111252}). 
}

Let
\beas 
\varrho_s^n(f)
&=& 
\sup_{\ul{l}_{s-1},\ul{z}_{s-1}}\bigg|
r_{s-1}(\ul{z}_{s-1})f^n_{s-1}(\ul{l}_{s-1},\ul{z}_{s-1})
-r_{s-1}(\ul{z}_{s-1})\wt{f}_{{s-1}}^n(\ul{l}_{s-1},\ul{z}_{s-1})\bigg|
\eeas
and 
\beas
\wh{\varrho}_s^n(f)
&=&
\sup_{\ul{l}_{s-1},\ul{z}_{s-1}}
\bigg| r_{s-1}(\ul{z}_{s-1})f^n_{s-1}(\ul{l}_{s-1},\ul{z}_{s-1})
- r_{s-1}(\ul{z}_{s-1})\wh{f}^n_{s-1}(\ul{l}_{s-1},\ul{z}_{s-1})
\bigg|
\eeas
for $s=1,...,S$. 
By definition, $\wh{\varrho}_S^n(f)\yeq\varrho_S^n(f)$.
{\fred 
Moreover,
\beas 
\wh{\varrho}_s^n(f)
&=&
\sup_{\ul{l}_{s-1},\ul{z}_{s-1}}\bigg|
r_{s-1}(\ul{z}_{s-1})f^n_{s-1}(\ul{l}_{s-1},\ul{z}_{s-1})
-r_{s-1}(\ul{z}_{s-1})\wh{f}_{{s-1}}^n(\ul{l}_{s-1},\ul{z}_{s-1})\bigg|
\nn\\&\leq&
\sup_{\ul{l}_{s-1},\ul{z}_{s-1}}\bigg|
r_{s-1}(\ul{z}_{s-1})f^n_{s-1}(\ul{l}_{s-1},\ul{z}_{s-1})-r_{s-1}(\ul{z}_{s-1})\wt{f}_{{s-1}}^n(\ul{l}_{s-1},\ul{z}_{s-1})\bigg|
\nn\\&&
+\sup_{\ul{l}_{s-1},\ul{z}_{s-1}}\bigg|
r_{s-1}(\ul{z}_{s-1})\wt{f}_{s-1}^{{\colorr n}}(\ul{l}_{s-1},\ul{z}_{s-1})-r_{s-1}(\ul{z}_{s-1})\wh{f}_{{s-1}}^n(\ul{l}_{s-1},\ul{z}_{s-1})\bigg|
\nn\\&\leq&
\varrho_s^n(f)
\nn\\&&
+
\sup_{\ul{l}_{s-1},\ul{z}_{s-1}}\bigg|
\int \frac{r_{s-1}(\ul{z}_{s-1})}{r_{s}(\ul{z}_{s})}
\big\{r_{s}(\ul{z}_{s})f^n_{s}\big(\ul{l}_{s},\ul{z}_{s}\big)
-r_{s}(\ul{z}_{s})\wh{f}^n_{s}\big(\ul{l}_{s},\ul{z}_{s}\big)\big\}
\nn\\&&\hspace{40pt}\times
\Psi^n_{s,\sfp,{\bf w}^n_s}(dz_s)
1_{\{l_s\in\Lambda^n_{s}\}}
\nu^n_{c_s}(dl_s,d{\bf w}^n_s)
q\big((\ul{l}_{s-1},\ul{z}_{s-1}),dc_{s}\big)
\bigg|
\eeas
\beas
&=&
\varrho_s^n(f)
+
\sup_{\ul{l}_{s-1},\ul{z}_{s-1}}\bigg|
\int (1+|z_s|^\gamma)
\big\{r_{s}(\ul{z}_{s})f^n_{s}\big(\ul{l}_{s},\ul{z}_{s}\big)
-r_{s}(\ul{z}_{s})\wh{f}^n_{s}\big(\ul{l}_{s},\ul{z}_{s}\big)\big\}
\nn\\&&\hspace{40pt}\times
\Psi^n_{s,\sfp,{\bf w}^n_s}(dz_s)
1_{\{l_s\in\Lambda^n_{s}\}}
\nu^n_{c_s}(dl_s,d{\bf w}^n_s)
q\big((\ul{l}_{s-1},\ul{z}_{s-1}),dc_{s}\big)
\bigg|
\nn\\&\leq&
\varrho_s^n(f)
\nn\\&&
+\wh{\varrho}_{s+1}^n(f)\sup_{\ul{l}_{s-1},\ul{z}_{s-1}}\int (1+|z_s|^\gamma)
|\Psi^n_{s,\sfp,{\bf w}^n_s}|(dz_s)
1_{\{l_s\in\Lambda^n_{s}\}}
\nu^n_{c_s}(dl_s,d{\bf w}^n_s)
q\big((\ul{l}_{s-1},\ul{z}_{s-1}),dc_{s}\big)
\nn\\&=&
\varrho_s^n(f)+\wh{\varrho}_{s+1}^n(f)U_{s+1}^n,
\eeas
where 
\bea\label{0404111727}
U_{s+1}^n
&=& 
\sup_{\ul{l}_{s-1},\ul{z}_{s-1}}
\int (1+|z_s|^\gamma)
|\Psi^n_{s,\sfp,{\bf w}^n_s}|(dz_s)
1_{\{l_s\in\Lambda^n_{s}\}}
\nu^n_{c_s}(dl_s,d{\bf w}^n_s)
q\big((\ul{l}_{s-1},\ul{z}_{s-1}),dc_{s}\big)
\eea
for $s=1,...,S-1$. 
}
Thus, 
\beas 
\wh{\varrho}_S^n(f) &=& \varrho_S^n(f), 
\nn\\
\wh{\varrho}_{S-1}^n(f) &\leq& \varrho_{S-1}^n(f)+\wh{\varrho}_{S}^n(f)U_{S}^n, 
\nn\\
\wh{\varrho}_{S-2}^n(f) &\leq& \varrho_{S-2}^n(f)+\wh{\varrho}_{S-1}^n(f)U_{S-1}^n, 
\nn\\
&\cdots&
\nn\\
|{\fred f^n_{0}-\wh{f}^n_{0}}|\yeq 
\wh{\varrho}_{1}^n(f) &\leq& \varrho_{1}^n(f)+\wh{\varrho}_{2}^n(f)U_{2}^n.
\eeas
{\tred 
Therefore, the following result has been shown. 
\begin{theorem}
Suppose that the condition (\ref{0404111252}) is satisfied. 
Then 
\bea\label{0404100456}
|{\fred f^n_{0}-\wh{f}^n_{0}}|
&\leq&
\sum_{s\in\bbS} \bigg(\varrho_s^n(f)\prod_{i=2}^sU^n_s\bigg)
\eea
for all $f\in\cald(\M,\gamma)$, $\M=(M_s)_{s\in\bbS}$, 
where $\prod_{i=2}^1U^n_s=1$. 
\end{theorem}
}
\end{en-text}


\section{Partial mixing and asymptotic expansion}\label{03112240118}
{\tred 
Theorem \ref{0404231847} gave the estimate (\ref{0411261319}) 
of the error caused by approximation of 
$P^n_s({\bf w}^n_s,dz_s)$ by $\Psi^n_{s,\sfp,{\bf w}^n_s}(dz_s)$. 
It is possible to assess this error bound by means of the theory of asymptotic expansion 
under partial mixing. 
We begin with a generalization of the theory to be incorporated with the sequential conditioning appearing in Section \ref{0311240905}. 
\begin{en-text}
\footnote{
The letter $T$ in this section denotes the total time or number of observations, e.g. $n_s$, 
not the number of stages in the previous sections. 
We keep $T$ for the reader's convenience since the proof in Section \ref{0311221323} deeply relies on the one in Yoshida \cite{yoshida2004partial}, where $T$ is used for the same role. 
}
\end{en-text}
}

\subsection{Asymptotic expansion for a partially mixing process}
In this section, we recall the asymptotic expansion scheme for a partial mixing process 
given by Yoshida \cite{yoshida2004partial}. 
We will present it in a simplified form of the theory though the paper is written in more generality. 
Moreover, this section includes a generalization of the original configuration, for this paper's use. 
For the reader's convenience in consulting the paper, we will basically keep the notation in Yoshida \cite{yoshida2004partial}. For example, the symbol ``$\calc$'' is used here 
to denote the conditioning $\sigma$-field for the partial mixing. 

{\fred On a probability space $(\Omega,\calf,P)$, we consider a family of sub $\sigma$-fields $(\calb_I)_{I\subset\bbR_+}$ such that 
$\calb_I\subset\calb_J$ for $I\subset J\subset\bbR_+$. 
Suppose that a sub $\sigma$-field $\calc$ of $\calf$ and $\calc$-measurable random numbers $\alpha(s,t|\calc)$ ($s,t\in\bbR_+, s\leq t)$ satisfy 
\beas 
1\ygeq\alpha(s,t|\calc)\ygeq
\sup\bigg\{\big|P_\calc[B_1\cap B_2]-P_\calc[B_1]P_\calc[B_2]\big|;\>
B_1\in\calb_{[0,s]}\vee\calc,B_2\in\calb_{[t,\infty)}\vee\calc\bigg\}.
\eeas
Here $P_\calc$ denotes a version of conditional probability given $\calc$. 
The random field $\alpha(s,t|\calc)$ is called a $\calc$-conditional $\alpha$-mixing coefficient. 
We suppose that $\calc$-measurable random numbers $\alpha(h|\calc)$ ($h\in\bbR_+$) satisfy 
\beas 
1\ygeq\alpha(h|\calc)\ygeq\sup\big\{\alpha(t,t+h'|\calc);\>h'\geq h,t\in\bbR_+\big\}
\eeas
for $h\in\bbR_+$.

We will consider a $\sfd$-dimensional process $Z=(Z_t)_{t\in\bbR_+}$ such that 
$Z_0$ is $\calb_{[0]}\vee\calc$-measurable and that $Z^s_t:=Z_t-Z_s$ is $\calb_{[s,t]}\vee\calc$-measurable 
for every $s,t\in\bbR_+$, $s\leq t$. 
The increment of $Z$ on the interval $I\subset\bbR_+$ is denoted by $Z_I$, i.e., $Z_I=Z_{\sup I}-Z_{\inf I}$. 
}

Given a positive integer $\sfp\geq3$, 
the $(\sfp-2)$-th asymptotic expansion formula for the distribution $T^{-1/2}Z_T$ 
is given by the $\calc$-conditional formula $\Psi_{T,\sfp,\calc}$. 
{\tred 
The formula $\Psi_{T,\sfp,\calc}$ is a $\calc$-measurable random signed measure 
specified as follows. 
The $\calc$-conditional cumulant functions $\chi_{T,r,\calc}(u)$ of $T^{-1/2}Z_T$ 
are defined by 
\bea\label{0404131247}
\chi_{T,r,\calc}(u)
&=& 
\bigg(\frac{\partial}{\partial\ep}\bigg)^r\log E_\calc\big[\exp\big(\tti\ep u\cdot T^{-1/2}Z_T\big)\big]\bigg|_{\ep=0}
\eea
for $u\in\bbR^\sfd$, where $E_\calc$ denotes the conditional expectation given $\calc$. 
Next, the $\calc$-measurable random functions ${P}_{T,r,\calc}(u)$ of $u\in\bbR^\sfd$
 are defined by 
the formal expansion 
\bea\label{0404181443} 
\exp\bigg(\sum_{r=2}^\infty\ep^{r-2}(r!)^{-1}\chi_{T,r,\calc}(u)\bigg)
&=& 
\exp\big(2^{-1}\chi_{T,2,\calc}(u)\big)\sum_{r=0}^\infty\ep^rT^{-r/2}
{P}_{T,r,\calc}\big(u;\chi_{T,\cdot,\calc}\big), 
\eea
$P_{T,0,\calc}\big(u;\chi_{T,\cdot,\calc}\big)=1$, 
and the function $\wh{\Psi}_{T,\sfp,\calc}(u)$ of $u\in\bbR^\sfd$ is defined as the partial sum 
\bea\label{0404131248}
\wh{\Psi}_{T,\sfp,\calc}(u)
&=& 
\exp\big(2^{-1}\chi_{T,2,\calc}(u)\big)\sum_{r=0}^{\sfp-2}T^{-r/2}{P}_{T,r,\calc}\big(u;\chi_{T,\cdot,\calc}\big). 
\eea
Finally the random signed measure $\Psi_{T,\sfp,\calc}$ is defined as 
the Fourier inversion of $\wh{\Psi}_{T,\sfp,\calc}$ : 
\bea\label{0404131249}
\Psi_{T,\sfp,\calc} &=& \calf^{-1}\big[\wh{\Psi}_{T,\sfp,\calc}\big].
\eea
We set $\Psi_{T,\sfp,\calc}$ {\fblue for the Dirac delta measure on $0$} if 
the quadratic form $\text{Var}_\calc[T^{-1/2}Z_T][u^{\otimes2}]=-\chi_{T,2,\calc}(u)$ degenerates. 
$[$ It should be remarked that, in the proof of Theorem \ref{0311221323}, 
the signed measure 
$\Psi_{T,\sfp,\calc}$ defined by the Fourier inversion (\ref{0404131249}) 
is only used on the event $\{s_T I_\sfd\leq \text{Var}_\calc[T^{-1/2}Z_T]\leq u_TI_\sfd\}$. 
So this proviso is not essential in the story. $]$
}

Introduce a $\calc$-measurable truncation functional $\Phi_T:\Omega\to\{0,1\}$. 
The error of the formula is evaluated by 
\bea\label{0311221401}
\Delta_T^\Phi(\sff)
&=& 
{\fblue
\bigg\| \Phi_T\bigg(E_\calc\big[\sff\big(T^{-1/2}Z_T\big)\big]-\Psi_{T,\sfp,\calc}[\sff]\bigg)\bigg\|_1}
\eea
for a $\calc$-measurable random field $\sff$ on $\bbR^\sfd$. 
{\fblue When working with (\ref{0311221401}), we consider the condition that 
\bea\label{0311221825} 
\limsup_{T\to\infty}\big\|\Phi_T|\Psi_{T,\sfp,\calc}|[1+|\cdot|^{\sfp_0}]\big\|_q &<& \infty
\eea
for some $q>1$,} 
where $\sfp_0=2\lfloor \sfp/2\rfloor$. 

In (\ref{0311221401}), we consider $\calc$-measurable random fields $\sff$ on $\bbR^\sfd$, 
generalizing the results in Yoshida \cite{yoshida2004partial}. 
This extension is necessary for the application to the batched bandit treated in this paper. 
Define a truncated modulus of continuity $\omega_2^\Phi(\sff;\ep,\nu)$ of the random field $\sff$ by 
\bea\label{0311221423}
\omega_2^\Phi(\sff;\ep,\nu)
&=& 
\bigg\{
E\bigg[\Phi_T\int_{\bbR^\sfd}\omega_\sff(x;\ep)^2\nu(dx)\bigg]\bigg\}^{1/2}, 
\eea
{\fred where $\omega_\sff(x;\ep)$ is a $\calc\times\bbB_\sfd$-measurable random variable satisfying 
$\omega_\sff(x;\ep)\geq\sup\big\{|\sff(x+h)-\sff(x)|;\>|h|\leq\ep\big\}$, 
and $\bbB_\sfd=\bbB[\bbR^\sfd]$, the $\sfd$-dimensional Borel $\sigma$-algebra.}
{\fblue We simply write $\omega_2^\Phi(\sff;\ep,d\nu/dx)$ for $\omega_2^\Phi(\sff;\ep,\nu)$ when $\nu$ is absolutely continuous.}

\begin{en-text}
{\fblue 
Among other possibilities, we also consider the measure of error 
\bea\label{0311221401}
\wt{\Delta}_T(\sff)
&=& 
\bigg\| \Phi_T\bigg(E_\calc\big[\sff\big(T^{-1/2}Z_T\big)\big]-\Psi_{T,\sfp,\calc}[\sff]\bigg)\bigg\|_1. 
\eea
For (\ref{0311221401}), the counterpart of (\ref{0311221825}) is the condition 
\bea\label{0412041715} 
\limsup_{T\to\infty}\big\|\Phi_T|\Psi_{T,\sfp,\calc}|[1+|\cdot|^{\sfp_0}]\big\|_q &<& \infty
\eea
for some $q>1$. 
\end{en-text}

}

%
%
\begin{en-text}
We assume that 
\bea\label{0211221828} 
P\big[\Phi_T=0\big]
&=& 
O(T^{-L})
\eea
as $T\to\infty$ for every $L>0$. 
\end{en-text}
%
%
%
Denote by $\check{\cale}(M,\gamma)$ 
the set of functions $\sff:\Omega\times\bbR^\sfd\to\bbR$ that are 
$\calc\times\bbB_\sfd$-measurable and satisfy
\beas
|\sff(\omega,x)| 
&\leq&
M\big(1+|x|)^\gamma
{\fblue\quad(x\in\bbR^\sfd)\quad a.s.}
\eeas
%
\begin{en-text}
For the sequence $\Phi=(\Phi_T)_{T\in\bbR_+}$ of $\calc$-measurable truncation functionals,
we denote by $\cale^\Phi(M,\gamma)$ \koko
\beas 
1_{\{\Phi_T(\omega)=0\}}
\omega.....\koko
\eeas
\end{en-text}

{\fred We will work with the following conditions.}

\bd
\im[{\bf [P1]}] 
There exists a positive constant $a$ such that 
$\|\alpha(h|\calc)\|_1\leq a^{-1}\exp(-ah)$ for all $h>0$. 
\ed

\bd
\im[{\bf [P2]}] 
There exists a positive number $h_0$ such that for every $L\in\bbN$, 
\beas 
E\big[\big(E_\calc[|Z_0|^{{\fred\sfp+1}}]\big)^L\big]+
\sup_{t,h:t\in\bbR_+\atop0\leq h\leq h_0}
E\bigg[\big(E_\calc\big[|Z^t_{t+h}|^{{\fred\sfp}+1}\big]\big)^L\bigg] &<& \infty
\eeas
and 
$E_\calc[Z_t]=0 $ a.s. for all $t\in\bbR_+$. 
\ed

{\fred 
To prove an error bound in asymptotic expansion, we need some regularity of the distribution of the functional in question. 
The so-called Cram\'er condition serves in the classical setting of independent variables to validate the Fourier inversion of the characteristic function of 
the sum. By the independency, the characteristic function is factorized into that of each increment and the Cram\'er condition provides a decay of each factor. 
For mixing processes, if the Markovian property is available, a similar factorization is feasible with the aid of the conditional characteristic function of each increment, 
and the Cram\'er condition is replaced by a conditional version of it. 
For functionals in stochastic analysis, a local non-degeneracy of the Malliavin covariance matrix of the increments of the functional undertakes this role 
(\cite{yoshida2004partial}). 
Indeed, a Malliavin calculus on a Wiener-Poisson space was used in \cite{yoshida2004partial} 
in applications to a stochastic differential equation with jumps, with the aid of a support theorem. 

The notion of reduction intervals is necessary. 
For a collection of intervals $(I_{\ell})_{\ell\in\{1,...,n'(T)\},\>{\fblue T>0}}$ 
is called 
(dense) reduction intervals associated with 
$\calc$ and $\big(\wh{\calc}({\ell})\big)_{\ell\in\{1,...,n'(T)\},\>{\fblue T>0}}$ 
if the following conditions are satisfied.
\begin{enumerate}[(i)]
\im $\wh{\calc}({\ell})$ ($\ell\in\{1,...,n'(T)\},\>{\fblue T>0}$) are sub $\sigma$-fields of $\calf$ with 
$\wh{\calc}({\ell})\supset\calc$. 
\im $I_{\ell}=[u_{\ell},v_{\ell}]$ satisfies 
$v_{\ell}\leq u_{\ell+1}$ for $\ell\in\{1,...,n'(T)-1\}$, 
$0<\inf_{T,\ell}\big(v_{\ell}-u_{\ell}\big)\leq\sup_{T,\ell}\big(v_{\ell}-u_{\ell}\big)<\infty$, and 
$\liminf_{T\to\infty}\big(n'(T)/T\big)>0$. 
\im For any subsequence $\ell_1,...,\ell_{n''(T)}$ of $1,...,n'(T)$, 
any bounded $\calb_{[0,T]\setminus\cup_{\ell\in\{\ell_1,...,\ell_{n''(T)}\}}I_{\ell}}\vee\calc$-measurable random variable ${\sf A}$ 
and any bounded $\calb_{I_{\ell''}}\vee{\fblue\calc}$-measurable random variable ${\sf A}_{\ell''}$, 
$\ell''\in\{\ell_1,...,\ell_{n''(T)}\}$, it holds that 
\bea\label{0409070839}
E_{\calc}\big[{\sf A}{\sf A}_{\ell_1}\cdots{\sf A}_{\ell_{n''(T)}}\big]
&=& 
E_{\calc}\bigg[{\sf A}E_{\wh{\calc}(\ell_1)}[{\sf A}_{\ell_1}]\cdots E_{\wh{\calc}(\ell_{n''(T)})} [{\sf A}_{\ell_{n''(T)}}]\bigg].
\eea
\end{enumerate}
In the above conditions, $I_{\ell}$ and $\wh{\calc}({\ell})$ may depend on $T$. 
The condition (\ref{0409070839}) is a conditional type of Markovian property. 
For example, if the underlying process of the system is Markovian, then 
we can take 
the $\sigma$-field ${\fblue\wh{\calc}(\ell)}$ consisting of the information of the process at the boundary of ${\fblue I_\ell}$. 

To the present case, the classical Cram\'er condition is extended in a somewhat abstract way by the following condition. 
}
\bd
\begin{en-text}
\im[{\bf [P3]}] 
For every $L>0$, there exist truncation functionals 
$\psi_j:(\Omega,\calf)\to([0,1],\bbB([0,1]))$ and constants $a',a\in(0,1)$ and $B>0$ such that 
$4a'<(a-1)^2$ and 
\bd
\im[(i)] 
$\ds P\bigg[n'(T)^{-1}\sum_j E_\calc\bigg[\sup_{u:|u|\geq B}
\big| E_{\wh{C}(j)}\big[e^{\tti u\cdot Z_{I(j)}}\psi_j\big]\big|\bigg]\geq  a'\bigg]
=o(T^{-L})$ as $T\to\infty$. 
\im[(ii)] 
$\ds P\bigg[n'(T)^{-1}\sum_jP_\calc[1-\psi_j]\geq a\bigg]=o(T^{-L})$ as $T\to\infty$. 
\ed
\ed
\end{en-text}
\im[{\bf [P3]}] 
For every $L>0$, there exist truncation functionals 
$\psi_j:(\Omega,\calf)\to([0,1],\bbB([0,1]))$, 
and there exist positive constants $\eta_1$, $\eta_2$, $\eta_3$ and $B$ 
such that $\eta_1+\eta_2<1$, $\eta_3<1$ and that 
\beas 
P\bigg[\sum_j {\fred P}_\calc\big[\wt{\Phi}(j)\leq 1-\eta_2\big]>\eta_3 n'(T)\bigg] 
&=& 
o(T^{-L})
\eeas
as $T\to\infty$, 
where 
\beas 
\wt{\Phi}(j) 
&=& 
E_{\wh{\calc}(j)}\big[\psi_j]\cdot 1_{\big\{\sup_{u:|u|\geq B}
\big|E_{\wh{\calc}(j)}\big[\psi_j\exp(\tti u\cdot Z_{I(j)})\big]\big| <\eta_1\big\}}.
\eeas
\ed
%
\begin{en-text}
Condition $[P3]$ is slihtly stronger than the usual condition in independent cases. 
As a matter of fact, it is possible to weaken this condition; 
see Remark 15 of Yoshida \cite{yoshida2004partial}, p.609. 
\end{en-text}
Condition $[P3]$ is Condition $[A3^\natural]$ mentioned by 
Remark 15 of Yoshida \cite{yoshida2004partial}, p.609. 

The following result is a generalization of Theorem 1 of Yoshida \cite{yoshida2004partial}, p.565.  
We will treat the conditionally exponentially mixing case to give priority to simplicity 
though slower mixing cases discussed in the above paper {\fred could be treated}.
{\fred 
Write $I^\varrho=\{t\in\bbR_+;\>\text{dist}(t,I)\leq\varrho\}$. 
}
\begin{theorem}\label{0311221323}
Let {\fred$\big(I(j)\big)_{j=1,...,n'(T)}$} be dense reduction intervals with $\big({\fred\wh{\calc}}(j)\big)_{j=1,...,n'(T)}$. 
Suppose that there exists a positive constant $\varrho$ such that 
${\fred\wh{\calc}(j)\subset\calb_{I(j)^\varrho}\vee\calc}$. 
Suppose that Conditions (\ref{0311221825}), 
$[P1]$, 
$[P2]$ and $[P3]$ are satisfied. 
Let $K,M\in(0,\infty)$, and 
let $s_T,u_T$ be any sequences of positive numbers with $\liminf_{T\to\infty}u_T/T^{c''}>0$ 
for some positive constant $c''$. 
Then there exist positive constants $c'$,  $M^*$,  {\fblue $\theta$} and $\delta^*$ such that 
\bea\label{0311221427}
\Delta_T^\Phi(\sff)
&\leq& 
M^*\bigg\{\bigg(P\bigg[\text{Var}_\calc\big[T^{-1/2}Z_T\big]<s_TI_\sfd,\>{\fblue\Phi_T=1}\bigg]\bigg)^\theta
\nn\\&&\hspace{30pt}
+u_T^{\gamma(1)}s_T^{-\gamma(2)}
\omega^\Phi_2\big(\sff;T^{-K},\phi(x;0,u_TI_\sfd)\big)
\bigg\}
\nn\\&&
+\ol{o}(T^{-(\sfp-2+\delta^*)/2})
\eea
uniformly in $\sff\in\check{\cale}(M,\sfp_0)$
as $T\to\infty$, 
if $s_T\geq T^{-c'}$ for large $T$, 
{\fred 
where $\gamma(1)=\big(3(\sfp-2)+\sfd\big)/2$ and $\gamma(2)=3(\sfp-2)\sfd+\sfd/2$, 
and $\ol{o}(...)$ stands for a term of order $o(...)$ uniformly in $\sff\in\check{\cale}(M,\sfp_0)$. 
}
\end{theorem}

{\fred 
\begin{remark}\label{0409171236}\rm
In application of Theorem \ref{0311221323} to 
a not so bad class of $\sff$ such as uniformly $\omega^\Phi_2\big(\sff;T^{-K},\phi(x;0,u_TI_\sfd)\big)=O(T^{-\ep K})$ for some positive number $\ep$, 
in order to achieve an error bound of order $o(T^{-(\sfp-2+\delta^*)/2})$ on the right-hand side of (\ref{0311221427}), 
we can set a number {\fblue $K$ as} ${\fblue\ep} K>\gamma(1)c''+1+(\sfp-2+1)/2$, $u_T=T^{c''}$, and $s_T=T^{-c'_*}$ 
with a positive constant $c'_*<c'$ such that $c'_*\gamma(2)<1$, 
and expect that the term $(P[....])^\theta$ on the right-hand side of (\ref{0311221427}) becomes as small as we like.
Such a class of $\sff$ is common and not restrictive at all in testing statistical hypotheses because the boundary of the critical region 
of the test specified by the discontinuities of $\sff$ is usually a very simple set thin in measure. 
\end{remark}}

To prove Theorem \ref{0311221323}, 
we follow the proof of Yoshida \cite{yoshida2004partial} but 
some technical modifications are necessary, as described in Section \ref{0409162318}.

{\fblue
It is possible to use other measures of the approximation error than (\ref{0311221401}). 
Let 
\beas\label{}
\Delta_T^{\prime\>\Phi}(\sff)
&=& 
\bigg\| E_\calc\big[\sff\big(T^{-1/2}Z_T\big)\big]-\Phi_T\Psi_{T,\sfp,\calc}[\sff]\bigg\|_1
\eeas
and 
\beas\label{}
\Delta_T(\sff)
&=& 
\bigg\| E_\calc\big[\sff\big(T^{-1/2}Z_T\big)\big]-\Psi_{T,\sfp,\calc}[\sff]\bigg\|_1. 
\eeas
Consider the following two conditions. 
For every $L>0$, 
\bea\label{0211221828} 
P\big[\Phi_T=0\big]
&=& 
O(T^{-L})\qquad(T\to\infty). 
\eea
For some $q>1$, 
\bea\label{0412060543}
\limsup_{T\to\infty}\big\||\Psi_{T,\sfp,\calc}|[1+|\cdot|^{\sfp_0}]\big\|_q &<& \infty. 
\eea

\begin{proposition}\label{0412160601}
\bd
\im[(i)] Suppose that (\ref{0211221828}) and the same conditions as Theorem \ref{0311221323} are satisfied. 
Then the inequality (\ref{0311221427}) holds 
if $\Delta_T^{\Phi}(\sff)$ is replaced by $\Delta_T^{\prime\>\Phi}(\sff)$. 
\im[(ii)] Suppose that (\ref{0211221828}), (\ref{0412060543}) and 
the same conditions as Theorem \ref{0311221323} except for (\ref{0311221825}). 
Then the inequality (\ref{0311221427}) holds 
if $\Delta_T(\sff)$ is replaced by $\Delta_T^{\prime\>\Phi}(\sff)$. 
\ed
\end{proposition}

Proposision \ref{0412160601} (i) validates the approximation to 
$E\big[f\big(T^{-1/2}Z_T\big)\big]$ by $E\big[\Phi_T\Psi_{T,\sfp,\calc}[\sff]\big]$ 
with the stabilizer $\Phi_T$. 
}

{\tred 
\subsection{Expansion of a transformed variable}\label{0404131032}
The expanded variable in the applications later discussed is defined as a transform of another additive functional for which asymptotic expansion is valid.
Suppose that $G_T$ is a $\sfd\times\sfd$ 
$\calc$-measurable random matrix 
for $T\in\bbR_+$. 
Let 
\bea\label{0404190439}
Z_T^*\yeq T^{-1/2}G_TZ_T. 
\eea
{\fblue 
Suppose that $G_T$ is invertible when $\Phi_T=1$, and that 
\bea\label{0404131039}
\sup_{T\in\bbR_+,\omega\in\Omega}
\big(\Phi_T|G_T|+\Phi_T|G_T^{-1}|\big)&<& \infty.
\eea
%
In the context of the batched bandit, $G_T$ is a functional of the assignments (see e.g. (\ref{0411231416})) 
and the stability of $G_T$ in (\ref{0404131039}) 
is naturally satisfied by the clipping. 

We will use a $\calc$
-measurable random signed measure $\Psi_{T,\sfp,\calc}^*$ 
to approximate the distribution of $Z_T^*$. 
It is defined in a similar fashon in parallel to 
the procedure from (\ref{0404131247}) to (\ref{0404131249}). 
More precisely, 
let 
\bea\label{0404131255}
\chi_{T,r,\calc}^*(v)
&=& 
\bigg(\frac{\partial}{\partial\ep}\bigg)^r\log E_\calc\big[\exp\big(\tti\ep v\cdot Z_T^*\big)\big]\bigg|_{\ep=0}
\eea
for $v\in\bbR^\sfd$. 
Then,
the function $\wh{\Psi}_{T,\sfp,\calc}^*(v)$ of $v\in\bbR^\sfd$ is defined as the sum 
\bea\label{0404140456}
\wh{\Psi}_{T,\sfp,\calc}^*(v)
&=& 
\exp\big(2^{-1}\chi_{T,2,\calc}^*(v)\big)\sum_{r=0}^{\sfp-2}T^{-r/2}{P}_{T,r,\calc}\big(v;\chi_{T,\cdot,\calc}^{{\fred*}}\big),
\eea
and the random signed measure $\Psi_{T,\sfp,\calc}^*$ is defined as 
the Fourier inversion of $\wh{\Psi}_{T,\sfp,\calc}^*$ : 
\bea\label{0404140457}
\Psi_{T,\sfp,\calc}^* &=& \calf^{-1}\big[\wh{\Psi}_{T,\sfp,\calc}^*\big].
\eea
Instead of (\ref{0311221401}), the error of approximation to $Z_T^*$ by $\Psi_{T,\sfp,\calc}^*$ 
is measured by 
\bea\label{0311221401a}
\Delta_T^{\Phi*}(\sff)
&=& 
\bigg\| \Phi_T\bigg(E_\calc\big[\sff\big(Z_T^*\big)\big]-\Psi_{T,\sfp,\calc}^*[\sff]\bigg)\bigg\|_1.
\eea
%
\begin{en-text}
Additionally to (\ref{0311221825}), we assume that 
\bea\label{0411231312} 
\limsup_{T\to\infty}\big\||\Psi_{T,\sfp,\calc}^*|[1+|\cdot|^{\sfp_0}]\big\|_q &<& \infty
\eea
for some $q>1$. 
\end{en-text}
%
%
\begin{theorem}\label{0404131140}
Suppose that the same conditions as Theorem \ref{0311221323} and (\ref{0404131039}) are satisfied.  
\begin{en-text}
there exist positive constants $c'$,  $M^*$ and $\delta^*$ such that 
the inequality (\ref{0311221427}) holds true 
but the constant $M^*$ is different from that of Theorem \ref{0311221323}. 
\end{en-text}
Then there exist positive constants $c'$,  $M^{**}$,  {\fblue $\theta$} and $\delta^*$ such that 
\bea\label{0311221427a}
\Delta_T^{\Phi*}(\sff)
&\leq& 
M^{**}\bigg\{\bigg(P\bigg[\text{Var}_\calc\big[T^{-1/2}Z_T\big]<s_TI_\sfd,\>{\fblue\Phi_T=1}\bigg]\bigg)^\theta
\nn\\&&\hspace{30pt}
+u_T^{\gamma(1)}s_T^{-\gamma(2)}
\omega^\Phi_2\big(\sff;T^{-K},\phi(x;0,u_TI_\sfd)\big)
\bigg\}
\nn\\&&
+\ol{o}(T^{-(\sfp-2+\delta^*)/2})
\eea
uniformly in $\sff\in\check{\cale}(M,\sfp_0)$, 
as $T\to\infty$, 
if $s_T\geq T^{-c'}$ for large $T$. 
\end{theorem}
{\fblue 
\begin{remark}\rm
Proof of Theorem \ref{0404131140} is in Section \ref{0411110631}. 
There we apply the error bound (\ref{0311221427}) of Theorem \ref{0311221323} to $T^{-1/2}Z_T$, and translate it to $Z_T^*$. 
Consequently, 
the estimate (\ref{0311221427a}) is valid along a subsequence ${\bf T}$ of $\bbR_+$, 
if (\ref{0404190439}) and (\ref{0404131039}) 
are satisfied for $T\in{\bf T}$ for given $(G_T)_{T\in{\bf T}}$. 
The variables $\Phi_T$ are assumed to be given for $T\in\bbR_+$ since they are involved in validation of the asymptotic expansion for $T^{-1/2}Z_T$, 
which is a limit theorem and assumes a kind of uniformity of summands. 
\end{remark}

\begin{remark}\rm
We are not assuming that $\Phi_T$ is near to $1$ in Theorem \ref{0311221323}. 
It is because Theorem \ref{0311221323} has adopted $\Delta^\Phi_T(\sff)$ as the measure of error and we do not pay $1-\Phi_T$ to detach $\Phi_T$. 
As a result, we can choose the functional $\Phi_T$ quite freely except for the constraint (\ref{0311221825}). 
This means we can set $\Phi_T=0$ for some $T$ arbitrarily. 
\end{remark}
}

\begin{remark}\rm 
Theorems \ref{0311221323} and \ref{0404131140} have been stated for the continuous parameter $T$. 
By simple embedding of a process $(V_n)_{n\in\bbZ_+}$ into a continuous time process as 
$(V_{\lfloor T \rfloor})_{T\in\bbR_+}$, we can apply these theorems to discrete time processes. 
It is also possible to reconstruct the theorems for a discrete time without changing the proof. 
%
\begin{en-text}
{\fblue 
(ii) The integrability condition in (\ref{0404131039}) is set to simply ensure 
$\limsup_{T\to\infty}\|E_\calc[|f(Z_T^*)|]\|_{q_*}<\infty$ for some $q_*>1$. 
(iii) 
Condition (\ref{0411231312}) adds non-degeneracy of $G_T$ to (\ref{0311221825}). 
}
\end{en-text}
\end{remark}

\begin{remark}\rm 
Like Proposition \ref{0412160601}, it is possible to measure the approximation error in slightly different ways. 
For example, the inequality (\ref{0311221427a}) with 
\beas 
\Delta_T^{\prime\Phi*}(\sff) 
&=& 
\bigg\|E_\calc\big[\sff\big(Z_T^*\big)\big]-\Phi_T\Psi_{T,\sfp,\calc}^*[\sff]\bigg\|_1
\eeas
in place of $\Delta_T^{\Phi*}(\sff)$ holds 
under the assumptions of Theorem \ref{0404131140} and (\ref{0211221828}), 
if 
\bea\label{0412060622}
\limsup_{T\to\infty}\|G_T\|_q<\infty
\eea
for some $q>\sfp_0$. We need (\ref{0412060622}) for $Z_T^*\in L^{\sfp_0}$. 
Moreover, if additionally 
\bea\label{0412160617}
\limsup_{T\to\infty}\big\|\Psi_{T,\sfp,\calc}^*[1+|\cdot|^{\sfp_0}]\big\|_q<\infty
\eea
for some $q>1$, then 
we obtain the inequality (\ref{0311221427a}) with 
\beas 
\Delta_T^{*}(\sff) 
&=& 
\bigg\|E_\calc\big[\sff\big(Z_T^*\big)\big]-\Psi_{T,\sfp,\calc}^*[\sff]\bigg\|_1
\eeas
in place of $\Delta_T^{\Phi*}(\sff)$. 
Condition (\ref{0412160617}) concerns the global non-degeneracy of $\text{Var}_\calc[Z^*_T]$, so that of $G_T$. 
\end{remark}
}

{\tred
\section{Asymptotic expansion applied to the backward approximation scheme for the bandit algorithm}\label{0311240905}
\subsection{Sequentially partial mixing process}\label{0404180545}\label{0404181101}
We go back to the model for bandits algorithms in Section \ref{0404101857}. 
The error bound of the backward approximation formula will be provided 
by the asymptotic expansion scheme under a partially mixing property, generalized in Section \ref{03112240118}. 

All random variables are assumed to be defined on some common probability space $(\Omega,\calf,P)$. 
To correctly apply the theory prepared so far, we begin with rebuilding an environment as follows. 
\begin{enumerate}[(1)]
\im $\bbS=\{1,...,S\}$, $n_s$ $(s\in\bbS)$ are positive integers depending on $n\in\bbN$ such that $\inf_{s\in\bbS}\liminf(n_s/n)>0$.
\im $\bbJ^\infty=\bbS\times\bbN$, 
$\bbJ^\infty_s=\{j=(s,i);\>i\in\bbN\}$, and $\bbJ^n_s=\{j=(s,i)\in\bbJ^\infty_s;\>i\leq n_s\}$. 
\im $c=(c_s)_{s\in\bbS}$, $c_s\in\mfkC_s$. $\mfkC=\prod_{s\in\bbS}\mfkC_s$. {\fblue Each $\mfkC_s$ is a measurable set.}
\im\label{0412060449} $(A_j)_{j=(s,i)\in\bbJ^\infty}$ is a collection of doubly indexed random vectors. 
$A_j$ is $\ol{k}_{s(j)}$-dimensional, where $s(j)=s$ for $j=(s,i)$. 
{\fblue$\A^n_s=(A_j)_{j\in\bbJ^n_s}$ and} $\A^\infty_s=(A_j)_{j\in\bbJ^\infty,\>s(j)=s}$. 
\im $(\ep_j)_{j=(s,i)\in\bbJ^\infty}$ is a collection of doubly indexed random vectors. 
$\ep_j$ is $\sfr_s$-dimensional when $s(j)=s$. 
\im Given a measurable space $\call_s$ for $s\in\bbS$, 
$L^n_s$ is an $\call_s$-valued random map measurable with respect to {\fblue$\sigma[\A^n_s]$.} 
\im 
{\fblue For $j\in\bbJ^n_s$,}
$H^s_j$ is a 
$\sfd_s\times\sfr_s$ {\fblue $\sigma[\A^n_s]$-measurable} random matrix. 
%

\im\label{0409171313} $Z^s_{T}=\sum_{i=1}^TH^s_{(s,i)}\ep_{(s,i)}$ for $T\in\bbN$. In particular, $Z^s_0=0$. 
\im {\fblue $G^s_{n_s}$ is a $\sfd_s\times\sfd_s$ $\sigma[\A^n_s]$-measurable random matrix for $(n,s)\in\bbN\times\bbS$.} 

\im $W^{n}_j={\fred n_s^{-1/2}}
G^s_{n_s}H^s_j$ for $n\in\bbN$ and {\fblue$j\in\bbJ^n_s$}.

\im\label{0404240240} ${\fred\bbZ^{n}_s}=\sum_{j\in\bbJ^n_s}{\fred W^{n}_j}\ep_j=G^s_{n_s}n_s^{-1/2}{\fred Z_{n_s}^s}$. 
%

\im\label{0412060448} 
{\fblue$\calg^\infty_s=\sigma\big[(L^n_{s'})_{n\in\bbN}, (\bbZ^n_{s'})_{n\in\bbN};\>s'\leq s\big]$} 
for $s\in\bbS$, and $\calg^\infty_0$ is the trivial $\sigma$-field. 

\im\label{0412060447} $\calc^s=\calg^\infty_{s-1}\vee\sigma[\A^\infty_s]$

\im\label{0404181100} $(\ep_j)_{j\in\bbJ^\infty_s}\indep\calc^s$ for every $s\in\bbS$.

\im $\{\calb^s_I\}_{I\subset{\fblue\bbZ_+}}$ 
is a family of sub $\sigma$-fields of $\calf$ such that 
$\calb^s_{I_1}\subset\calb^s_{I_2}$ for $I_1\subset I_2\subset{\fblue\bbZ_+}$. 

\im\label{0404180543} $\ep_{s,i}$ is $\calb^s_{[i]}\vee\calc^s$-measurable for $s\in\bbS$ and $i\in\bbN$. 
\im $\alpha^s(h|\calc^s):\Omega\to[0,1]$ is a $\calc^s$-measurable function for $h\in\bbR_+$ such that 
\beas 
\alpha^s(h|\calc^s)
&\geq&
\sup_{h'\geq h\atop t\in{\fblue\bbZ_+}}
\sup_{B_1\in\calb^s_{[0,t]}\vee\calc^s\atop B_2\in\calb^s_{[t,\infty)}\vee\calc^s}
\big|P_{\calc^s}[B_1\cap B_2]-P_{\calc^s}[B_1]P_{\calc^s}[B_2]\big|\quad a.s.
\eeas
\im\label{0404240241} 
{\fblue 
$\Phi_{s}^{n_s}{\fblue=}1_{\{L^n_s\in\Lambda^n_{s}\}}$ 
for a measurable set $\Lambda^n_{s}\subset\call_s$ for 
$(n,s)\in\bbN\times\bbS$ 
}

\begin{en-text}
\beas 
\sup_{n\in\bbN,s\in\bbS,\omega\in\Omega}\big(\Phi^n_{s,c}|G^s_{n_s}|+\Phi^n_{s,c}|(G^s_{n_s})^{-1}|\big)&<&\infty.
\eeas
\end{en-text}
\end{enumerate}

{\fred 
Condition (\ref{0404181100}) is corresponding to Condition (\ref{0311191900}). 
Under this condition, each $\calc^s$-conditional cumulant is a constant not depending on $\calc^s$. 
The number of summands is $n_s$ in Stage $s\in\bbS$, and $(n_s)_{n\in\bbN}$ is a sequence driven by $n\in\bbN$. 
A bit complicated but we need a finer index $T\in\bbZ_+\subset\bbR_+$ in each stage to correctly apply the results in Section \ref{03112240118}, 
finally embedding $(n_s)_{n\in\bbN}$ to $\bbN$.}
{\vred See (\ref{0411231416}) 
for an example of the decompositions (\ref{0409171313}) and (\ref{0404240240}).} 
We assume the following conditions. 
\bd\im[{\bf [B1]}] 
There exists a positive constant $a$ such that 
\beas 
\big\|\alpha^s(h|\calc^s)\big\|_1&\leq& a^{-1}e^{-ah}\qquad(h>0,\>s\in\bbS).
\eeas
\ed
\bd\im[{\bf [B2]}] For every $L>0$, 
\beas 
\max_{s\in\bbS}\sup_{j\in\bbJ^\infty_s}E\bigg[\big|E_{\calc^s}[|H^s_j\ep_j|^{\sfp+1}]\big|^L\bigg] &<& \infty
\eeas
and $E_{\calc^s}[H^s_j\ep_j]=0$ for $j\in\bbJ^\infty_s$, $s\in\bbS$. 
\ed
\halflineskip

For $s\in\bbS$, 
a collection of intervals $(I^s_{T,\ell})_{\ell\in\{1,...,n'(T)\},\>T\in\bbN}$ 
is called 
(dense) reduction intervals associated with 
$\calc^s$ and $\big(\wh{\calc}^s_{T,\ell}\big)_{\ell\in\{1,...,n'(T)\},\>T\in\bbN}$ 
if the following conditions are satisfied.
\begin{enumerate}[(i)]
\im $\wh{\calc}^s_{T,\ell}$ ($\ell\in\{1,...,n'_s(T)\},\>T\in\bbN$) are sub $\sigma$-fields of $\calf$ with 
$\wh{\calc}^s_{T,\ell}\supset\calc^s$. 
\im $I^s_{T,\ell}=[u^s_{T,\ell},v^s_{T,\ell}]$ satisfies 
$v^s_{T,\ell}\leq u^s_{T,\ell+1}$ for $\ell\in\{1,...,n'_s(T)-1$, 
$0<\inf_{T,\ell}\big(v^s_{T,\ell}-u^s_{T,\ell}\big)\leq\sup_{T,\ell}\big(v^s_{T,\ell}-u^s_{T,\ell}\big)<\infty$, and 
$\liminf_{T\to\infty}\big(n'_s(T)/T\big)>0$.
\im For 
any subsequence $\ell_1,...,\ell_{n''_s(T)}$ of $1,...,n'_s(T)$, 
any bounded $\calb_{[0,T]\setminus\cup_{\ell\in\{\ell_1,...,\ell_{n''_s(T)}\}}I^s_{T,\ell}}\vee\calc^s$-measurable random variable ${\sf A}$ 
and any bounded $\calb^s_{I^s_{T,\ell''}}\vee\calc^s$-measurable random variable ${\sf A}_{\ell''}$, 
$\ell''\in\{\ell_1,...,\ell_{n''_s(T)}\}$, it holds that 
\bea\label{0404171411}
E_{\calc^s}\big[{\sf A}{\sf A}_{\ell_1}\cdots{\sf A}_{\ell_{n''_s(T)}}\big]
&=& 
E_{\calc^s}\bigg[{\sf A}E_{\wh{\calc}^s_{T,\ell_1}}[{\sf A}_{\ell_1}]\cdots E_{\wh{\calc}^s_{T,\ell_{n''_s(T)}}} [{\sf A}_{\ell_{n''_s(T)}}]\bigg].
\eea
\end{enumerate}
\begin{en-text}
The condition (\ref{0404171411}) is a conditional type of Markovian property. 
For example, if the underlying process of the system is Markovian, then 
we can take 
the $\sigma$-field $\wh{\calc}^s_{T,\ell}$ consisting of the information of the process at the boundary of $I^s_{T,\ell}$. 
\end{en-text}

{\fred 
We consider the following condition corresponding to the classical Cram\'er condition. }
\bd\im[{\bf [B3]}] 
For every $L>0$, 
there exist measurable maps $\psi^s_{T,\ell}:\Omega\to[0,1]$ 
($s\in\bbS$, $T\in\bbN$ and every $\ell\in\{1,....,n'_s(T)\}$), and 
there exist positive constants $\eta_1$, $\eta_2$, $\eta_3$ and $B$ such that $\eta_1+\eta_2<1$, $\eta_3<1$ and that 
\bea\label{0504081753}
P\bigg[\sum_{\ell=1}^{n'_s(T)} {\fred P}_{\calc^s}\big[\wt{\Phi}^s_{T,\ell}\leq1-\eta_2\big]>\eta_3n'_s(T)\bigg]
&=& O(T^{-L})
\eea
as $T\to\infty$, where 
\bea\label{0504081754}
\wt{\Phi}^s_{T,\ell}
&=&
E_{\wh{\calc}^s_{T,\ell}}[\psi^s_{T,\ell}]\cdot
1_{
\bigg\{\sup_{u\in\bbR^{\sfd_s}\atop|u|\geq B}
\big|E_{\wh{\calc}^s_{T,\ell}}\big[\psi^s_{T,\ell}\exp\big\{\tti u\cdot \big(Z^s_{v^s_{T,\ell}}-Z^s_{u^s_{T,\ell}}\big)\big\}\big]\big|<\eta_1\bigg\}
}.
\eea
\ed
\halflineskip

Asymptotic expansion for a stochastic differential equation under a random environment is discussed 
by Yoshida \cite{yoshida2004partial}. 
{\fblue There, such a collective, conditional type Cram\'er condition like $[B3]$ is used to ensure regularity of the distribution of the targeted variable.
The Malliavin calculus worked effectively to verify it for jump diffusion processes in a random environment.} 
On the other hand, as we will see later, this condition becomes quite simple for conditionally independent experiments. 
\begin{en-text}
class $\cald^s$
\beas 
r_{s-1}(\ol{z}_{s-1})f^n_{s-1}
\eeas
\end{en-text}

Additionally, we need the following condition. 
\bd
\im[{\bf [B4]}] 
\bd\im[(i)] 
There exists a positive constant $\delta$ such that 
\beas
\text{Var}_{\calc^s}[n_s^{-1/2}Z_{n_s}^s]>\delta I_{\sfd_s}\qquad a.s. \text{ on }\{L^n_s\in\Lambda^n_{s}\}
\eeas
for {\fblue $(s,n)\in\bbS\times\bbN$}. 
\im[(ii)] 
{\fblue 
$G^s_{n_s}$ is invertible when $\Phi^{n_s}_s=1$, and 
\beas 
\sup_{n\in\bbN,s\in\bbS,\omega\in\Omega}\big({\fblue\Phi^{n_s}_s}|G^s_{n_s}|+{\fblue\Phi^{n_s}_s}|(G^s_{n_s})^{-1}|\big)&<&\infty.
\eeas
}
\ed
\ed
{\fblue 
Condition $[B4]$ (i) is a global (in time) non-degeneracy of $n_s^{-1/2}Z^s_{n_s}$. 
This condition is necessary because it does not follow from 
the non-degeneracy and $L^p$-boundedness of the summands of $Z^s_{n_s}$; consider a telescoping series.   
In $[B4]$, 
we assume the properties only for the subsequence $(n_s)_{n\in\bbN}$, not for $T\in\bbN$. 
It is sufficient because the error bound estimate for the asymptotic expansion 
(i.e., the right-hand side of the inequality) is only used along $(n_s)_{n\in\bbN}$. 
See also the remarks after Theorem \ref{0404131140}, and 
the proof of Theorem \ref{0404171827}. } 

{\fblue When considering the batched bandit,}
at Stage $s$, we use the $\calc^s$-measurable random signed measure 
$\Psi^n_{s,\sfp,{\bf W}^n_s}$ in Section \ref{0404190428}. 
In the environment of this section, 
the asymptotic expansion $\Psi_{n_s,\sfp,\calc^s}^*$ for ${\fred\bbZ^{n}_s}$ serves as $\Psi^n_{s,\sfp,{\bf W}^n_s}$; 
recall that 
$\Psi_{T,\sfp,\calc}^*$ was generically defined by (\ref{0404140457}) for $Z_T^*$ of (\ref{0404190439}). 
More precisely, for 
\bea\label{0404190451}
\chi_{n_s,r,\calc^s}(u)
&=& 
\bigg(\frac{\partial}{\partial\ep}\bigg)^r\log E_{\calc^s}\big[\exp\big(\tti\ep u\cdot {\fblue\bbZ^{n}_s}\big)\big]\bigg|_{\ep=0}
\eea
for $u\in\bbR^{\sfd_s}$, 
the functions ${P}_{n_s,r,\calc^s}\big(u;\chi_{n_s,\cdot,\calc^s})$ are determined by the formal expansion 
\bea\label{0404181443bis} 
\exp\bigg(\sum_{r=2}^\infty\ep^{r-2}(r!)^{-1}\chi_{n_s,r,\calc^s}(u)\bigg)
&=& 
\exp\big(2^{-1}\chi_{n_s,2,\calc^s}(u)\big)\sum_{r=0}^\infty\ep^rn_s^{-r/2}
{P}_{n_s,r,\calc^s}\big(u;\chi_{n_s,\cdot,\calc^s}).
\eea
{\fred We remark that 
$\chi_{n_s,r,\calc^s}(u)$ is a function of $\W^n_s=(W^n_j)_{j\in\bbJ^n_s}$ under Condition (\ref{0404181100}). 
Therefore, for each $(n_s,r)$, there exists a measurable function 
$\ol{P}_{n_s,r,{\fred{\bf w}^{n}_s}}(u)$ of ${\fred{\bf w}^{n}_s}$ such that 
\beas
{P}_{n_s,r,\calc^s}\big(u;\chi_{n_s,\cdot,\calc^s})&=&\ol{P}_{n_s,r,{\fred{\bf w}^{n}_s}}(u)\big|_{{\fred{\bf w}^{n}_s}={\fred\W^{n}_s}}.
\eeas
}
\noindent
Define $\wh{\Psi}^n_{s,\sfp,{\bf w}^n_s}$ by 
\bea\label{0404190454}
\wh{\Psi}^n_{s,\sfp,{\fred{\bf w}^{n}_s}}(u)
&=&
\exp\big(2^{-1}\chi_{n_s,2,\calc^s}(u)\big)\sum_{r=0}^{\sfp-2}n_s^{-r/2}{\fred \ol{P}_{n_s,r,{\fred{\bf w}^{n}_s}}(u)},
\eea
where $\chi_{n_s,2,\calc^s}(u)$ is interpreted as a measurable function of ${\bf w}^n_s$. 
The random signed measure $\Psi^n_{s,\sfp,{\fred{\bf w}^{n}_s}}$ on $\bbR^{\sfd_s}$ is defined by 
the Fourier inversion of $\wh{\Psi}^n_{s,\sfp,{\fred{\bf w}^{n}_s}}$ : 
\bea\label{0404190509}
\Psi^n_{s,\sfp,{\fred{\bf w}^{n}_s}} &=& \calf^{-1}\big[\wh{\Psi}^n_{s,\sfp,{\fred{\bf w}^{n}_s}}\big].
\eea
In other words, $\Psi^n_{s,\sfp,{\fred{\bf w}^{n}_s}}$ is absolutely continuous on the event 
${\sf E}_{n_s,s}=\{\chi_{n_s,2,\calc^s}$ is non-degenerate$\}$, and the local density on it is 
\bea\label{0404190540}
\frac{d\Psi^n_{s,\sfp,{\fred{\bf w}^{n}_s}}}{dz}(z)
&=& 
\sum_{r=0}^{\sfp-2}n_s^{-r/2}{\fred \ol{P}_{n_s,r,{\fred{\bf w}^{n}_s}}(-\tti \partial_z)}^\circledast
\phi\big(z;0,\text{Var}_{\calc^s}[{\fred\bbZ^{n}_s}]\big)
\eea
for $z\in\bbR^{\sfd_s}$, where $\circledast$ makes the adjoint operator 
of ${\fred \ol{P}_{n_s,r,{\fred{\bf w}^{n}_s}}(-\tti \partial_z)}$. 
Specifying $\Psi^n_{s,\sfp,{\bf w}^n_s}$ 
outside of ${\sf E}_{n_s,s}$ is not important but we may set $\Psi^n_{s,\sfp,{\bf w}^n_s}=\delta_0$, 
the delta measure.

A representation of the local density of $\Psi^n_{s,\sfp,{\bf w}^n_s}$ is presented by (\ref{0404190540}). 
We write
\beas 
\lambda_{n_s,r,\calc^s}[(\tti u)^{\otimes r}]
&=& 
n_s^{(r-2)/2}
\chi_{n_s,r,\calc^s}(u)
\qquad(u\in\bbR^{\sfd_s}). 
\eeas
{\fred 
For notational simplicity, hereafter we abuse the symbol $\chi_{n_s,r,\calc^s}(u)$ for its representation as a function of ${\fred{\bf w}^{n}_s}$,  
like ${P}_{n_s,r,\calc^s}\big(u;\chi_{n_s,\cdot,\calc^s})$ for $\ol{P}_{n_s,r,{\fred{\bf w}^{n}_s}}(u)$. 
This rule also applies to $\lambda_{n_s,r,\calc^s}$.}
Then the formula (\ref{0404190540}) gives a local density representation of $\Psi^n_{s,\sfp,{\fred{\bf w}^{n}_s}}$ as 
\bea\label{0404190604}
\frac{d\Psi^n_{s,\sfp,{\fred{\bf w}^{n}_s}}}{dz}(z)
&=&
\bigg\{1
+\frac{1}{6}n_s^{-1/2}\lambda_{n_s,3,\calc^s}[(-\partial_z)^{\otimes 3}]
\nn\\&&\hspace{20pt}
+n_s^{-1}\bigg(\frac{1}{24}\lambda_{n_s,4,\calc^s}[(-\partial_z)^{\otimes 4}]
+\frac{1}{72}(\lambda_{n_s,3,\calc^s}[(-\partial_z)^{\otimes 3}])^2\bigg)+\cdots\bigg\}
\nn\\&&\hspace{10pt}\times
\phi\big(z;0,\text{Var}_{\calc^s}[{\fred\bbZ^{n}_s}]\big).
\eea
With the tensors consisting of the ${\sfd_s}^r$ Hermite polynomials 
\beas 
h_r(z;\Sigma) &=& (-1)^r\phi(z;0,\Sigma)^{-1}\partial_z^r\phi(z;0,\Sigma), 
\eeas
we can rewrite (\ref{0404190604}) as 
\bea\label{0404190639}
\frac{d\Psi^n_{s,\sfp,{\fred{\bf w}^{n}_s}}}{dz}(z)
&=&
\bigg\{1
+\frac{1}{6}n_s^{-1/2}\lambda_{n_s,3,\calc^s}\big[h_3(z;\text{Var}_{\calc^s}[{\fred\bbZ^{n}_s}])\big]
\nn\\&&\hspace{20pt}
+n_s^{-1}\bigg(\frac{1}{24}\lambda_{n_s,4,\calc^s}\big[h_4(z;\text{Var}_{\calc^s}[{\fred\bbZ^{n}_s}])\big]
+\frac{1}{72}(\lambda_{n_s,3,\calc^s})^{\otimes2}\big[h_6(z;\text{Var}_{\calc^s}[{\fred\bbZ^{n}_s}])\big]\bigg)
\nn\\&&\hspace{20pt}
+\cdots\cdots\bigg\}\phi\big(z;0,\text{Var}_{\calc^s}[{\fred\bbZ^{n}_s}]\big).
\eea

{\fred 
\noindent
We once again remark that the $\calc^s$-conditional cumulants and hence 
the right-hand side is a measurable function of ${\fred{\bf w}^{n}_s}$. }

{\fblue 
To simplify the error bound, we can further restrict $\cald(\M,\gamma)$ to 
a subclass $\wh{\cald}(M,\gamma)$. 
For a positive number $c_0$, let $u_n=n^{c_0}$. 
For some positive constant $\ep_0$, we suppose
\bea\label{0409171214}
\sup_{\sff\in\wh{\cald}(M,\gamma)}\max_{s\in\bbS}
\sup_{n\in\bbN}\>
\Xi^n_s({\vred(\wh{g}_f)^n_s},r)
&=&
O(r^{\ep_0})\qquad(r\down0),
\eea
\noindent
where 
\beas 
\Xi^n_s({\vred(\wh{g}_f)^n_s},r)
&=& 
\bigg(E\bigg[\Phi^{n_s}_s\int_{\bbR^{\sfd_s}}\omega^n_s({\vred(\wh{g}_f)^n_s},z_s,r)^2
\phi(z_s;0,u_nI_{\sfd_s})dz_s\bigg]\bigg)^\half
\eeas
for a measurable function $\omega^n_s((\wh{g}_f)^n_s,z_s,r)$ such that 
\beas 
\omega^n_s({\vred(\wh{g}_f)^n_s},z_s,r)
&\geq&
\sup_{h_s\in\bbR^{\sfd_s}:|h_s|\leq r}\big|{\vred(\wh{g}_f)^n_s}(\ul{L}^n_s,\ul{\bbZ}^n_{s-1},z_s+h_s)
-{\vred(\wh{g}_f)^n_s}(\ul{L}^n_s,\ul{\bbZ}^n_{s-1},z_s)\big|. 
\eeas
\begin{en-text}
\beas 
\Xi^n_s(f,r)
&=& 
\sup\bigg\{ r_{s-1}(\ul{z}_{s-1})
\omega_2^{\Phi^\cdot_{s}}\big(f^n_{s}(\ul{l}_s,\ul{z}_{s-1},\cdot);r,N(0,{\sf U}^s_{n_s}I_{\sfd_s})\big);\>
\nn\\&&\hspace{50pt}
\ul{l}_{s-1}\in\Pi_{s'=1}^{s-1}\Lambda^{n}_{s'},\>\ul{z}_{s-1}\in\bbR^{\sum_{s'=1}^{s-1}{\sfd_{s'}}}\bigg\}. 
\eeas
\end{en-text}
For example, if the functions $f$ are bounded and have ``good'' discontinuities like critical regions in testing problems, it is a simple matter to check Condition (\ref{0409171214}).
}

{\fred 
\begin{en-text}
In order to verify the condition (\ref{0311221825}), 

we are assuming, for every $s\in\bbS$, 
\bea\label{0409171321} 
\limsup_{T\in\bbN,T\to\infty}\big\||\Psi_{T,\sfp,\calc^s}|[1+|\cdot|^{\sfp_0}]\big\|_q &<& \infty\qquad(s\in\bbS)
\eea
for some $q>1$, 
where $\Psi_{T,\sfp,\calc^s}$ is the signed measure of asymptotic expansion constructed 
for $T^{-1/2}Z^s_T$ given in (\ref{0409171313}).
}
%
{\fblue Additionally to (\ref{0409171321}), we assume that 
\bea\label{0411241119} 
\limsup_{T\to\infty}\big\||\Psi^n_{s,\sfp,{\bf w}^n_s}|
[1+|\cdot|^{\sfp_0}]\big\|_q &<& \infty\qquad(s\in\bbS)
\eea
for some $q>1$, where $\Psi^n_{s,\sfp,{\bf w}^n_s}$ 
denotes the signed measure of asymptotic expansion for 
$\bbZ^{n}_s$ of (\ref{0404240240}). 
We need these conditions because we do not put a global integrability condition for the inverses of 
$\text{Var}_\calc[T^{-1/2}Z^s_T]$ and $G^s_{n_s}$. 
However, these conditions are easily satisfied in the batched bandit thanks to ordinarily used clipping.}
\end{en-text}
%

By the above consideration, 
we obtain the following result, 
since the elements appearing in the error bound are well controlled along the subsequence $(n_s)$ in $\bbN$, 
thanks to the assumptions. 
}
{\fblue 
\begin{theorem}\label{0404171827}
Suppose that Conditions $[B1]$-$[B4]$ are fulfilled as well as 
(\ref{0409171214}). 
Let $M_s>0$ for $s\in\bbS$, and ${\fblue\gamma\in[0,\sfp_0]}$. 
Then both $f^n_0$ and $\wh{f}^n_0$ are well defined, and 
for some constants $M^*$ and $\delta_1>0$, 
$\E_{n}=f^n_{0}$ admits the following estimates:
\bea\label{04041734}
|{\fred f^n_{0}-\wh{f}^n_{0}}|
&\leq&
{\fblue M^*n^{-(\sfp-2+\delta_1)/2}+\sum_{s=2}^SV^{\delta_1,n}_{s-1}(f)}
\eea
for all $f\in\wh{\cald}(\M,\gamma)$, $\M=(M_s)_{s\in\bbS}$. 
\end{theorem}

We can remove the term 
\beas 
\sum_{s=2}^SV^{\delta_1,n}_{s-1}(f)
&=&
\sum_{s=2}^S\big\|1_{\{R^n_{s-1}>(n_s)^{\delta_1}\}}R^n_{s-1} \triangle^n_{s-1}(f)\big\|_1
\eeas
on the right-hand side of (\ref{0411270559}) from the expression if 
$\limsup_{n\in\bbN}\|(R^n_{s-1})^q \triangle^n_{s-1}(f)\|_1<\infty$ for $q= 1+(\sfp-2+\delta_1)/(2\delta_1)$. 
Then we obtain 
\bea\label{0411270646}
\sup_{f\in\wh{\cald}(\M,\gamma)}
|{\fred f^n_{0}-\wh{f}^n_{0}}|
&\leq&
M^\dagger n^{-(\sfp-2+\delta_1)/2}
\eea
for some constant $M^\dagger$. 
\qed\halflineskip

\noindent
{\it Proof of Theorem \ref{0404171827}.} 
{\fblue 
With $\calc^s$ for $\calc$, 
we repeatedly apply Theorem \ref{0404131140} at each stage $s$ to the sequence 
$\big(\bbZ^n_s\big)_{n\in\bbN}$ of (\ref{0404240240}) made from the complete sequence 
$\big(Z^s_T\big)_{T\in\bbN}$ (or the sequence embedded into $T\in\bbR_+$) of (\ref{0409171313}). 
{\fblue
We define the functional $\Phi^T_s$ by 
\bea\label{0412050210}
\Phi^T_s 
&=& 
\left\{\begin{array}{cl}
\Phi^{T}_s&(T\in(n_s)_{n\in\bbN})\vspace{2mm}\\
0&(T\not\in(n_s)_{n\in\bbN}). 
\end{array}\right. 
\eea
We recall that $\Phi_{s}^{n_s}{\fblue=}1_{\{L^n_s\in\Lambda^n_{s}\}}$ from (\ref{0404240241}). 
Then 
\bea\label{0412150245}
\limsup_{T\in\bbN,T\to\infty}\big\|\Phi_s^T|\Psi_{T,\sfp,\calc^s}|[1+|\cdot|^{\sfp_0}]\big\|_q &<& \infty\qquad(s\in\bbS)
\eea
for any $q>1$, 
by the definition of $\Phi^T_s$ and Condition $[B4]$ (i) and (ii), 
with the aid of the $L^\inftym$-boundedness of the $\calc^s$-conditional cumulants of $(T^{-1/2}Z_T^s)_{T\in\bbN}$ up to the $\sfp$-th order (pp.597-598 of \cite{yoshida2004partial}). 
Then, (\ref{0412150245}) verifies (\ref{0404131039}) for the sequence $(T^{-1/2}Z_T^s)_{T\in\bbN}$. 
%
%
%
\begin{en-text}
By Theorem \ref{0311221323}, 
we validate 
the asymptotic expansion of $T^{-1/2}Z^s_T$ along $T\in\bbN$ in each step $s\in\bbS$ by using Conditions $[B1]$, $[B2]$, $[B3]$ and $[B4]$ (iii). 

Then 
\end{en-text}
We will make 
Theorem \ref{0404131140} validate the asymptotic expansion of $\bbZ^n_s$ (along $n_s$) by using $[B4]$, 
and next back-propagate these expansions through $\wh{f}^n_s$. 
}

Set $r_{s}=\prod_{s'=1}^{{\fred s}}(1+|z_{s'}|)^{-\sfp_0}$.
Under $[B4]$ (i), $\text{Var}_{\calc^s}[n_s^{-1/2}Z^s_{n_s}]>\delta I_{\sfd_s}$ a.s. on $\{L^n_s\in\Lambda^n_s\}$. 
The $\calc^s$-conditional cumulants of $n_s^{-1/2}Z^s_{n_s}$ is in $L^\inftym$ under $[B1]$ and $[B2]$, as proved in Yoshida \cite{yoshida2004partial}, pp.597-598. 
Therefore, $1_{\{L^n_s\in\Lambda^n_s\}}|\Psi^n_{s,\sfp,\calc^s}|[(1+|z_s|)^\gamma]\in L^\inftym$. 
Moreover, 
\bea\label{0411270504}&&
E\big[(R^n_{s-1})^{1+\eta}\>\Pi_{s'=1}^{s-1}1_{\{L^n_{s'}\in\Lambda^n_{s'}\}}\big]
\nn\\&=&
E\big[(R^n_{s-2})^{1+\eta}E_{\calc^{s-1}}[(1+|\bbZ^n_{s-1}|)^{\sfp_0(1+\eta)}]\big]
\nn\\&\leq&
E\big[(R^n_{s-2})^{(1+\eta)^2}\>\Pi_{s'=1}^{s-2}1_{\{L^n_{s'}\in\Lambda^n_{s'}\}}\big]^{1/(1+\eta)} 
\nn\\&&\times
E\bigg[1_{\{L^n_{s-1}\in\Lambda^n_{s-1}\}}\big(E_{\calc^{s-1}}[(1+|\bbZ^n_{s-1}|)^{\sfp_0(1+\eta)}]\big)^{(1+\eta)/\eta}\bigg]^{\eta/(1+\eta)}. 
\eea
The second factor on the right-hand side of (\ref{0411270504}) is finite for each $n$ due to $[B2]$ and the boundedness of $G^s_{n_s}$ given by $[B4]$ (ii), 
if we take a sufficiently small positive number $\eta$. 
The first factor on the right-hand side is the same form of the expectation on the left-hand side. 
So, for sufficiently small $\eta$, we can show $E\big[(R^n_{s-1})^{1+\eta}\>\Pi_{s'=1}^{s-1}1_{\{L^n_{s'}\in\Lambda^n_{s'}\}}\big]<\infty$ by induction. 
Consequently, 
\beas 
E\big[R^n_{s-1}\> |\Psi^n_{s,\sfp,{\bf W}^n_s}|[1+|z_s|^\gamma]\>\Pi_{s'=1}^{s}1_{\{L^n_{s'}\in\Lambda^n_{s'}\}}\big] &<& \infty
\eeas
for $\gamma\in[0,\sfp_0]$, which ensures (\ref{0404111252}) for each $n$, in the present situation. 
Therefore, the functions $\wh{f}^n_{s-1}({\sred\ul{L}^n_{s-1},\>}\ul{\bbZ}^n_{s-1})$ are inductively well defined. 
In a similar way, since $f\in\wh{\cald}(\M,\gamma)\subset\cald(\M,\gamma)$, 
the conditions (\ref{0411261348}) for $s=S$ can be used to verify (\ref{0411261109}). 
In particular, each $f^n_s$ 
exists and is determined $P^{(\ul{L}^n_s,\ul{\bbZ}^n_s)}$-a.s.

Since Condition (\ref{0311191900}) is satisfied by (\ref{0412060447}), (\ref{0412060448}) and (\ref{0412060449}), 
we obtain, from Theorem \ref{0404231847}, 
\bea\label{0411270559}
\big|f^n_0-\wh{f}^n_0\big|
&\leq&
\sum_{s=1}^S(n_s)^{\delta_1}\Delta^n_s\big((\wh{g}_f)^n_s\big)+\sum_{s=2}^SV^{\delta_1,n}_{s-1}(f)
\eea
for $f\in\wh{\cald}(\M,\gamma)$, for any constant $\delta_1>0$. 
 
We apply Theorem \ref{0404131140} with $\delta$ for $s_T=s_n$. 
Since $|E_{s-1,\A^n_s}[\cdot]|=\big|E_{s-1,\A^n_s}\big[E_{s-1,\A^\infty_s}[\cdot]\big]\big|\leq E_{s-1,\A^n_s}\big[|E_{s-1,\A^\infty_s}[\cdot]|\big]$, and 
$(\wh{g}_f)^n_s(\ul{L}^n_{s},\ul{z}_{s}\big)=\Phi^{n_s}_s(\wh{g}_f)^n_s(\ul{L}^n_{s},\ul{z}_{s}\big)$ by (\ref{0404240241}), 
we obtain
\beas
\Delta^n_s\big((\wh{g}_f)^n_s\big)
&=&
\bigg\|E_{s-1,\A^n_s}\big[(\wh{g}_f)^n_s(\ul{L}^n_{s},\ul{\bbZ}^n_{s}\big)\big]
-\Psi^n_{s,\sfp,{\bf W}^n_s}\big[(\wh{g}_f)^n_s(\ul{L}^n_{s},\ul{\bbZ}^n_{s-1},\cdot\big)\big]\bigg\|_1
\nn\\&\leq&
\bigg\|E_{s-1,\A^\infty_s}\big[(\wh{g}_f)^n_s(\ul{L}^n_{s},\ul{\bbZ}^n_{s}\big)\big]
-\Psi^n_{s,\sfp,{\bf W}^n_s}\big[(\wh{g}_f)^n_s(\ul{L}^n_{s},\ul{\bbZ}^n_{s-1},\cdot\big)\big]\bigg\|_1
\nn\\&=&
\bigg\|\Phi^{n_s}_s\bigg(E_{s-1,\A^\infty_s}\big[(\wh{g}_f)^n_s(\ul{L}^n_{s},\ul{\bbZ}^n_{s}\big)\big]
-\Psi^n_{s,\sfp,{\bf W}^n_s}\big[(\wh{g}_f)^n_s(\ul{L}^n_{s},\ul{\bbZ}^n_{s-1},\cdot\big)\big]\bigg)\bigg\|_1
\nn\\&=&
\Delta^{\Phi*}_{n_s}\big((\wh{g}_f)^n_s(\ul{L}^n_{s},\ul{\bbZ}^n_{s-1},\cdot\big)\big), 
\eeas
where $\Delta^{\Phi*}_{n_s}$ is corresponding to $\bbZ^n_s=n_s^{-1/2}G^s_{n_s}Z^s_{n_s}$ in place of 
$Z_T^*\yeq T^{-1/2}G_TZ_T$ of (\ref{0404190439}). 
Due to the property (\ref{0411261348}), 
the random functions $(\wh{g}_f)^n_s(\ul{L}^n_{s},\ul{\bbZ}^n_{s-1},\cdot\big)$ in $\cale_s(M_s,\gamma)$ admit the estimate (\ref{0311221427a}) of Theorem \ref{0404131140}, and hence 
\bea\label{0412050613}
\sup_{\sff\in\wh{\cald}(\M,\gamma)}\Delta^{\Phi*}_{n_s}\big((\wh{g}_f)^n_s(\ul{L}^n_{s},\ul{\bbZ}^n_{s-1},\cdot\big)\big)
&\leq& 
M^{**}\bigg(P\bigg[\text{Var}_\calc\big[n_s^{-1/2}Z^s_{n_s}\big]<\delta I_{\sfd_s},\>{\fblue\Phi^{n_s}_s=1}\bigg]\bigg)^\theta
\nn\\&&\hspace{30pt}
+M^{**}n_s^{c''\gamma(1)}\delta^{-\gamma(2)}
\sup_{f\in\wh{\cald}(\M,\gamma)}\Xi^n_s\big((\wh{g}_f)^n_s,n_s^{-K}\big)
\bigg\}
\nn\\&&
+o(n_s^{-(\sfp-2+3\delta_1)/2})
\eea
for some positive constant $\delta_1$. 
By (\ref{0409171214}), we can make 
\beas 
n^{c''\gamma(1)}\sup_{f\in\wh{\cald}(\M,\gamma)}\Xi^n_s\big((\wh{g}_f)^n_s,n_s^{-K}\big)
&=& 
O(n^{-(\sfp-2+3\delta_1)/2})
\eeas
by choosing a sufficiently large $K$. 
The first term on the right-hand side of (\ref{0412050613}) vanishes thanks to $[B4]$ (i). 
Consequently we obtain 
\bea\label{0311270609}
\sup_{f\in\wh{\cald}(\M,\gamma)}
\Delta^n_s\big((\wh{g}_f)^n_s\big)
&=& 
O\big(n^{-(\sfp-2+3\delta_1)/2}\big).
\eea
Now (\ref{0411270559}) gives (\ref{04041734}). 
\qed
\halflineskip

Theorem \ref{0404171827} and (\ref{0411270646}) suggest use of ${\fred \wh{f}^n_{0}}$ to approximate ${\fred\E_{n}}$. 
\begin{en-text}
The $U_{s+1}^n$ is given in (\ref{0404111727}). 
Usually $U_{s+1}^n$ is uniformly bounded in $n$ or grows so slowly as to be controlled by the factor $n^{-\delta_1/2}$ 
in the error bound of (\ref{04041734}). 
It depends on how to cut off a bad event by the truncation function $1_{\{L^n_s\in\Lambda^n_{s}\}}$. 
\end{en-text}
{\fred 
Integrating what we obtained so far, 
for the functionals $\bbY^n_s=Y^n_s(L^n_s,\bbZ^n_s)$ of (\ref{0404200532}), 
we can approximate 
the expectation $\ol{{\bf E}}_n \yeq E\big[f\big({\sred\ul{L}_S^n,{\tred\ul{\bbY}^n_S}}\big)\big]$ 
at (\ref{0311191157}) by using the expectations ${\bf E}_{n}$ at (\ref{0311191836}), 
based on Theorem \ref{0311191131}. 
As shown in Theorem \ref{0404171827}, 
the expectation ${\bf E}_{n}= f^n_{0}$ is approximated by $\wh{f}^n_{0}$ 
defined by the backward scheme in (\ref{0311211418b}): 
\bea\label{0404200621} 
\wh{f}^n_{s-1}({\sred\ul{l}_{s-1},\>}\ul{z}_{s-1})
&=&
\int 
\wh{f}^n_{s}\big({\sred\ul{l}_{s},\ul{z}_{s}}\big)
{\fred \Psi^n_{s,\sfp,{\fred{\bf w}^{n}_s}}(dz_s)}
1_{\{l_s\in\Lambda^n_{s}\}}
\nu^n_{c_s}(dl_s,d{\bf w}^n_s)
q\big((\ul{l}_{s-1},\ul{z}_{s-1}),dc_{s}\big)
\nn\\&&
\eea
for $s\in\bbS$, 
where $\Psi_{n_s,p,\calc^s}$ is the random signed measure constructed along (\ref{0404190451})-(\ref{0404190509}) 
for $\bbZ^{n}_s$, having the representation (\ref{0404190639}). 
}

{\fblue 
\begin{remark}\label{0412010715}\rm 
As already mentioned in Remark \ref{0412010652}, 
the variable $\A^\infty_s$ has no special role, but it is only used for specifying measurability by the $\sigma$-field $\sigma[\A^\infty_s]$. 
We can incorporate covariates  into $\A^\infty_s$, if they exist. 
Therefore the setting we adopt here is fairly general. 
\end{remark}

We have validated the approximation of $\ol{\E}$ by the backward scheme $\wh{f}^n_0$, 
for the sequentially partial mixing processes. 
This method accommodates dependent systems at each stage. 
On the other hand, if they have conditional independency, i.e., perfect mixing, 
the conditions can fairly be simplified, as seen below. 
}

\begin{en-text}
\begin{remark}\rm 
Condition (\ref{0404111252}) is satisfied by $[B2]$ and $[B4]$. 
\end{remark}
\end{en-text}

}

{\tred 
\subsection{Sequentially independent experiments}\label{0404181058}
We consider the environment specified by (1)-{\vred(\ref{0404181100})} and (\ref{0404240241}) in Section \ref{0404180545}. 
Additionally, suppose that 
the sequence $(\ep_j)_{j\in\bbJ^\infty_s}$ is 
{\vred independent and satisfies
\begin{en-text}
\bea\label{0404200459} 
\sup_{j\in\bbJ^\infty_s}E_{\calc^s}[|\ep_j|^{\sfp+1}]<\infty\text{ and }
E_{\calc^s}[\ep_j]\yeq0\qquad(j\in\bbJ^\infty_s)
\eea
for every $s\in\bbS$. 
\end{en-text}
\bea\label{0404200459} 
\sup_{j\in\bbJ^\infty}E[|\ep_j|^{\sfp+1}]<\infty\text{ and }
E[\ep_j]\yeq0\qquad(j\in\bbJ^\infty). 
\eea
}
Condition $[B2]$ is rephrased as 
\bd
\im[{\bf [B2$'$]}] For every $L>0$, 
\beas 
\max_{s\in\bbS}\sup_{j\in\bbJ^\infty_s}E\big[|H^s_j|^L\big] &<& \infty. 
\eeas
\ed
\halflineskip

We replace $[B3]$ by a Cram\'er condition under the conditional probability: 
\bd
\im[{\bf [B3$'$]}]
There exist constants $B_0$, {\vred $\varepsilon>0$ and $\eta<1$} such that 
\begin{en-text}
\beas 
\max_{s\in\bbS}
\sup_{u\in\bbR^{\sfd_s}:|u|\geq B_0} {\fblue\sup_{j\in\bbJ^\infty_s}}\big|E_{\calc^s}[\exp(\tti u\cdot \ep_{{\fblue j}})]\big|
&<& 1. 
\eeas
\end{en-text}
\bea\label{0504081755}
\max_{s\in\bbS}
\sup_{u\in\bbR^{\sfd_s}:|u|\geq B_0} {\fblue\sup_{j\in\bbJ^\infty_s}}\big|{\vred E}[\exp(\tti u\cdot \ep_{{\fblue j}})]\big|
&<& 1,
\eea
{\vred 
and 
\bea\label{0504081756}
\max_{s\in\bbS}
P\bigg[\sum_{i=1}^T
1_{\big\{\lambda_{min}\big((H^s_{(s,i)})^\star H^s_{(s,i)}\big)< \varepsilon\big\}}>\eta\> T
\bigg]
&=&O(T^{-L})\qquad(T\to\infty)
\eea
for every $L>0$, 
where $\lambda_{min}$ denotes the minimum eigenvalue of a symmetric matrix.

}
\ed

Now we apply Theorem \ref{0404171827} to obtain 
\begin{theorem}\label{0404180641}
Let $\M=(M_s)_{s\in\bbS}\in(0,\infty)^S$ and ${\fblue\gamma\in[0,\sfp_0]}$. 
Suppose that $(\ep_j)_{j\in\bbJ^\infty_s}$ is 
{\vred an independent sequence}
satisfying (\ref{0404200459}) and that 
Conditions $[B2']$, $[B3']$, $[B4]$ {\fblue and (\ref{0409171214})
} 
are fulfilled. 
Then the estimate (\ref{04041734}) is valid for all $f\in\wh{\cald}(\M,\gamma)$ 
for {\fred some constants $\delta_1>0$ and} $M^*$. 
\end{theorem}
\proof 
{\fblue 
We verify the conditions of Theorem \ref{0404171827}. 
Condition $[B1]$ is satisfied because the sequence $(\ep_j)_{j\in\bbJ^\infty_s}$ is $\calc^s$-conditionally independent {\vred under (\ref{0404181100})}. 
Condition $[B2]$ is met under $[B2']$ and (\ref{0404200459}). 
Take $\calb^s_I$ so that $\calb_{[i]}^s=\sigma[\ep_{(s,i)}]$. 
Choose $u^s_{T,\ell}$ and $v^s_{T,\ell}$ in $[B3]$ as $u^s_{T,\ell}=\ell-1$ and $v^s_{T,\ell}={\vred\ell}$. 
The intervals $I^s_{T,\ell}=[\ell-1,\ell]$ form dense reduction intervals 
{\vred associated with $\wh{\calc}^s_{T,\ell}=\calc^s$.} 
A choice of $\psi^s_{T,\ell}$ is 
\beas 
\psi^s_{T,\ell}
&=& 
1_{\big\{\lambda_{min}\big((H^s_j)^\star H^s_j\big)\geq \varepsilon\big\}}, \qquad j=(s,\ell). 
\eeas
%
%
{\vred 
Wtih (\ref{0504081755}), we see 
the function $\wt{\Phi}^s_{T,\ell}$ in (\ref{0504081754}) of $[B3]$ is now 
\beas 
\wt{\Phi}^s_{T,\ell}
&=&
1_{\big\{\lambda_{min}\big((H^s_j)^\star H^s_j\big)\geq \varepsilon\big\}}
\eeas
if $B$ and $\eta_1$ are suitably chosen.
Therefore, Condition (\ref{0504081753}) is reduced to (\ref{0504081756}).}
Thus, $[B3']$ implies $[B3]$. 
We apply Theorem  \ref{0404171827} to obtain the result. 
\begin{en-text}
\beas 
\psi^s_{T,\ell}
&=& 
1_{\big\{\lambda_{min}\big((G^s_TH^s_j)^\star G^s_TH^s_j\big)\geq \varepsilon_1\big\}}, \qquad j=(s,\ell), 
\eeas
for some small positive constant $\varepsilon_1$, where $\lambda_{min}$ denotes the minimum eigenvalue of a symmetric matrix. 
\end{en-text}
\qed
}
\halflineskip
}

{\fblue 
If there exists a positive constant $\delta'$ such that 
\beas 
\text{Var}[\ep_j]>\delta'I_{\sfd_r}\qquad a.s.\quad(j\in\bbJ^\infty_s) 
\eeas
for every $s\in\bbS$, 
then $[B4]$ (i) is equivalent to the condition: 
there exists a positive constant $\delta$ such that 
\beas
n_s^{-1}\sum_{j\in\bbJ^n_s}H^s_j(H^s_j)^\star\geq\delta I_{\sfd_s}\qquad a.s. \text{ on }\{L^n_s\in\Lambda^n_{s}\}
\eeas
for $(s,c,n)\in\bbS\times\mfkC\times\bbN$. 
}

{\tred 
\section{
Linear and nonlinear statistics of hierarchically conditionally i.i.d. sequences}\label{0411110433}
We consider the setting in Section \ref{0404181058}. 
Additionally to Condition (\ref{0404181100}) of Section \ref{0404181101} (originally presented as (\ref{0311191900})), 
we assume that given $\calg^\infty_{s-1}$, the sequence 
$(\ep_j)_{j\in\bbJ^\infty_s}$ is a ($\calc^s$-conditionally) independent and identically distributed sequence. 
%
{\fblue The variables $(A_j)_{j\in\bbJ^n_s}$ describe the assignments in Stage $s$ as written just before (\ref{0311180317}).}
An example of $(A_j)_{j\in\bbJ^\infty_s}$ is 
a $\calg^\infty_{s-1}$-conditionally independent and identically distributed sequence with 
$
\call\{A_j|\calg^\infty_{s-1}\}\yeq \text{Multinomial}(1,\pi_s)
$
for $j\in\bbJ^n_s$, where $\pi_s\in\{p=(p_k)\in(0,1)^{\ol{k}_s};\>\sum_{k\in\calk_s}p_k=1\}$. 
However, any distribution of $(A_j)_{j\in\bbJ^\infty_s}$ is applicable in general. 
Even  $\calg^\infty_{s-1}$-conditional independency between $\{A_j\}_{j\in\bbJ^\infty_s}$ is not necessary. 
So it is possible to treat an assignment mechanism that keeps at least a fixed number of observations for each arm, for example. 
Suppose that 
{\fblue
\bea\label{0404181102} 
E[|\ep_j|^{\sfp+1}]=\beta_{s,\sfp+1},\ 
E[\ep_j]\yeq0\text{ and }\text{Var}[\ep_j]\yeq{\fblue \Sigma_s}\qquad(j\in\bbJ^\infty_s), 
\eea}
where $\beta_{s,\sfp+1}$ is a finite constant and 
{\fblue $\Sigma_s$ is a positive definite matrix}, for each $s\in\bbS$. 
Constancy of these conditional moments takes place since $(\ep_j)_{j\in\bbJ^\infty_s}$ is independent of $\calc^s$ 
{\fred (Condition (\ref{0404181100}) of Section \ref{0404181101}). 
In particular, the $\calc^s$-conditional cumulants of $\ep_j$ up to the order $\sfp+1$ are constants independent of $\calc^s$.}
{\fblue 
In this section, we will apply the backward expansion scheme validated in Section \ref{0311240905} 
to the batched bandit with assignment variables $A_j$. 
We will derive asymptotic expansion formulas for a linear functional and a nonlinear functional of the underlying noise sequence,  
in Sections \ref{0404200354} and \ref{0404200515}, respectively. 
and finally in Section \ref{0411101429} for the batched OLS of Zhang et al. \cite{zhang2020inference}. 
Due to replacement of $\sigma_s^2$ by its estimator, the resulting variable involves a nonlinear transform of $\bbZ^n_s$ 
in the first-order correction term of the stochastic expansion. 
The traditional approach by the Bhattacharya-Ghosh map gives an expansion of the density of the nonlinear functional, as mentioned in Section \ref{0412020311}. 
On the other hand, it is possible to numerically compute the expected value, which is the final goal in applications of the batched bandit, 
without the transform. A method of importance sampling will be discussed in Section \ref{0409171651}. 
}
\begin{en-text} 
though we admit more complicated cases, e.g. the higher-order of moments depend on $\calc^s$. 
It is also possible to relax the condition (\ref{0404181102}) as previously treated in the general setting. 
\end{en-text}

\begin{en-text}
Let $\calw^n_s=\calg^n_{t-1}\vee\sigma[\A^n_{t}]$ and 
$\calc^s=\vee_{n\in\bbN,t\in\bbT}\calw^n_s$. 
Recall that we are assuming that the $\sigma$-fields $\calw^n_s$ ($n\in\bbN$) have been 
constructed so that they are independent. 
The expansion of the $\calg_t$-conditional distribution of 
\beas
\bbZ^n_s &=& \sum_{j\in\bbJ^n_s}W^n_j\ep_j 
\eeas
of (\ref{0312071101}) is given as follows. 
The $\calc^s$-conditional cumulant functions $\chi_{n_s,r,\calc^s}$ of $\bbZ^n_s$ are 
defined by 
\beas 
\chi_{n_s,r,\calc^s}(u) 
&=& 
(\partial_\eta)^r_0\log E\big[\exp(\tti\eta u_t\cdot \bbZ_T^n)\big]
\eeas
for $u_t\in\bbR^{\sfd_t}$. 
The function $\wt{P}_{n_s,r,\calc^s}$ is defined by the formal expansion 
\bea\label{0312071538}
\exp\bigg(\sum_{r=2}^\infty\frac{1}{r!}\eta^{r-2}\chi_{n_s,r,\calc^s}(u)\bigg)
&=& 
\exp\bigg(\half\chi_{n_s,2,\calc^s}(u)\bigg)
+\sum_{r=1}^\infty\eta^rn_s^{-r/2}\wt{P}_{n_s,r,\calc^s}(u).
\eea
Then we make the expansion formula $\Psi_{n_s,rp\calc^s}$ by 
\beas 
\Psi_{n_s,r,\calc^s} &=& \calf^{-1}\big[\wh{\Psi}_{n_s,rp\calc^s}\big], 
\eeas
the Fourier inversion of $\wh{\Psi}_{n_s,p,\calc^s}$ given by 
\beas 
\wh{\Psi}_{n_s,p,\calc^s}(u) 
&=& 
\exp\bigg(\half\chi_{n_s,2,\calc^s}(u)\bigg)
+\sum_{r=1}^{p-2}n_s^{-r/2}\wt{P}_{n_s,r,\calc^s}(u).
\eeas
\end{en-text}

\subsection{Asymptotic expansion for a linear functional}\label{0404200354}
\begin{en-text}
{\color{gray}[0.3]
We consider the setting in Section \ref{0312071021}, that is, 
additionally to Condition (\ref{0311191900}), 
we assume that given $\calg^n_{t-1}$, 
each of 
$(\ep_j)_{j\in\bbJ^n_s}$ and $(A_j)_{j\in\bbJ^n_s}$ is  
an independent and identically distributed sequence, and 
$
\call\{A_j|\calg^n_{t-1}\}\yeq \text{Multinomial}(1,\pi_t)
$
for $j\in\bbJ^n_s$, 
except for the conditional Gaussianity of 
$
\call\{\ep_j|\calg^n_{t-1}\}
$
for $j\in\bbJ^n_s$. 

Let $\calw^n_s=\calg^n_{t-1}\vee\sigma[\A^n_{t}]$ and 
$\calc^s=\vee_{n\in\bbN,t\in\bbT}\calw^n_s$. 
Recall that we are assuming that the $\sigma$-fields $\calw^n_s$ ($n\in\bbN$) have been 
constructed so that they are independent. 
The expansion of the $\calg_t$-conditional distribution of 
\beas
\bbZ^n_s &=& \sum_{j\in\bbJ^n_s}W^n_j\ep_j 
\eeas
of (\ref{0312071101}) is given as follows. 
}
The $\calc^s$-conditional cumulant functions $\chi_{n_s,r,\calc^s}$ of $\bbZ^n_s$ are 
defined by 
\beas 
\chi_{n_s,r,\calc^s}(u) 
&=& 
(\partial_\eta)^r_0\log E\big[\exp(\tti\eta u_t\cdot \bbZ_T)\big]
\eeas
for $u_t\in\bbR^{\sfd_t}$. 
The function $\wt{P}_{n_s,r,\calc^s}$ is defined by the formal expansion 
\bea\label{0312071538}
\exp\bigg(\sum_{r=2}^\infty\frac{1}{r!}\eta^{r-2}\chi_{n_s,r,\calc^s}(u)\bigg)
&=& 
\exp\bigg(\half\chi_{n_s,2,\calc^s}(u)\bigg)
+\sum_{r=1}^\infty\eta^rn_s^{-r/2}\wt{P}_{n_s,r,\calc^s}(u).
\eea
Then we make the expansion formula $\Psi_{n_s,\sfp,\calc^s}$ by 
\beas 
\Psi_{n_s,\sfp,\calc^s} &=& \calf^{-1}\big[\wh{\Psi}_{n_s,\sfp,\calc^s}\big], 
\eeas
the Fourier inversion of $\wh{\Psi}_{n_s,\\sfp,calc^s}$ given by 
\beas 
\wh{\Psi}_{n_s,\sfp,\calc^s}(u) 
&=& 
\exp\bigg(\half\chi_{n_s,2,\calc^s}(u)\bigg)
+\sum_{r=1}^{p-2}n_s^{-r/2}\wt{P}_{n_s,r,\calc^s}(u).
\eeas
\end{en-text}

In Section \ref{0404200354}, each $\ep_j$ is one-dimensional. 
Let $N^n_s=\big(N^n_{s,k_s}\big)_{k_s\in\calk_s}$ with 
$N^n_{s,k_s}=\sum_{j\in\bbJ^n_s}A_{j,k_s}$ for $k_s\in\calk_s$, $s\in\bbS$. 
Suppose that 
\bea\label{0312071154}
W^n_j=w_{n,s}(N^n_s/n_s)A_j,
\eea
where $w_{n,s}:[0,1]^{\ol{k}_s}\to\bbR^{\ol{k}_s}\otimes\bbR^{\ol{k}_s}$, the set of $\ol{k}_t\times\ol{k}_s$-matrices, 
is a measurable map, 
and the function $w_{n,s}$ is given by 
\bea\label{0404181341}
w_{n,s}(x_1,...,x_{\ol{k}_s})
&=&
\text{diag}\big(g_1(x_1)x_1^{-1/2}n_{\sred s}^{-1/2},...,
g_{\ol{k}_s}(x_{\ol{k}_s})x_{\ol{k}_s}^{-1/2}n_{\sred s}^{-1/2}\big)
\eea
for some positive measurable functions $g_1,...,g_{\ol{k}_s}$. 
Moreover, we assume that each 
$w_{n,s}(N^n_s/n_s)$ is 
{\fblue well defined (i.e., $N^n_{s,k_s}>0$ a.s.) and} 
invertible whenever $L^n_s\in\Lambda^n_{s}$. 

We have
\beas 
\bbZ^n_s &=& \sum_{j\in\bbJ^n_s}W^n_j\ep_j 
\nn\\&=& 
\bigg(
{\sred 
g_1(N^n_{s,1}/n_s)(N^n_{s,1})^{-1/2}\sum_{j\in\bbJ^n_s}A_{j,1}\ep_j,
...,g_{\ol{k}_s}(N^n_{s,\ol{k}_s}/n_s)(N^n_{s,\ol{k}_s})^{-1/2}\sum_{j\in\bbJ^n_s}
A_{j,\ol{k}_s}\ep_j
}
\bigg)^\star.
\eeas
By the conditional independency of $\{\ep_j\}_{j\in\bbJ^n_s}$, 
\beas 
E_{\calc^s}\big[\exp(\tti\eta {\fblue u\cdot \bbZ_s^n})\big]
&=&
E_{\calc^s}\bigg[\exp\bigg(\sum_{k\in\calk_s}\eta u_k
{\sred 
g_k(N^n_{s,k}/n_s)(N^n_{s,k})^{-1/2}\sum_{j\in\bbJ^n_s}A_{j,k}\ep_j
}
\bigg)\bigg]
\nn\\&=&
\prod_{j\in\bbJ^n_s}
E_{\calc^s}\bigg[\exp\bigg(\eta \sum_{k\in\calk_s}u_k
{\sred 
g_k(N^n_{s,k}/n_s)(N^n_{s,k})^{-1/2}A_{j,k}\ep_j
}
\bigg)\bigg]
\eeas
for $u=(u_k)_{k=1,...,\ol{k}_s}\in\bbR^{\ol{k}_s}$ ($\sfd_s=\ol{k}_s$) and $\eta\in\bbR$, 
and hence
\beas 
\chi_{n_s,r,\calc^s}(u) 
&=&
(\partial_\eta)_0^r\log E\big[\exp(\tti\eta u\cdot \bbZ^{n_s}_s)\big]
\nn\\&=& 
\sum_{j\in\bbJ^n_s}
\tti^r
 \kappa_{r,\calc^s}
 \bigg(\sum_{k\in\calk_s}u_k
g_k(N^n_{s,k}/n_s)(N^n_{s,k})^{-1/2}A_{j,k}\ep_j\bigg)
\nn\\&&\hspace{80pt}
\quad(\kappa_{r,\calc^s}:\text{ the }r\text{-th }\calc^s\text{-conditional cumulant})
\nn\\&=& 
\sum_{j\in\bbJ^n_s}\tti^r\bigg(\sum_{k\in\calk_s}u_k
g_k(N^n_{s,k}/n_s)(N^n_{s,k})^{-1/2}A_{j,k}\bigg)^r
\kappa_{r,\calc^s}(\ep_{(s,1)})
\nn\\&=& 
\sum_{j\in\bbJ^n_s}\sum_{k\in\calk_s}(\tti u_k)^r
\big\{g_k(N^n_{s,k}/n_s)\big\}^r(N^n_{s,k})^{-r/2}A_{j,k}\>
\kappa_{r,\calc^s}(\ep_{(s,1)})
\nn\\&=& 
\sum_{k\in\calk_s}(\tti u_k)^r
\big\{g_k(N^n_{s,k}/n_s)\big\}^r(N^n_{s,k})^{-(r-2)/2}\>
\kappa_{r,\calc^s}(\ep_{(s,1)})
\nn\\&=&
\sum_{k\in\calk_s} n_s^{-(r-2)/2}\lambda_{n_s,r,\calc^s,k}(\tti u_k)^r,
\eeas
where the coefficients
\beas 
\lambda_{n_s,r,\calc^s,k}
&=& 
\big\{g_k(N^n_{s,k}/n_s)\big\}^r(N^n_{s,k}/n_s)^{-(r-2)/2}\>
\kappa_{r,\calc^s}(\ep_{(s,1)})
\eeas
are random. 
Remark that the conditional variance of $\bbZ^{n_s}_s$ is a diagonal matrix 
due to the orthogonality between $(A_{j,k})_{k\in\calk_s}$. 

For example, 
if the functions $g_{k_s}$ are identically equals to $1$, as 
in the case of the batched OLS by Zhang et al. \cite{zhang2020inference}, we obtain 
\beas
\chi_{n_s,2,\calc^s}(u) 
&=&
\sum_{k\in\calk_s}(\tti u_k)^2\kappa_{2,\calc^s}(\ep_{(s,1)})
\nn\\
\chi_{n_s,r,\calc^s}(u) 
&=&
\sum_{k\in\calk_s}(\tti u_k)^r
(N^n_{s,k})^{-(r-2)/2}\>
\kappa_{r,\calc^s}(\ep_{(s,1)})
\quad(r\geq3),
\eeas
and in particular, 
\beas 
\lambda_{n_s,2,\calc^s,k} &=& \kappa_{2,\calc^s}(\ep_{(s,1)})=:\sigma_s^2.
\eeas

For the general functions $g_k$, the formula (\ref{0404190639}) shows 
the function $d\Psi^n_{s,\sfp,{\fred{\bf w}^{n}_s}}/dz(z)$ 
($z=(z_1,...,z_{\ol{k}_s})$) is 
{\fblue 
\bea\label{0404200632} &&
\prod_{k\in\calk_s} 
\phi(z_k;\sigma_s^2)
\times
\prod_{k\in\calk_s} 
\bigg\{1
+ \frac{1 }{6}n_s^{-1/2}\lambda_{n_s,3,\calc^s,k}h_3(z_k; \sigma_s^2)
\nn\\&&\hspace{30pt}
+n_s^{-1}\bigg(\frac{1}{24} \lambda_{n_s,4,\calc^s,k}h_4(z_k;\sigma_s^2)
+\frac{1}{72} \lambda_{nn_s,3,\calc^s,k}^2h_6(z_k;\sigma_s^2)
\bigg)+\cdots
\bigg\}, 
\eea
}
the summation taken 
up to order $n_s^{-(p-2)/2}$.

\begin{en-text}
\subsection{An example of the formula (\ref{0404200632})}
Fill ! \koko
\end{en-text}

\subsection{Asymptotic expansion for a normalized estimator}\label{0404200515}
In this section, we assume that $W^n_j$ takes the form 
{\sblue 
\bea\label{0403271023}
W^n_j&=&
\begin{pmatrix}
w_{n,s}(N^n_s/n_s)n_s^{-1/2}A_{j}& {\bf 0}\\ 
0& n_s^{-1/2}
\end{pmatrix}
\qquad(j\in\bbJ^n_s)
\eea
where $w_{n,s}:[0,1]^{\ol{k}_s}\to\bbR^{\ol{k}_s}\otimes\bbR^{\ol{k}_s}$, the set of $\ol{k}_s\times\ol{k}_s$-matrices, 
is a measurable map, and the function $w_{n,s}$ is given by} 
(\ref{0404181341}). 
}
As before, $N^n_s=\big(N^n_{s,k_s}\big)_{k_s\in\calk_s}$ with 
$N^n_{s,k_s}=\sum_{j\in\bbJ^n_s}A_{j,k_s}$ for $k_s\in\calk_s$, $s\in\bbS$. 
We assume that each 
$w_{n,s}(N^n_s/n_s)$ is invertible whenever $L^n_s\in\Lambda^n_{s}$. 
We consider the error term $\ep_j$ set by 
\beas 
\ep_j&=& 
\begin{pmatrix}
\dot{\ep}_j \y \dot{\ep}_j ^2-\sigma_s^2
\end{pmatrix}
\eeas
when $j\in\bbJ^\infty_s$, where $\dot{\ep}_j$ is a one-dimensional random variable {\fblue independent of $\calc^s$ and 
satisfying $E[\dot{\ep}_j]=0$ and $\text{Var}[\dot{\ep}_j]=\sigma_s^2$, a positive constant.}
Then 
\bea\label{0404242321}
\bbZ^n_s 
&=&
 \sum_{j\in\bbJ^n_s}W^n_j\ep_j 
\yeq
\big({(\dot{\bbZ}^n_s})^\star,\ddot{\bbZ}^n_s\big)^\star
\eea
for
\bea\label{0404242322}
\dot{\bbZ}^n_s
&=&
\bigg(
{\sred 
g_1(N^n_{s,1}/n_s)(N^n_{s,1})^{-1/2}\sum_{j\in\bbJ^n_s}A_{j,1}{\sblue\dot{\ep}_j,}
...,g_{\ol{k}_s}(N^n_{s,\ol{k}_s}/n_s)(N^n_{s,\ol{k}_s})^{-1/2}\sum_{j\in\bbJ^n_s}
A_{j,\ol{k}_s}{\sblue\dot{\ep}_j}
}\bigg)^\star
\eea
and 
\bea\label{0404242323}
\ddot{\bbZ}^n_s
&=&
{\sblue n_s^{-1/2}\sum_{j\in\bbJ^n_s}(\dot{\ep}_j^2-\sigma_s^2)}.
\eea

By the conditional independency of $\{\ep_j\}_{j\in\bbJ^n_s}$, 
\beas &&
E_{\calc^s}\big[\exp(\tti\eta u\cdot \bbZ^n_s)\big]
\nn\\&=&
E_{\calc^s}\bigg[\exp\bigg(\sum_{k\in\calk_s}\tti\eta u_k
{\sred 
g_k(N^n_{s,k}/n_s)(N^n_{s,k})^{-1/2}\sum_{j\in\bbJ^n_s}A_{j,k}{\sblue\dot{\ep}_j}
+\tti\eta u_0n_s^{-1/2}\sum_{j\in\bbJ^n_s}(\dot{\ep}_j^2-\sigma_s^2)
}
\bigg)\bigg]
\nn\\&=&
\prod_{j\in\bbJ^n_s}
E_{\calc^s}\bigg[\exp\bigg(\tti\eta \sum_{k\in\calk_s}u_k
{\sred 
g_k(N^n_{s,k}/n_s)(N^n_{s,k})^{-1/2}\sum_{j\in\bbJ^n_s}A_{j,k}{\sblue\dot{\ep}_j}
{\sblue +\tti\eta u_0n_s^{-1/2}(\dot{\ep}_j^2-\sigma_s^2)}
}
\bigg)\bigg]
\eeas
{\sblue for $u=\big((u_k)_{k\in\calk_s},u_0\big)\in\bbR^{\ol{k}_s+1}$ and $\eta\in\bbR$,} 
and hence
\beas 
\chi_{n_s,r,\calc^s}(u) 
&=&
(\partial_\eta)_0^r\log E_{\calc^s}\big[\exp(\tti\eta u\cdot \bbZ_s^n)\big]
\nn\\&=& 
\sum_{j\in\bbJ^n_s}
\tti^r
 \kappa_{\calc^s,r}
 \bigg(\sum_{k\in\calk_s}u_k
g_k(N^n_{s,k}/n_s)(N^n_{s,k})^{-1/2}A_{j,k}{\sblue\dot{\ep}_j}
{\sblue + u_0n_s^{-1/2}(\dot{\ep}_j^2-\sigma_s^2)}
\bigg)
\nn\\&&\hspace{100pt}
\quad(\kappa_{\calc^s,r}:\text{ the }r\text{-th } \calc^s\text{-conditional cumulant})
\eeas
Special cases are 
\beas &&
\kappa_{\calc^s,2}
 \bigg(\sum_{k\in\calk_s}u_k
g_k(N^n_{s,k}/n_s)(N^n_{s,k})^{-1/2}A_{j,k}{\sblue\dot{\ep}_j}
{\sblue + u_0n_s^{-1/2}(\dot{\ep}_j^2-\sigma_s^2)}
\bigg)
\nn\\&=&
E_{\calc^s}\bigg[
\bigg(\sum_{k\in\calk_s}u_k
g_k(N^n_{s,k}/n_s)(N^n_{s,k})^{-1/2}A_{j,k}{\sblue\dot{\ep}_j}
{\sblue + u_0n_s^{-1/2}(\dot{\ep}_j^2-\sigma_s^2)}
\bigg)^2
\bigg]
\nn\\&=&
\sum_{k\in\calk_t}u_k^2
g_k(N^n_{s,k}/n_s)^2(N^n_{s,k})^{-1}A_{j,k}\sigma_s^2\quad(\kappa[\dot{\ep}]=:\sigma_s^2)
\nn\\&&
+2\sum_{k\in\calk_s}u_ku_0
g_k(N^n_{s,k}/n_s)(N^n_{s,k})^{-1/2}A_{j,k}n_s^{-1/2}
E[\dot{\ep}_1^3]
\nn\\&&
+u_0^2n_s^{-1}
\big(E[\dot{\ep}_1^4]-\sigma_s^4\big)
\eeas
and 
\beas &&
\kappa_{\calc^s,3}
 \bigg(\sum_{k\in\calk_s}u_k
g_k(N^n_{s,k}/n_s)(N^n_{s,k})^{-1/2}A_{j,k}{\sblue\dot{\ep}_j}
{\sblue + u_0n_s^{-1/2}\sum_{j\in\bbJ^n_s}(\dot{\ep}_j^2-\sigma_s^2)}
\bigg)
\nn\\&=&
E_{\calc^s}\bigg[
\bigg(\sum_{k\in\calk_s}u_k
g_k(N^n_{s,k}/n_s)(N^n_{s,k})^{-1/2}A_{j,k}{\sblue\dot{\ep}_j}
{\sblue + u_0n_s^{-1/2}(\dot{\ep}_j^2-\sigma_s^2)}
\bigg)^3\bigg]
\nn\\&=&
E_{\calc^s}\bigg[
\bigg(\sum_{k\in\calk_s}u_k
g_k(N^n_{s,k}/n_s)(N^n_{s,k})^{-1/2}A_{j,k}{\sblue\dot{\ep}_j}
\bigg)^3\bigg]
\nn\\&&
+3E_{\calc^s}\bigg[
\bigg(\sum_{k\in\calk_s}u_k
g_k(N^n_{s,k}/n_s)(N^n_{s,k})^{-1/2}A_{j,k}{\sblue\dot{\ep}_j}\bigg)^2
{\sblue u_0n_s^{-1/2}(\dot{\ep}_j^2-\sigma_s^2)}\bigg]
\nn\\&&
+3E_{\calc^s}\bigg[
\bigg(\sum_{k\in\calk_s}u_k
g_k(N^n_{s,k}/n_s)(N^n_{s,k})^{-1/2}A_{j,k}{\sblue\dot{\ep}_j}\bigg)
\bigg({\sblue u_0n_s^{-1/2}(\dot{\ep}_j^2-\sigma_s^2)}
\bigg)^2\bigg]
\nn\\&&
+E_{\calc^s}\bigg[
\bigg({\sblue u_0n_s^{-1/2}(\dot{\ep}_j^2-\sigma_s^2)}
\bigg)^3\bigg]
\nn\\&=&
\sum_{k\in\calk_s}u_k^3 g_k(N^n_{s,k}/n_s)^3(N^n_{s,k})^{-3/2}A_{j,k}
E[{\sblue\dot{\ep}_j}^3]
\nn\\&&
+3\sum_{k\in\calk_s}u_k^2u_0 g_k(N^n_{s,k}/n_s)^2(N^n_{s,k})^{-1}n_s^{-1/2}A_{j,k}
\big(E\big[{\sblue\dot{\ep}_j}^4\big]-\sigma_s^4\big)
\nn\\&&
+3\sum_{k\in\calk_s}u_ku_0^2
g_k(N^n_{s,k}/n_s)(N^n_{s,k})^{-1/2}A_{j,k}n_s^{-1}
\big(E\big[{\sblue\dot{\ep}_j}^5\big]-2\sigma_s^2E\big[{\sblue\dot{\ep}_j}^3\big]\big)
\nn\\&&
+u_0^3n_s^{-3/2}
\big(E\big[{\sblue\dot{\ep}_j}^6\big]-3\sigma_s^2E\big[{\sblue\dot{\ep}_j}^4\big]
+2\sigma_s^6\big).
\eeas
Therefore, 
\bea\label{0404210326} 
\chi_{n_s,2,\calc^s}(u) 
&=&
\sum_{j\in\bbJ^n_s}
\tti^2
 \kappa_{\calc^s,2}
 \bigg(\sum_{k\in\calk_s}u_k
g_k(N^n_{s,k}/n_s)(N^n_{s,k})^{-1/2}A_{j,k}{\sblue\dot{\ep}_j}
{\sblue + u_0n_s^{-1/2}(\dot{\ep}_j^2-\sigma_s^2)}
\bigg)
\nn\\&=&
\sum_{k\in\calk_s}(\tti u_k)^2
g_k(N^n_{s,k}/n_s)^2\sigma_s^2
\nn\\&&
+2\sum_{k\in\calk_s}(\tti u_k)(\tti u_0)
g_k(N^n_{s,k}/n_s)(N^n_{s,k})^{1/2}n_s^{-1/2}E[\dot{\ep}_1^3]
\nn\\&&
+(\tti u_0)^2
\big(E[\dot{\ep}_1^4]-\sigma_s^4\big)
\eea
and
\bea\label{0404210327}
\chi_{n_s,3,\calc^s}(u) 
&=&
\sum_{j\in\bbJ^n_s}
\tti^3
 \kappa_{\calc^s,3}
 \bigg(\sum_{k\in\calk_s}u_k
g_k(N^n_{s,k}/n_s)(N^n_{s,k})^{-1/2}A_{j,k}{\sblue\dot{\ep}_j}
{\sblue + u_0n_s^{-1/2}(\dot{\ep}_j^2-\sigma_s^2)}
\bigg)
\nn\\&=&
\sum_{k\in\calk_s}(\tti u_k)^3 g_k(N^n_{s,k}/n_s)^3(N^n_{s,k})^{-1/2}
E\big[{\sblue\dot{\ep}_1}^3\big]
\nn\\&&
+3\sum_{k\in\calk_s}(\tti u_k)^2(\tti u_0) g_k(N^n_{s,k}/n_s)^2n_s^{-1/2}
\big(E\big[{\sblue\dot{\ep}_1}^4\big]-\sigma_s^4\big)
\nn\\&&
+3\sum_{k\in\calk_s}(\tti u_k)(\tti u_0)^2
g_k(N^n_{s,k}/n_s)(N^n_{s,k})^{1/2}n_s^{-1}
\big(E\big[{\sblue\dot{\ep}_1}^5\big]-2\sigma_s^2E\big[{\sblue\dot{\ep}_1}^3\big]\big)
\nn\\&&
+(\tti u_0)^3n_s^{-1/2}
\big(E\big[{\sblue\dot{\ep}_1}^6\big]-3\sigma_s^2E\big[{\sblue\dot{\ep}_1}^4\big]
+2\sigma_s^6\big).
\eea

Let us consider normalized variable
\beas
\mfkT^n_s
&=& 
\frac{\dot{\bbZ}^n_s}{\wh{\sigma}_s}
\eeas
for $s\in\bbS$, where $\wh{\sigma}_s^2$ is a $\calg^n_s$-measurable statistic. 
%
Suppose that 
$\wh{\sigma}_s^2$ admits a stochastic expansion 
\bea\label{0412020825}
\wh{\sigma}_s^2 
&=& 
n_s^{-1}\sum_{j\in\bbJ^n_s}\dot{\ep}_j^2+R^s_{n_s}
\eea
with $R^s_{n_s}=O_{L^p}(n_s^{-1})$. 
Then $\mfkT^n_s$ has a stochastic expansion 
\beas 
\mfkT^n_s
&=&
\sigma_s^{-1}\dot{\bbZ}^n_s-\frac{1}{2\sigma_s^3}\dot{\bbZ}^n_s\big(\wh{\sigma}_s^2-\sigma_s^2)+\cdots
\nn\\&=&
\sigma_s^{-1}\dot{\bbZ}^n_s-\frac{1}{2\sigma_s^3n_s^{1/2}}\ddot{\bbZ}^n_s\dot{\bbZ}^n_s+\wt{R}^s_{n_s},
\eeas
where $\wt{R}^s_{n_s}=O_{L^\inftym}(n_s^{-1})$. 
Remark that $\ddot{\bbZ}^n_s$ is a scalar random variable, while $\dot{\bbZ}^n_s$ is $\ol{k}_s$-dimensional. 

We set 
\bea\label{0404231626} 
Y^n_s(z_s) &=& \sigma_s^{-1}\dot{z}_s-\frac{1}{2\sigma_s^3n_s^{1/2}}\ddot{z}_s\dot{z}_s
\eea
for $z_s=(\dot{z}_s,\ddot{z}_s)$. 
The map $Y^n_s$ is $\bbR^{\ol{k}_s}$-valued. 
Then 
\bea\label{0404242257}
\mfkT^n_s
&=& 
Y^n_s(\bbZ^n_s)+\wt{R}^s_{n_s}.
\eea

If we are interested in a linear combination 
\bea\label{0404200852} 
\mfkT^n 
&=& 
\sum_{s\in\bbS}\dot{L}^n_s\mfkT^n_s
\eea
of $\mfkT^n_s$ with coefficient matrices $\dot{L}^n_s$, each of which is a function of the components of $L^n_s$, then 
computation of the expectation $E[g(\mfkT^n)]$ results in 
that of $E\big[f(\ul{L}^n_S,\ul{\mfkT}^n_S)\big]$ for 
\beas 
f(\ul{l}_S,\ul{y}_S) 
&=& 
g\bigg(\sum_{s\in\bbS}\dot{l}_s,y_s\bigg), 
\eeas
where  the argument $\dot{l}_s$ is for $\dot{L}^n_s$, and 
$y_s$ for $\mfkT^n_s$. 
Moreover, $E\big[f(\ul{L}^n_S,\ul{\mfkT}^n_S)\big]$ can be approximated by $E\big[f(\ul{L}^n_S,\ul{\bbY}^n_S)\big]$.  
Then the error caused by the shift $\wt{R}^s_{n_s}$ 
{\fblue can be included in a measure-theoretic modulus of continuity associated with $f$ on the event $\big\{|\wt{R}^s_{n_s}|\leq n^{-\xi}\big\}$, $\xi\in(1/2,1)$, 
while the contribution of the event $\big\{|\wt{R}^s_{n_s}|>n^{-\xi}\big\}$ is negligible. 
We refer the reader to Sakamoto and Yoshida \cite{SakamotoYoshida2004} for explanation of this idea. 
Therefore, the error caused by removing $\wt{R}^s_{n_s}$ is negligible if $f$ is sufficiently regular in measure. 
This is easy to understand if we consider an indicator function of a rectangle $\{\cdot\leq a\}$ for $a\in\bbR^{\sfd_s}$ since 
$\sup_n\big\|\Psi^n_{s,\sfp,{\bf w}^n_s}1_{[\{\cdot\leq a+h\}]}-\Psi^n_{s,\sfp,{\bf w}^n_s}1_{[\{\cdot\leq a\}]}\big\|_1=O(|h|)$ as $h\to0$ for $h\in\bbR^{\sfd_s}$.}
%
Finally, the expectation $E\big[f(\ul{L}^n_S,\ul{\bbY}^n_S)\big]$ can be approximated 
by using the backward scheme (\ref{0404200621}) incorporating the asymptotic expansion method.

\subsection{Reward model (\ref{0311180317}): a test statistic as $\mfkT^n$ of (\ref{0404200852})}\label{0411101429}
\subsubsection{{\fblue Expressing $\wh{f}^n_0$}}
Zhang et al. \cite{zhang2020inference} studied a batched bandit in which 
the reward (effect) $R_j$ for the action $A_j$ expressed by (\ref{0311180317}): 
\beas
R_j
&=&
A_j^\star\beta_{s(j)}+{\sblue\dot{\ep}_j}
\quad
(j\in\bbJ^n).
\eeas
When the bandits have two arms (i.e., $\ol{k}_s=2$), they proposed 
using the test statistic
\beas 
\mfkU^n 
&=& 
S^{-1/2}\sum_{s\in\bbS}\frac{(N^n_{s,1}N^n_{s,2})^{1/2}}{n_s^{1/2}\wh{\sigma}_s}\big(\wh{\Delta}_s-a\big)
\eeas
for testing the null hypothesis that the treatment has no effect, i.e., $H_0$: $a=0$, 
where the batched OLS $\wh{\Delta}_s$ at Stage $s$ is defined by
\beas 
\wh{\Delta}_s
&=& 
\frac{\sum_{j\in\bbJ^n_s}A_{j,1}R_j}{\sum_{j\in\bbJ^n_s}A_{j,1}}
-\frac{\sum_{j\in\bbJ^n_s}A_{j,2}R_j}{\sum_{j\in\bbJ^n_s}A_{j,2}}.
\eeas

Consider the functional 
$
\mfkT^n_s
=
\dot{\bbZ}^n_s/\wh{\sigma}_s
$
with
\beas
\dot{\bbZ}^n_s
&=&
\bigg(
{\sred 
(N^n_{s,1})^{-1/2}\sum_{j\in\bbJ^n_s}A_{j,1}\dot{\ep}_j,
(N^n_{s,2})^{-1/2}\sum_{j\in\bbJ^n_s}A_{j,2}\dot{\ep}_j
}\bigg)^\star
\nn\\&=&
\begin{pmatrix}
(N^n_{s,1}/n_s)^{-1/2}&0\vspace{1mm}\\
0&(N^n_{s,2}/n_s)^{-1/2}
\end{pmatrix}
\begin{pmatrix}
n_s^{-1/2}\sum_{j\in\bbJ^n_s}A_{j,1}\dot{\ep}_j\\
n_s^{-1/2}\sum_{j\in\bbJ^n_s}A_{j,2}\dot{\ep}_j
\end{pmatrix},
\eeas
$g_k(x)=1$ for $k\in\calk_s=\{1,2\}$, and 
{\fred 
\beas
\wh{\sigma}_s^2 
&=& 
\frac{1}{n_s-1}\sum_{j\in\bbJ^n_s}\big(R_j-A_j^\star\wh{\beta}_s\big)^2
\eeas
with 
\beas 
\wh{\beta}_s
&=&
\bigg(\frac{\sum_{j\in\bbJ^n_s}A_{j,1}R_j}{\sum_{j\in\bbJ^n_s}A_{j,1}}, \frac{\sum_{j\in\bbJ^n_s}A_{j,2}R_j}{\sum_{j\in\bbJ^n_s}A_{j,2}}\bigg)^\star. 
\eeas
The estimate $\wh{R}^s_{n_s}=O_{L^p}(n_s^{-1})$ is valid under suitable moment conditions in the representation (\ref{0412020825}) of $\wh{\sigma}^2_s$. 
{\vred 
We remark that 
\bea\label{0411231416}
\begin{pmatrix}
(N^n_{s,1}/n_s)^{-1/2}&0\vspace{1mm}&0\\
0&(N^n_{s,2}/n_s)^{-1/2}&0\\
0&0&1
\end{pmatrix}
\begin{pmatrix}
n_s^{-1/2}\sum_{j\in\bbJ^n_s}A_{j,1}\dot{\ep}_j\\
n_s^{-1/2}\sum_{j\in\bbJ^n_s}A_{j,2}\dot{\ep}_j\\
n_s^{-1/2}\sum_{j\in\bbJ^n_s}(\dot{\ep}_j^2-\sigma_s^2)
\end{pmatrix}
&=:&
G^s_{n_s}\>n_s^{-1/2}Z_{n_s}^s
\eea
for
\beas 
Z^s_{n_s}
&=&
\sum_{j\in\bbJ^n_s}H^s_j\ep_j
\eeas
with 
\beas
H^s_j
\yeq
\begin{pmatrix}
A_{j,1}&0\\
A_{j,2}&0\\ 
0&1
\end{pmatrix}
\quad{\text{and}}\quad
\ep_j
\yeq
\begin{pmatrix}
\dot{\ep}_j\\
\dot{\ep}_j^2-\sigma_s^2
\end{pmatrix}.
\eeas
}
}

Under $H_0$: $a=0$, we have 
\beas 
\mfkT^n
&:=&
\mfkU^n 
\nn\\&=& 
S^{-1/2}\sum_{s\in\bbS}\frac{(N^n_{s,1}N^n_{s,2})^{1/2}}{n_s^{1/2}\wh{\sigma}_s}
\bigg(\frac{\sum_{j\in\bbJ^n_s}A_{j,1}R_j}{\sum_{j\in\bbJ^n_s}A_{j,1}}
-\frac{\sum_{j\in\bbJ^n_s}A_{j,2}R_j}{\sum_{j\in\bbJ^n_s}A_{j,2}}\bigg)
\nn\\&=&
S^{-1/2}\sum_{s\in\bbS}\frac{(N^n_{s,1}N^n_{s,2})^{1/2}}{n_s^{1/2}\wh{\sigma}_s}
\bigg(\frac{\sum_{j\in\bbJ^n_s}A_{j,1}\dot{\ep}_j}{\sum_{j\in\bbJ^n_s}A_{j,1}}
-\frac{\sum_{j\in\bbJ^n_s}A_{j,2}\dot{\ep}_j}{\sum_{j\in\bbJ^n_s}A_{j,2}}\bigg)
\nn\\&=&
S^{-1/2}\sum_{s\in\bbS}\frac{(N^n_{s,1}N^n_{s,2})^{1/2}}{n_s^{1/2}\wh{\sigma}_s}
\bigg(\frac{(N^n_{s,1})^{1/2}}{\sum_{j\in\bbJ^n_s}A_{j,1}}
(N_{s,1}^n)^{-1/2}\sum_{j\in\bbJ^n_s}A_{j,1}\dot{\ep}_j
\nn\\&&\hspace{130pt}
-\frac{(N^n_{s,22})^{1/2}}{\sum_{j\in\bbJ^n_s}A_{j,2}}
(N_{s,2}^n)^{-1/2}\sum_{j\in\bbJ^n_s}A_{j,2}\dot{\ep}_j
\bigg)
\nn\\&=&
S^{-1/2}\sum_{s\in\bbS}\frac{(N^n_{s,1}N^n_{s,2})^{1/2}}{n_s^{1/2}}
\big((N^n_{s,1})^{-1/2},\>-(N^n_{s,2})^{-1/2}\big)\>
\mfkT^n_s
\nn\\&=&
\sum_{s\in\bbS}\dot{L}^n_s\mfkT^n_s
\eeas
for 
\bea\label{0404242259}
\dot{L}^n_s
&=&
S^{-1/2}\big((N^n_{s,2}/n_s)^{1/2},\>-(N^n_{s,1}/n_s)^{1/2}\big). 
\eea
Therefore, we can apply the asymptotic expansion for $\mfkT^n$ of 
(\ref{0404200852}) to $\mfkU^n$, 
{\fblue suppose that $N^n_{s,k}/n_s$ are bounded from below on the support of the probability. }

Suppose that 
\beas 
E[\dot{\ep}_j]\yeq0,\quad
E[\dot{\ep}_j^2]\yeq\sigma_s^2,\quad
E[\dot{\ep}_j^r]\yeq\mu_{s,r}\quad(r=3,...,6)
\eeas
for $j\in\bbJ^\infty_s$, for some constants $\sigma_s^2$ and $\mu_{s,r}$. 
From (\ref{0404210326}) and (\ref{0404210327}), for $g_k(x)=1$, we obtain 
\bea\label{0404210341} 
\chi_{n_s,2,\calc^s}(u) 
&=&
\sum_{k\in\calk_s}(\tti u_k)^2
\sigma_s^2
+2\sum_{k\in\calk_s}(\tti u_k)(\tti u_0)
(N^n_{s,k})^{1/2}n_s^{-1/2}\mu_{s,3}
\nn\\&&
+(\tti u_0)^2
\big(\mu_{s,4}-\sigma_s^4\big)
\eea
and
\bea\label{0404210342}
\chi_{n_s,3,\calc^s}(u) 
&=&
\sum_{k\in\calk_s}(\tti u_k)^3 (N^n_{s,k})^{-1/2}
\mu_{s,3}
\nn\\&&
+3\sum_{k\in\calk_s}(\tti u_k)^2(\tti u_0) n_s^{-1/2}
\big(\mu_{s,4}-\sigma_s^4\big)
\nn\\&&
+3\sum_{k\in\calk_s}(\tti u_k)(\tti u_0)^2
(N^n_{s,k})^{1/2}n_s^{-1}
\big(\mu_{s,5}-2\sigma_s^2\mu_{s,3}\big)
\nn\\&&
+(\tti u_0)^3n_s^{-1/2}
\big(\mu_{s,6}-3\sigma_s^2\mu_{s,4}
+2\sigma_s^6\big).
\eea

For the random matrix 
\beas 
V^{n_s}_s 
&=& 
\begin{pmatrix}
\sigma_s^2&0&(N^n_{s,1})^{1/2}n_s^{-1/2}\mu_{s,3}\\
0&\sigma_s^2&(N^n_{s,2})^{1/2}n_s^{-1/2}\mu_{s,3}\\
(N^n_{s,1})^{1/2}n_s^{-1/2}\mu_{s,3}&(N^n_{s,2})^{1/2}n_s^{-1/2}\mu_{s,3}&\mu_{s,4}-\sigma_s^4
\end{pmatrix},
\eeas
the Hermite polynomials $H_{\alpha_1,\alpha_2,\alpha_3}$ are defined by 
\beas 
H_{\alpha_1,\alpha_2,\alpha_3}(z_1,z_2,z_0;V^{n_s}_s )
&=& 
\phi\big((z_1,z_2,z_0);0,V^{n_s}_s \big)^{-1}
\nn\\&&\hspace{10pt}\times
(-\partial_{z_{\alpha_1}})(-\partial_{z_{\alpha_2}})(-\partial_{z_{\alpha_3}})
\phi\big((z_1,z_2,z_0);0,V^{n_s}_s \big)
\eeas
for $(\alpha_1,\alpha_2,\alpha_3)\in\{1,2,0\}^3$.

Now, 
the first-order asymptotic expansion formula for $\call\{\bbZ^n_s|\calc^s\}$ is obtained 
from (\ref{0404190639}) as 
\bea\label{0404210421}
\frac{d\Psi^n_{s,\sfp,{\fred{\bf w}^{n}_s}}}{dz}(z)
&=&
\phi\big((z_1,z_2,z_0);0,V^{n_s}_s \big)
\nn\\&&
+\frac{1}{6}n_s^{-1/2}\phi\big((z_1,z_2,z_0);0,V^{n_s}_s \big)
\nn\\&&\times
\bigg\{
\sum_{k\in\calk_s} (N^n_{s,k})^{-1/2}n_s^{1/2}\mu_{s,3}H_{k,k,k}(z_1,z_2,z_0;V^{n_s}_s )
\nn\\&&\hspace{20pt}
+3\sum_{k\in\calk_s}
\big(\mu_{s,4}-\sigma_s^4\big)H_{k,k,0}(z_1,z_2,z_0;V^{n_s}_s )
\nn\\&&\hspace{20pt}
+3\sum_{k\in\calk_s}
(N^n_{s,k})^{1/2}n_s^{-1/2}
\big(\mu_{s,5}-2\sigma_s^2\mu_{s,3}\big)H_{k,0,0}(z_1,z_2,z_0;V^{n_s}_s )
\nn\\&&\hspace{20pt}
+
\big(\mu_{s,6}-3\sigma_s^2\mu_{s,4}+2\sigma_s^6\big)H_{0,0,0}(z_1,z_2,z_0;V^{n_s}_s )
\bigg\}
\eea
for $z=(z_1,z_2,z_0)\in\bbR^3$.

For $\dot{L}^n_s$ of (\ref{0404242259}), let us consider the probability 
\beas
P\big[\mfkT^n\geq x] 
&=&
P\bigg[\sum_{s\in\bbS}\dot{L}^n_s\mfkT^n_s\geq x\bigg]
\qquad(\because (\ref{0404200852}))
\nn\\&=& 
P\bigg[\sum_{s\in\bbS}\dot{L}^n_s\big(Y^n_s(\bbZ^n_s)+\wt{R}^s_{n_s}\big)\geq x\bigg]
\qquad(\because(\ref{0404242257}))
\nn\\&=& 
P\bigg[\sum_{s\in\bbS}\dot{L}^n_sY^n_s(\bbZ^n_s)\geq x\bigg]+\ol{O}(n^{-\xi})
\eeas
where $\ol{O}(n^{-\xi})$ denotes a term that is of order $O(n^{-\xi})$ uniformly in $x\in\bbR$. 
{\fblue This estimate is possible by including the random location shift $\wt{R}^s_{n_s}$ into $x$ as $x\pm n^{-\xi}$ 
and by getting back after the asymptotic expansion having an integral representation,  
if $P[|\wt{R}^s_{n_s}|>n^{-\xi}]=O(n^{-\xi})$. 
}

In this situation, $f$ of (\ref{0404242304}) is 
\beas 
f\big(\ul{l}_S,\ul{y}_S\big)
&=&
1_{\big\{
\sum_{s\in\bbS}\dot{l}_sy_s
\geq x\big\}}
\qquad \big(\ul{l}_S=\big(l_s=\big(\dot{l}_s,\ddot{l}_s)\big)_{s\in\bbS},\>\ul{y}_S=(y_s)_{s\in\bbS}\big)
\eeas
and set {\fblue$\ul{L}^n_S=\big((\dot{L}^n_s,\ddot{L}^n_s)\big)_{s\in\bbS}$ with $\dot{L}^n_s$ of (\ref{0404242259}) and $\ddot{L}^n_s$ specified later,}
as well as $\ul{\bbY}_S^n=\big(\bbY^n_s\big)_{s\in\bbS}$ with 
\beas
\bbY_s^n 
&=& 
Y^n_s(\bbZ^n_s)
\eeas
for the function $Y^n_s$ of (\ref{0404231626}) 
and $\bbZ^n_s$ of (\ref{0404242321}) accompanied with (\ref{0404242322}) for $g_k=1$ and (\ref{0404242323}).

Let us illustrate the backward approximation in the case $S=2$ though the description is quite the same for a larger value of $S$. 
We set
\beas 
\wh{f}^n_{2}(\ul{l}_2,\ul{z}_2)
&=& 
1_{\big\{\sum_{s=1}^2\dot{l}_sY^n_s(z_s)\geq x\big\}}.
\eeas
{\fred 
From (\ref{0404200621}), 
\beas&&
\wh{f}^n_{1}({\sred\ul{l}_{1},\>}\ul{z}_{1})
\nn\\&=&
\int 
\wh{f}^n_{2}\big({\sred\ul{l}_{2},\ul{z}_{2}}\big)
{\fred \Psi^n_{2,\sfp,{\fred{\bf w}^{n}_2}}(dz_2)}
1_{\{l_2\in\Lambda^n_{2}\}}
\nu^n_{c_2}(dl_2,d{\fred{\bf w}^{n}_2})
q\big((\ul{l}_{1},\ul{z}_{1}),dc_{2}\big)
\nn\\&=&
\int 
1_{\big\{\sum_{s=1}^2\dot{l}_sY^n_s(z_s)\geq x\big\}}
{\fred \Psi^n_{2,\sfp,{\fred{\bf w}^{n}_2}}(dz_2)}
1_{\{l_2\in\Lambda^n_{2}\}}
\nu^n_{c_2}(dl_2,d{\fred{\bf w}^{n}_2})
q\big((\ul{l}_{1},\ul{z}_{1}),dc_{2}\big)
\nn\\&=&
\int
\int_{\dot{z}_2\in\bbR^2\atop\ddot{z}_2\in\bbR}
1_{\bigg\{\sum_{s=1}^2\dot{l}_s\big(\sigma_s^{-1}\dot{z}_s-\frac{1}{2\sigma_s^3n_s^{1/2}}\ddot{z}_s\dot{z}_s\big)\geq x\bigg\}}
{\fred \Psi^n_{2,\sfp,{\fred{\bf w}^{n}_2}}(d\dot{z}_2,d\ddot{z}_2)}
\nn\\&&\hspace{180pt}\times
1_{\{l_2\in\Lambda^n_{2}\}}
\nu^n_{c_2}(dl_2,d{\fred{\bf w}^{n}_2})
q\big((\ul{l}_{1},\ul{z}_{1}),dc_{2}\big)
\nn\\&=&
\int
\int_{\dot{z}_2\in\bbR^2\atop\ddot{z}_2\in\bbR}
1_{\bigg\{\dot{l}_2\big(\sigma_2^{-1}\dot{z}_2-\frac{1}{2\sigma_2^3n_2^{1/2}}\ddot{z}_2\dot{z}_2\big)\geq x
-\dot{l}_1\big(\sigma_1^{-1}\dot{z}_1-\frac{1}{2\sigma_1^3n_1^{1/2}}\ddot{z}_1\dot{z}_1\big)\bigg\}}
{\fred \Psi^n_{2,\sfp,{\fred{\bf w}^{n}_2}}(d\dot{z}_2,d\ddot{z}_2)}
\nn\\&&\hspace{180pt}\times
1_{\{l_2\in\Lambda^n_{2}\}}
\nu^n_{c_2}(dl_2,d{\fred{\bf w}^{n}_2})
q\big((\ul{l}_{1},\ul{z}_{1}),dc_{2}\big). 
\eeas
If we apply the transform of the density, the last expression serves in Stage 2. 
Our scheme propagates this formula backward to 
\bea\label{0412030305}
\wh{f}^n_{0}
&=&
\int 
\wh{f}^n_{1}\big({\sred\ul{l}_{1},\ul{z}_{1}}\big)
{\fred \Psi^n_{1,\sfp,{\fred{\bf w}^{n}_1}}(dz_1)}
1_{\{l_1\in\Lambda^n_{1}\}}
\nu^n_{c_1}(dl_1,d{\fred{\bf w}^{n}_1})
\nn\\&=&%
\int
\int_{\dot{z}_1\in\bbR^2\atop\ddot{z}_1\in\bbR}
\int
\int_{\dot{z}_2\in\bbR^2\atop\ddot{z}_2\in\bbR}
1_{\bigg\{\sum_{s=1}^2\dot{l}_s\big(\sigma_s^{-1}\dot{z}_s-\frac{1}{2\sigma_s^3n_s^{1/2}}\ddot{z}_s\dot{z}_s\big)\geq x\bigg\}}
{\fred \Psi^n_{2,\sfp,{\fred{\bf w}^{n}_2}}(d\dot{z}_2,d\ddot{z}_2)}
\nn\\&&\hspace{60pt}\times
1_{\{l_2\in\Lambda^n_{2}\}}
\nu^n_{c_2}(dl_2,d{\bf w}^n_2)
q\big((\ul{l}_{1},\ul{z}_{1}),dc_{2}\big)
\nn\\&&\hspace{60pt}\times
{\fred \Psi^n_{1,\sfp,{\fred{\bf w}^{n}_1}}(d\dot{z}_1,d\ddot{z}_1)}
1_{\{l_1\in\Lambda^n_{1}\}}
\nu^n_{c_1}(dl_1,d{\bf w}^n_1). 
\eea
}
\begin{en-text}
By Fubini's theorem, 
{\fred 
\beas
\wh{f}^n_{0}
&=&
\int\int
\int_{\dot{z}_1\in\bbR^2\atop\ddot{z}_1\in\bbR}
\int_{\dot{z}_2\in\bbR^2\atop\ddot{z}_2\in\bbR}
\int_{c_2\in\mfkC_2}
1_{\bigg\{\sum_{s=1}^2\dot{l}_s\big(\sigma_s^{-1}\dot{z}_s-\frac{1}{2\sigma_s^3n_s^{1/2}}\ddot{z}_s\dot{z}_s\big)\geq x\bigg\}}
q\big((l_{1},z_{1}),dc_{2}\big)
\nn\\&&\hspace{100pt}\times
{\fred \Psi^n_{2,\sfp,{\fred{\bf w}^{n}_2}}(d\dot{z}_2,d\ddot{z}_2)}
{\fred \Psi^n_{1,\sfp,{\bf w}^{n_1}_2}(d\dot{z}_1,d\ddot{z}_1)}
\nn\\&&\hspace{100pt}\times
1_{\{l_2\in\Lambda^n_{2,c}\}}
\nu^n_{c_2}(dl_2,d{\bf w}^n_2)
1_{\{l_1\in\Lambda^n_{1,c}\}}
\nu^n_{c_1}(dl_1,d{\bf w}^n_1)
\eeas
\end{en-text}
\begin{en-text}
An example is the $\ep$-Greedy algorithm having $\mfkC_2=\{c_2^{(1)},c_2^{(2)}\}$ with
\beas 
q\big((\ul{l}_{1},\ul{z}_{1}),dc_{2}\big)
&=& 
1_{\{{\tt h}_1(\ul{l}_{1},\ul{z}_{1})\geq0\}}\delta_{c_2^{(1)}}(dc_2)
+1_{\{{\tt h}_1(\ul{l}_{1},\ul{z}_{1})<0\}}\delta_{c_2^{(2)}}(dc_2)
\eeas
for some function ${\tt h}_1$. 
On the other hand, the Thompson sampling is realized as 
\bea\label{0408251130}
q\big((\ul{l}_{1},\ul{z}_{1}),dc_{2}\big)
&=& 
1_{\{{\tt h}_1(\ul{l}_{1},\ul{z}_{1})\leq a_1\}}\delta_{c_2^{(1)}}(dc_2)
+1_{\{{\tt h}_1(\ul{l}_{1},\ul{z}_{1})>a_2\}}\delta_{c_2^{(2)}}(dc_2)
\nn\\&&
+1_{\{a_1<{\tt h}_1(\ul{l}_{1},\ul{z}_{1})\leq a_2\}}\delta_{C_2(l_1,z_1)}(dc_2)
\eea
with some constants $a_1,a_2$ and some $\mfkC_2$-valued function $C_2$ of $(l_1,z_1)$. 
Obviously, the right-hand side of (\ref{0408251130}) can be written as 
$\delta_{C_2'(l_1,z_1)}(dc_2)$ with a $\mfkC_2$-valued function $C_2'$ of $(l_1,z_1)$ defined by 
\beas 
C_2'(l_1,z_1)
&=&
\left\{\begin{array}{cl}
c_2^{(1)}&({\tt h}_1(l_1,z_1)\leq a_1)\y
C_2(l_1,z_1)&(a_1<{\tt h}_1(\ul{l}_{1},\ul{z}_{1})\leq a_2)\y
c_2^{(2)}&({\tt h}_1(\ul{l}_{1},\ul{z}_{1})>a_2). 
\end{array}\right.
\eeas
\end{en-text}
}

In the $\ep$-Greedy algorithm, 
the strategy is determined by the sign of the statistic
\beas 
\Xi^n_1 
&=&
(N^n_{1,1})^{-1}\sum_{j\in\bbJ^n_1}A_{j,1}R_j-(N^n_{1,2})^{-1}\sum_{j\in\bbJ^n_1}A_{j,2}R_j
\eeas
Under the null hypothesis $H_0$: $a=0$, 
\beas 
n_1^{1/2}\Xi^n_1 
&=&
n_1^{1/2}(N^n_{1,1})^{-1}\sum_{j\in\bbJ^n_1}A_{j,1}\dot{\ep}_j-n_1^{1/2}{\tred(} N^n_{1,2})^{-1}\sum_{j\in\bbJ^n_1}A_{j,2}\dot{\ep}_j
\nn\\&=&
\ddot{L}^n_1\cdot\dot{\bbZ}^n_1,
\eeas
where 
\beas 
\ddot{L}^n_1
&=&
\big((N^n_{1,1}/n_1)^{-1/2},{\tred -}(N^n_{1,2}/n_1)^{-1/2}\big).
\eeas
Let {\fblue $\ddot{L}^n_2=0$. Thus, the $\ep$-Greedy algorithm (\ref{0412030158}) selects the strategy $c_2\in{\mathfrak C}_2$ in Stage $2$ with the function 
$
{\tt h}_1(\ul{l}_1,\ul{z}_1)
=
\ddot{l}_1\cdot\dot{z}_1.
$
}
\begin{en-text}
\beas 
L^n_1&=&(\dot{L}^n_1,\ddot{L}^n_1),\\
L^n_2&=&(\dot{L}^n_2,\ddot{L}^n_2)\yeq(\dot{L}^n_2,0),\\
l_1&=&(\dot{l}_1,{\tred\ddot{l}_1}).\\
\eeas
\end{en-text}
\begin{en-text}
{\fred 
Then 
\bea\label{0404260239}
\wh{f}^n_{0}
&=&
\int\int
\int_{\dot{z}_1\in\bbR^2\atop\ddot{z}_1\in\bbR}
\int_{\dot{z}_2\in\bbR^2\atop\ddot{z}_2\in\bbR}
\int_{c_2\in\mfkC_2}
1_{\bigg\{\sum_{s=1}^2\dot{l}_s\big(\sigma_s^{-1}\dot{z}_s-\frac{1}{2\sigma_s^3n_s^{1/2}}\ddot{z}_s\dot{z}_s\big)\geq x\bigg\}}
q\big((l_{1},z_{1}),dc_{2}\big)
\nn\\&&\hspace{100pt}\times
{\fred \Psi^n_{2,\sfp,{\fred{\bf w}^{n}_2}}(d\dot{z}_2,d\ddot{z}_2)}
{\fred \Psi^n_{1,\sfp,{\bf w}^{n}_1}(d\dot{z}_1,d\ddot{z}_1)}
\nn\\&&\hspace{100pt}\times
1_{\{l_2\in\Lambda^n_{2}\}}
\nu^n_{c_2}(dl_2,d{\bf w}^n_2)
1_{\{l_1\in\Lambda^n_{1,c}\}}
\nu^n_{c_1}(dl_1,d{\bf w}^n_1)
\eea
For example, if ${\tt h}_1(l_1,z_1)=\ddot{l}_1\dot{z}_1$ in the $\ep$-Greedy, then one should compute the components
\bea\label{0408250817}
\wh{f}^n_{0,c}
&=&
\int\int
\int_{\dot{z}_1\in\bbR^2\atop\ddot{z}_1\in\bbR}
\int_{\dot{z}_2\in\bbR^2\atop\ddot{z}_2\in\bbR}
1_{\bigg\{\sum_{s=1}^2\dot{l}_s\big(\sigma_s^{-1}\dot{z}_s-\frac{1}{2\sigma_s^3n_s^{1/2}}\ddot{z}_s\dot{z}_s\big)\geq x\bigg\}}
1_{\big\{(-1)^{c_2+1}\ddot{l}_1\dot{z}_1>0\big\}}
\nn\\&&\hspace{100pt}\times
{\fred \Psi^n_{2,\sfp,{\fred{\bf w}^{n}_2}}(d\dot{z}_2,d\ddot{z}_2)}
{\fred \Psi^n_{1,\sfp,{\bf w}^{n}_1}(d\dot{z}_1,d\ddot{z}_1)}
\nn\\&&\hspace{100pt}\times
1_{\{l_2\in\Lambda^n_{2,c}\}}
\nu^n_{c_2}(dl_2,d{\bf w}^n_2)
1_{\{l_1\in\Lambda^n_{1,c}\}}
\nu^n_{c_1}(dl_1,d{\bf w}^n_1)
\eea
for $c=(c_1,c_2)$ with $c_2\in\mfkC_2$ and a fixed $c_1\in\mfkC_1$. 
}
\end{en-text}

Introduce the transform 
\bea\label{0404260247}
y_s \yeq 
\begin{pmatrix}\dot{y}_s\\ \ddot{y}_s\end{pmatrix}
\yeq
\begin{pmatrix}\sigma_s^{-1}\dot{z}_s\\ \sigma_s^{-2}\ddot{z}_s\end{pmatrix}. 
\eea
Then the formula (\ref{0412030305}) has the expression 
{\fred
\bea\label{0404260240}
\wh{f}^n_{0}
&=&
\int\int
\int_{\dot{y}_1\in\bbR^2\atop\ddot{y}_1\in\bbR}
\int_{\dot{y}_2\in\bbR^2\atop\ddot{y}_2\in\bbR}
1_{\bigg\{\sum_{s=1}^2\dot{l}_s\big(\dot{y}_s-\frac{1}{2n_s^{1/2}}\ddot{y}_s\dot{y}_s\big)\geq x\bigg\}}
\nn\\&&\hspace{100pt}\times
{\fred \wt{\Psi}^n_{2,\sfp,{\fred{\bf w}^{n}_2}}(d\dot{y}_2,d\ddot{y}_2)}
1_{\{l_2\in\Lambda^n_{2}\}}
\nu^n_{c_2}(dl_2,d{\bf w}^n_2)
\nn\\&&\hspace{100pt}\times
q\big((l_{1},\sigma_1^{-1}y_{1}),dc_{2}\big)
{\fred \wt{\Psi}^n_{1,\sfp,{\fred{\bf w}^{n}_1}}(d\dot{y}_1,d\ddot{y}_1)}
1_{\{l_1\in\Lambda^n_{1}\}}
\nu^n_{c_1}(dl_1,d{\bf w}^n_1)
\nn\\&&
\eea
}
where 
{\fred$\wt{\Psi}^n_{s,\sfp,{\fred{\bf w}^{n}_s}}(d\dot{y}_s,d\ddot{y}_s)$} 
is the asymptotic expansion transformed by (\ref{0404260247}) 
from {\fred$\Psi^n_{s,\sfp,{\fred{\bf w}^{n}_s}}(d\dot{z}_s,d\ddot{z}_s)$.

It should be remarked that each integral with respect to $\nu^n_{c_s}(dl_s,d{\bf w}^n_s)$  
can be simplified 
to a more concise integral 
by using aggregate statistics appearing in 
$\Psi^n_{s,\sfp,{\fred{\bf w}^{n}_s}}(d\dot{y}_s,d\ddot{y}_s)$ and 
$\wt{\Psi}^n_{s,\sfp,{\fred{\bf w}^{n}_s}}(d\dot{y}_s,d\ddot{y}_s)$ than the integral with respect to $(l_s,{\bf w}^{n_s})$. 
See the formula (\ref{0404210421}). 
} 

{\fred 
\subsubsection{Transform of the asymptotic expansion}\label{0412020311}
We will consider an analytic method for numerical computation of $\wh{f}^n_{0}$. 
From (\ref{0404260240}), 
\bea
\wh{f}^n_{0}
&=& 
\int\int
\int_{\dot{y}_1\in\bbR^2\atop\ddot{y}_1\in\bbR}
\cali_2(\dot{y}_1,\ddot{y}_1)
1_{\{l_2\in\Lambda^n_{2}\}}
\nu^n_{c_2}(dl_2,d{\bf w}^n_2)
\nn\\&&\hspace{50pt}\times
q\big((l_{1},\sigma_1^{-1}y_{1}),dc_{2}\big)
\psi_1({\fred{\bf w}^{n}_1},\dot{y}_1,\ddot{y}_1)d\dot{y}_1d\ddot{y}_1
1_{\{l_1\in\Lambda^n_{1}\}}
\nu^n_{c_1}(dl_1,d{\bf w}^n_1),\label{0408250810}
\eea
where 
\beas 
\cali_2(\dot{y}_1,\ddot{y}_1)
&=& 
\int_{\dot{y}_2\in\bbR^2\atop\ddot{y}_2\in\bbR}
1_{\bigg\{\sum_{s=1}^2\dot{l}_s\big(\dot{y}_s-\frac{1}{2n_s^{1/2}}\ddot{y}_s\dot{y}_s\big)\geq x\bigg\}}
\psi_2({\fred{\bf w}^{n}_2},\dot{y}_2,\ddot{y}_2)d\dot{y}_2d\ddot{y}_2 
\eeas
for given $\ul{l}_2$ and $\ul{{\bf w}}^n_2$, with
\beas
\psi_1({\fred{\bf w}^{n}_1},\dot{y}_1,\ddot{y}_1)
&=& 
\frac{{\fred d\wt{\Psi}^n_{1,\sfp,{\fred{\bf w}^{n}_1}}}}
{dy_1}(\dot{y}_1,\ddot{y}_1),
\nn\\
\psi_2({\fred{\bf w}^{n}_2},\dot{y}_2,\ddot{y}_2)
&=& 
\frac{{\fred d\wt{\Psi}^n_{2,\sfp,{\fred{\bf w}^{n}_2}}}}
{dy_2}(\dot{y}_2,\ddot{y}_2)
\eeas
for $\ul{l}_2$ and $\ul{{\bf w}}^n_2$. 
\begin{en-text}
The formula (\ref{0408250810}) is rewritten as 
\bea
\cali
&=& 
\int_{\dot{y}_1\in\bbR^2\atop\ddot{y}_1\in\bbR}
\cali_2(\dot{y}_1,\ddot{y}_1)
\wt{q}\big((l_1,\sigma_1^{-1}y_1),c_2)
\psi_1(\dot{y}_1,\ddot{y}_1)d\dot{y}_1d\ddot{y}_1
\label{0405030357}
\eea
with 
\beas 
\cali_2(\dot{y}_1,\ddot{y}_1)
&=&
\int_{\dot{y}_2\in\bbR^2\atop\ddot{y}_2\in\bbR}
1_{\bigg\{\sum_{s=1}^2\dot{l}_s\big(\dot{y}_s-\frac{1}{2n_s^{1/2}}\ddot{y}_s\dot{y}_s\big)\geq x\bigg\}}
\psi_2(\dot{y}_2,\ddot{y}_2)d\dot{y}_2d\ddot{y}_2. 
\eeas
\end{en-text}
}
\begin{en-text}
{\tred 
Let $x>0$. 
For numerical computation of $\wh{f}^n_{0,c}$, we need to compute the value of the integral 
\bea
\cali
&=& 
\int_{\dot{y}_1\in\bbR^2\atop\ddot{y}_1\in\bbR}
\int_{\dot{y}_2\in\bbR^2\atop\ddot{y}_2\in\bbR}
1_{\bigg\{\sum_{s=1}^2\dot{l}_s\big(\dot{y}_s-\frac{1}{2n_s^{1/2}}\ddot{y}_s\dot{y}_s\big)\geq x\bigg\}}
1_{\big\{(-1)^{c_2+1}\ddot{l}_1\dot{y}_1>0\big\}}
\nn\\&&\hspace{100pt}\times
\psi_2(\dot{y}_2,\ddot{y}_2)d\dot{y}_2d\ddot{y}_2 
\psi_1(\dot{y}_1,\ddot{y}_1)d\dot{y}_1d\ddot{y}_1
\nn\\&=&
\int_{\dot{y}_1\in\bbR^2\atop\ddot{y}_1\in\bbR}
\cali_2(\dot{y}_1,\ddot{y}_1)
1_{\big\{(-1)^{c_2+1}\ddot{l}_1\dot{y}_1>0\big\}}
\psi_1(\dot{y}_1,\ddot{y}_1)d\dot{y}_1d\ddot{y}_1
\label{0405030357}
\eea
for given $c$, $\ul{l}_2$ and $\ul{{\bf w}}^n_2$, where 
\beas
\psi_1(\dot{y}_1,\ddot{y}_1)
&=& 
\frac{{\fred d\wt{\Psi}^n_{1,\sfp,{\fred{\bf w}^{n}_1}}}}
{dy_1}(\dot{y}_1,\ddot{y}_1)
\nn\\
\psi_2(\dot{y}_2,\ddot{y}_2)
&=& 
\frac{{\fred d\wt{\Psi}^n_{2,\sfp,{\fred{\bf w}^{n}_2}}}}
{dy_2}(\dot{y}_2,\ddot{y}_2)
\eeas
for $\ul{l}_2$ and $\ul{{\bf w}}^n_2$, 
and 
\beas 
\cali_2(\dot{y}_1,\ddot{y}_1)
&=&
\int_{\dot{y}_2\in\bbR^2\atop\ddot{y}_2\in\bbR}
1_{\bigg\{\sum_{s=1}^2\dot{l}_s\big(\dot{y}_s-\frac{1}{2n_s^{1/2}}\ddot{y}_s\dot{y}_s\big)\geq x\bigg\}}
\psi_2(\dot{y}_2,\ddot{y}_2)d\dot{y}_2d\ddot{y}_2. 
\eeas
\end{en-text}
The integral $\cali_2(\dot{y}_1,\ddot{y}_1)$ is rewritten as 
\bea\label{0405030219}
\cali_2(\dot{y}_1,\ddot{y}_1)
&=& 
\int_{\dot{y}_2\in\bbR^2\atop\ddot{y}_2\in\bbR}
1_{\bigg\{\dot{l}_2\big(\dot{y}_2-\frac{1}{2n_2^{1/2}}\ddot{y}_2\dot{y}_2\big)
\geq U_1(\dot{y}_1,\ddot{y}_1) \bigg\}}
\psi_2({\fred{\bf w}^{n}_2},\dot{y}_2,\ddot{y}_2)d\dot{y}_2d\ddot{y}_2
\eea
with 
\beas 
U_1(\dot{y}_1,\ddot{y}_1)
&=&
x-\dot{l}_1\big(\dot{y}_1-\frac{1}{2n_1^{1/2}}\ddot{y}_1\dot{y}_1\big).
\eeas
We change variables by 
\bea\label{0405030235}
\begin{pmatrix}\dot{w} \\ \ddot{w}\end{pmatrix}
&=&
F(\dot{y}_2,\ddot{y}_2)
\yeq
\begin{pmatrix}\wt{l}_2\big(\dot{y}_2-\frac{1}{2n_2^{1/2}}\ddot{y}_2\dot{y}_2\big) \\ \ddot{y}_2\end{pmatrix},
\eea
where {\fblue we choose }
\bea\label{0405030401}
\wt{l}_2 
&=& 
\begin{pmatrix}\dot{l}_2 \\ \dot{l}_2^\perp \end{pmatrix}\qquad(2\times2)
\eea
{\fblue so that $\wt{l}_2$ becomes {\vred nondegenerate}.
If we plug a value of $\dot{L}_2$ of (\ref{0404242259}), then we may set $\dot{l}_2^\perp$ 
to the corresponding value of $\big((N^n_{2,1}/n_2)^{1/2},(N^n_{2,2}/n_2)^{1/2}\big)$, for example.}
\begin{en-text}
\beas 
 \dot{l}_2^\perp 
 &=& 
 \big((N^n_{2,1}/n_2)^{1/2},(N^n_{2,2}/n_2)^{1/2}\big)
\eeas
for example. 
\end{en-text}
Write $\dot{w}=(\dot{w}^{(1)},\dot{w}^{(2)})$. 
The change of variables (\ref{0405030235}) applied to (\ref{0405030219}) yields
\bea
\cali_2(\dot{y}_1,\ddot{y}_1)
&=& 
\int_{\dot{y}_2\in\bbR^2\atop\ddot{y}_2\in\bbR}
1_{\bigg\{\dot{l}_2\big(\dot{y}_2-\frac{1}{2n_2^{1/2}}\ddot{y}_2\dot{y}_2\big)
\geq U_1(\dot{y}_1,\ddot{y}_1) \bigg\}}
\psi_2({\fred{\bf w}^{n}_2},\dot{y}_2,\ddot{y}_2)d\dot{y}_2d\ddot{y}_2
\label{0405030310}\\&=&
\int_{\dot{y}_2\in\bbR^2\atop\ddot{y}_2\in\bbR}
1_{\big\{\dot{w}^{(1)}\geq U_1(\dot{y}_1,\ddot{y}_1) \big\}}
\psi_2({\fred{\bf w}^{n}_2},F^{-1}(\dot{w},\ddot{w}))\big|\det J_{F^{-1}}(\dot{w},\ddot{w})\big|d\dot{w}d\ddot{w}
\nn\\&=&
\int_{\dot{y}_2\in\bbR^2\atop\ddot{y}_2\in\bbR}
1_{\big\{\dot{w}^{(1)}\geq U_1(\dot{y}_1,\ddot{y}_1) \big\}}
\psi_2\big({\fred{\bf w}^{n}_2},\tilde{l}_2^{-1}\dot{w}+2^{-1}n_2^{-1/2}\ddot{w}\tilde{l}_2^{-1}\dot{w},\ddot{w}\big)
\nn\\&&\hspace{100pt}\times
\big|\det\big((1+2^{-1}n_2^{-1/2}\ddot{w})\wt{l}_2^{-1}\big)\big|d\dot{w}d\ddot{w}+o(n_s^{-1/2-\delta})
\label{0405030311}
\nn\\&=:&
\wt{\cali}_2(\dot{y}_1,\ddot{y}_1)+o(n_s^{-1/2-\delta})
\eea
as $n_s\to\infty$ for some positive constant $\delta$. 
Here $J_{F^{-1}}$ is the Jacobi matrix of the map $F^{-1}$. 
The passage from (\ref{0405030310}) to (\ref{0405030311}) is intuitive because 
the map $F^{-1}$ is not well defined on the whole space. 
However, the approximation $\wt{\cali}_2(\dot{y}_1,\ddot{y}_1)$ at (\ref{0405030311}) to $\cali_2(\dot{y}_1,\ddot{y}_1)$ is valid 
and the order of the error is uniform in $(\dot{y}_1,\ddot{y}_1)$. 
More precisely, first we restrict the domain of the integral to 
$\big\{(\dot{y}_2,\ddot{y}_2);\>|(\dot{y}_2,\ddot{y}_2)|<n_2^{1/4}\big\}$, then 
the map $F$ becomes a diffeomorphism and the above computation is validated for large $n_2$.  
The contribution of the integral over the cut off area is negligible. 
As a matter of fact, this approach was taken by Bhattacharya and Ghosh \cite{BhattacharyaGhosh1978}, 
cf. Sakamoto and Yoshida \cite{SakamotoYoshida2004,SakamotoYoshida2009}, 
Yoshida \cite{yoshidaSuritokeigaku}. 
The approximation formula $\wt{\cali}_2(\dot{y}_1,\ddot{y}_1)$ at (\ref{0405030311}) can further be simplified by 
expanding $\psi_2$ and $\det$, and neglecting the terms of order $n_2^{-1}$ or smaller. 
It should be remarked that thus simplified $\wt{\cali}_2(\dot{y}_1,\ddot{y}_1)$ is a smooth function of $(\dot{y}_1,\ddot{y}_1)$ and has a closed form. 

For the computation of the integral of $\cali_2(\dot{y}_1,\ddot{y}_1)$ of (\ref{0405030219}), 
we may apply a linear transform of $(\dot{y}_1,\ddot{y}_1)$ with a matrix $\wt{l}_1$ like (\ref{0405030401}). 
Since the function $\cali_2(\dot{y}_1,\ddot{y}_1)$ is in $C_b^\infty(\bbR^3)$, 
an approximation by a power series in $(\dot{y}_1,\ddot{y}_1)$ should work well. 
Then the integral of $\cali_2(\dot{y}_1,\ddot{y}_1)$ can be calculated 
with a closed form of approximation. 
This approach is feasible for $S$ larger than two. 
}

%
{\fred 
\subsubsection{Importance sampling}\label{0409171651}
Let 
\beas 
\calj(\dot{y}_1,\ddot{y}_1,\dot{y}_2,\ddot{y}_2)
&=&
\int 1_{\bigg\{\sum_{s=1}^2\dot{l}_s\big(\dot{y}_s-\frac{1}{2n_s^{1/2}}\ddot{y}_s\dot{y}_s\big)\geq x\bigg\}}
\psi_2({\fred{\bf w}^{n}_2},\dot{y}_2,\ddot{y}_2)
1_{\{l_2\in\Lambda^n_{2}\}}
\nu^n_{c_2}(dl_2,d{\bf w}^n_2)
\nn\\&&\hspace{70pt}\times
q\big((l_{1},\sigma_1^{-1}y_{1}),dc_{2}\big)
\psi_1({\fred{\bf w}^{n}_1},\dot{y}_1,\ddot{y}_1)
1_{\{l_1\in\Lambda^n_{1}\}}
\nu^n_{c_1}(dl_1,d{\bf w}^n_1).
\eeas
}\noindent
To carry out the integration of $\wh{f}^n_0$, we apply the importance sampling. 
Let $x>0$. 
Prepare constants $a_s\in\bbR^2$ for $s\in\bbS=\{1,2\}$. For example, 
\beas 
a_s\yeq\big(\sum_{s\in\bbS}|\dot{l}_s|^2\big)^{-1}
{\tred\begin{pmatrix}\dot{l}_s^{(1)}\\ \dot{l}_s^{(2)}\end{pmatrix}},
\eeas
where $\dot{l}_s=(\dot{l}_s^{(1)},\dot{l}_s^{(2)})$. 
Suppose that $(\xi_s^i)_{i\in\bbN}$ is a three-dimensional i.i.d. sequence such that 
\beas 
\xi_s^i\yeq\begin{pmatrix}\dot{\xi}_s^i\\ \ddot{\xi}_s^i\end{pmatrix}
\sim N_3\left(\mu_s,\>\Sigma_s\right),\qquad\mu_s\yeq\begin{pmatrix}xa_s\\ 0\end{pmatrix},
\eeas
for $s\in\bbS$, 
and that $(\xi_1^i)_{i\in\bbN}$ and $(\xi_2^i)_{i\in\bbN}$ are independent. 
The matrix $\Sigma_s$ is a positive-definite matrix bigger than the covariance matrix of the normal distribution $N_3(0,{\sf V}_s)$ of the principal part of 
$\psi_s$. For example, $\Sigma_s=p\>{\sf V}_s$ for some constant $p>1$. 
We approximate {\fred$\wh{f}^n_0$} by 
{\fred 
\beas 
\cali_N
&=& 
\frac{1}{N}
\sum_{i=1}^N
\frac{
\calj(\dot{\xi}^i_1,\ddot{\xi}^i_1,\dot{\xi}^i_2,\ddot{\xi}^i_2)
}{
\phi(\dot{\xi}^i_2,\ddot{\xi}^i_2;\mu_2,\Sigma_2) 
\phi(\dot{\xi}^i_1,\ddot{\xi}^i_1;\mu_1,\Sigma_1)
}
\eeas
}
with large $N$. 
\begin{en-text}
{\fred In the case of (\ref{0408250817}), what should be computed is 
\beas 
\cali_{N,c_2}
&=& 
\frac{1}{N}
\sum_{i=1}^N
1_{\bigg\{\sum_{s=1}^2\dot{l}_s\big(\dot{\xi}^i_s-\frac{1}{2n_s^{1/2}}\ddot{\xi}^i_s\dot{\xi}^i_s\big)\geq x\bigg\}}
1_{\big\{(-1)^{c_2+1}\ddot{l}_1\dot{\xi}^i_1>0\big\}}
\frac{
\psi_2(\dot{\xi}^i_2,\ddot{\xi}^i_2) 
\psi_1(\dot{\xi}^i_1,\ddot{\xi}^i_1)
}{
\phi(\dot{\xi}^i_2,\ddot{\xi}^i_2;\mu_2,\Sigma_2) 
\phi(\dot{\xi}^i_1,\ddot{\xi}^i_1;\mu_1,\Sigma_1)
}
\eeas
for each $c_2\in\mfkC_2$. 
}
\end{en-text}
This computation is feasible much faster than the raw simulation of the batched bandits model. 

{\fred 
It should be remarked that our method based on the asymptotic expansion is a kind of invariance principle 
in that the approximation formulas depend on the first several moments, that are  
estimable from the data, while simulation methods need complete information about the distribution of the error.}
The measures {\fred$\wt{\Psi}^n_{s,\sfp,{\fred{\bf w}^{n}_s}}(d\dot{y}_s,d\ddot{y}_s)$}
involve the moments of $\dot{\ep}_j$ in the correction terms. 
In practice, we substitute them by their estimated value. 
It is possible to use a reduced formula that neglects the terms having the moments of $\dot{\ep}_j$ 
of order larger than four because these moments appear in higher-order terms than the first-order asymptotic expansion.

\section{Simulation study}\label{0411080600}
In this section we provide numerical results based on our asymptotic expansion formula. 
The setting we focus on here is the simplest case of two stage adaptive experiments, which captures the essence of the asymptotic expansion scheme considered in our theory. In clinical trials, this corresponds to the case where the first stage is the exploratory, and the second stage is confirmatory design. 

We first present baseline results. 
Each batch allocates $n_s=50$ individuals for $s\in\{1,2\}$. 
The noise $\dot{\ep}_j$ follows the {\fred standardized} gamma distribution $\Gamma(k,\theta)$ with $k=3$ and $\theta=2$, where  
the probability density function of $\Gamma(k,\theta)$ is $ f(x) =\Gamma ( k)^{-1} \theta^{-k}x^{k-1} e^{-x/\theta}1_{\{x>0\}}$. 
The treatment assignment probabilities to assign treatments at the first stage are uniformly $(0.5,0.5)$ to the two arms, 
and the $\ep$-greedy policy with clipping $0.2$ is applied to the second stage. 
For all our numerical experiments, to prevent non-singularity of the design matrix of the BOLS estimator, we ensured zero probability of either arm having zero arm allocation, and instead shifted that probability to being allocated one arm. 
{\fred We set the strategy $c_s$ so that the minimum sample size of each arm is not less than $5$.}
In short, we modeled a truncated binomial assignment function.  
To carry out numerical integration in the asymptotic expansion formula, the importance sampling proposed in Section \ref{0409171651} was used. 
Table \ref{tablegamma} compares the $2.5$, $5$, $95$, $97.5$ percentiles obtained by the different methods. 
\begin{table}[H]
  \centering
  \caption{Comparison between the quantiles: gamma distribution}
  \vspace{3mm}
  \begin{tabular}{lcccc}
    \hline
    Method\textbackslash Probability  & 0.025  &  0.050 & 0.950 & 0.975  \\
    \hline \hline
    Monte Carlo  & -2.11 & -1.78  & 1.59& 1.90\\
    Asymptotic expansion  & -2.13    & -1.81& 1.58&1.89\\
    Normal approximation  & -1.96  &-1.64 &1.65& 1.96 \\
    \hline
  \end{tabular}
  \vspace{3mm}\\
  \begin{minipage}[b]{10cm}
  \begin{small}
Centered $\Gamma(3,2)$ as noise distribution, $50$ samples per batch, 
  first stage uniform sampling, 
  second stage epsilon greedy policy with clipping $0.2$
 \end{small}
 \end{minipage}
 \label{tablenormal}
\end{table}
\noindent
As we can see, because of the skewness of $\Gamma(3,2)$, 
the normal approximations for the $2.5$ and $5$ percentiles are not precise,  
while the asymptotic expansion based method is correcting the apparent skewness of the distribution.

Next, it may be of some interest to check how our estimator behaves when the underlying distribution is normal. 
According to the theory, there should be no significant difference between the normal approximation and the asymptotic expansion. 
When $\dot{\ep}_j\sim N(0,1)$, 
as a matter of fact, 
Table \ref{tablenormal} confirms this guess to show that both asymptotic expansion and normal approximation perform almost identically. 
\begin{table}[H]
  \centering
   \caption{Comparison between the quantiles: normal distribution}
  \vspace{3mm}
  \begin{tabular}{lcccc}
    \hline
    Method\textbackslash Probability  & 0.025  &  0.050 & 0.950 & 0.975  \\
    \hline \hline
    Monte Carlo  & -1.98 & -1.66  & 1.66& 1.95\\
    Asymptotic expansion  & -1.97  & -1.65& 1.65&1.98\\
    Normal approximation  & -1.96  &-1.64 &1.65& 1.96 \\
    \hline
  \end{tabular}
 \vspace{3mm}\\
  \begin{minipage}[b]{10cm}
  \begin{small}
  $N(0,1)$ as noise distribution, $50$ samples per batch, 
  first stage uniform sampling, 
  second stage epsilon greedy policy with clipping $0.15$
 \end{small}
 \end{minipage}
\label{tablegamma}
\end{table}

Finally, as the distribution of $\dot{\ep}_j$, we considered the mixture of normal distributions with density $0.7\phi(x;0,1)+0.3\phi(x;3,4)$ standardized 
to mean $0$ and variance $1$. 
%
%
Table \ref{tablemixednormal} summarizes the results. 
\begin{table}[H]
  \centering
  \caption{Comparison between the quantiles: normal mixture}
  \vspace{3mm}
  \begin{tabular}{lcccc}
    \hline
    Method\textbackslash Probability  & 0.025  &  0.050 & 0.950 & 0.975  \\
    \hline \hline
    Monte Carlo  & -2.31 & -1.93  & 1.64& 2.01\\
    Asymptotic expansion  & -2.10  & -1.76& 1.65&1.96\\
    Normal approximation  & -1.96  &-1.64 &1.65& 1.96 \\
    \hline
  \end{tabular}
  \vspace{3mm}\\
  \begin{minipage}[b]{10cm}
  \begin{small}
Standardized normal mixture as noise distribution, 50 samples per batch, first stage assignment probabilities $= ( 0.2,0.8) $, second stage uniform sampling 
 \end{small}
 \end{minipage}
\label{tablemixednormal}
\end{table}
The nominal type I error by the normal approximation is the worst of the three, with $3.2$ percent inflation. 
This is the case when the noise distribution is the normal mixture, 
but perhaps more importantly, note that the first stage assignment is not uniformly distributed. 
Although this might not necessarily be a common strategy deployed in practice, 
we predict this problematic type I error inflation comes not only from the skewness of the underlying distribution, 
but also the asymmetry of assignment probabilities between the first arm and the second arm under the null hypothesis (no margin case).  
In the case of no margin, the two arms are exactly identical, and if the first stage is assigned symmetrically, 
the arm difference estimate (BOLS statistic) is likely to have the skewness to cancel out, a 'symmetrization' phenomena. 
This may imply that when one is about to conduct an adaptive experiment, 
and is actually suspecting that the arm difference is quite small, 
uniform sampling is recommended because that may lead to higher-order approximation. 
We think theoretically validating this  hypothesis is an interesting future direction for research.

All in all, although this is just a preliminary of the numerical validation, our method seems to be useful at least for relatively small sample size with higher-order cumulants departing from normality.

\section{Proof of Theorems \ref{0311221323} and \ref{0404131140}}\label{0411110634}
\subsection{A sketch of the proof of Theorem \ref{0311221323}}\label{0409162318}
{\fred 
(I) 
Fix some positive constants $\delta$ and $\wt{\delta}$ satisfying $\delta<\wt{\delta}$. 
We consider a sequence $\big(t_{T,j})_{j=1,...,N_T}$ for every $T\in\bbR_+$ such that 
$0=t_{T,0}<t_{T,1}<\cdots<t_{T,N_T}=T$ and 
$\delta<t_{T,j}-t_{T,j-1}<\wt{\delta}$ ($j=1,...,N_T-1$) and $t_{T,N_T}-t_{T,N_T-1}<\wt{\delta}$. 
Let $\wt{I}_j=[t_{T,j-1},t_{T,j}]$ ($j=1,...,N_T$). 
By definition, $Z_{\wt{I}_j}=Z_{t_{T,j}}-Z_{t_{T,j-1}}$ for $j=1,...,N_T$. Let $Z_{\wt{I}_0}=Z_0$.}

Let $\varphi:\bbR^\sfd\to[0,1]$ be a measurable function such that 
$\varphi(x)=1$ if $|x|\leq1/2$, and $\varphi(x)=0$ if $|x|\geq1$. 
The function 
$\varphi_T$ is defined by $\varphi_T(x)=x\varphi(x/(2T^\beta))$, where $\beta\in(0,1/2)$ 
is a constant; see pp.585-586 of Yoshida \cite{yoshida2004partial}. 
Let $Z'_T=\sum_{j=0}^{N_T}\varphi_T(Z_{\wt{I}_j})$ (p.601) and 
\beas 
e_T 
&=& 
T^{-1/2}\bigg\{E_\calc[\varphi_T(Z_0)]
+\sum_{j=1}^{N_T}E_\calc\big[\varphi_T\big(Z_{\wt{I}_j}\big)\big]\bigg\} 
\eeas
as was done on p.596. 
{\fred 
Let $\wt{Z}_{T,j}=\varphi_T(Z_{\wt{I}_j})-E_\calc\big[\varphi_T(Z_{\wt{I}_j})\big]$ for $j=0,1,...,N_T$. 
Furthermore, let $\wt{Z}_T=\sum_{j=1}^{N_T}\wt{Z}_{T,j}$ and $S_T^*=T^{-1/2}\wt{Z}_T$. 
Set $H_T(u,\calc)=E_\calc\big[e^{iu\cdot S_T^*}\big]$ for $u\in\bbR^\sfd$. 
For $I=(j_1,...,j_r)\in\{0,1,...,N_T\}^r$, we denote $\wt{Z}_I=\wt{Z}_{T,j_1}\otimes\cdots\otimes\wt{Z}_{T,j_r}$. 
}

{\fblue 
The $\calc$-conditional characteristic function of $S_T^*$ is denoted by $H_T(u,\calc)$, i.e., 
$H_T(u,\calc)=E_\calc[e^{\tti u\cdot S_T^*}]$ for $u\in\bbR^\sfd$. 
The proof uses the complex-valued conjugate conditional (CVCC) expectation $E_\calc[X](V)$ defined by 
$E_\calc[X](V)=E^0_\calc[X](V)/E^0_\calc[1](V)$ when $E^0_\calc[1](V)\not=0$, where 
$E^0_\calc[X](V)=E_\calc[Xe^{\tti V}]$ for real random variables $X$ and $V$, $\tti=\sqrt{-1}$. 
As defined on p. 565 of \cite{yoshida2004partial}, 
$\omega(\sff;\ep,\nu)=\int\omega_\sff(x;\ep)d\nu$ for a Borel measure $\nu$ on $\bbR^\sfd$.

The strategy of the proof of Theorem 1 in \cite{yoshida2004partial} is as follows. 
\begin{enumerate}[(1)] 
\im\label{0411201701} Conditional moment estimate for $E_\calc[\wt{Z}_I](u\cdot S_T^*)$ (Lemma 12, p.609)
\im\label{0411201702} Estimate for the CVCC covariance $\text{Cov}\big[\wt{Z}_{I_1},\wt{Z}_{I_2}\big](u\cdot S_T^*)$ (Lemma 13, p.611)
\im\label{0411201703} Estimate for the CVCC cumulant $\kappa_\calc\big[S_T^*,...,S_T^*\big](u\cdot S_T^*)$ 
(Lemma 14, p.611; $\Leftarrow$ Lemma 12 + Lemma 13); the bound with a factor $\theta_T(u,\omega)$.
\im\label{0411201704} Estimate of the factor $\theta_T(u,\omega)$ (Lemma 15, p.613)
\im\label{0411201705} Estimate for the CVCC cumulant $\kappa_\calc\big[S_T^*,...,S_T^*\big](u\cdot S_T^*)$ 
(Lemma 5, p.591; proof p.616 $\Leftarrow$ Lemma 14 + Lemma 15) 
\im\label{0411201706} Estimate of the gap between the conditional cumulants $\kappa_\calc[S_T^*,...,S_T^*]$ and $\kappa_\calc[S_T,...,S_T]$ 
for $S_T=T^{-1/2}Z_T$ (Lemma 6, p.591; proof p.616) 
\im\label{0411201707} Estimate of the gap between $H_T(u,\calc)$ and $\wh{\Psi}_{T,\sfp,\calc}(u)$ 
(Lemma 7, p.591; proof p. 617 $\Leftarrow$ Lemma 6 + Lemma 5)
\im\label{0411201708} $L^p$-estimate of $\partial_u^lH_T(u,\calc)$ for large $u$, 
under Condition ``$[A1^\sharp]$'' on p.592 and Condition ``$[A3^\flat]$'' on p.591 
(Lemma 10, p.594)
\im\label{0411201709} Estimate like (\ref{0311221522}) of Lemma \ref{0311230140} below, for deterministic $\sff$, without $\Phi_T$ 
(Lemma 11, p.601; $\Leftarrow$ Lemma 5 + Lemma 6)

\im\label{0411201712} ``$[A3]$'' $\Iku$ ``$[A3^\flat]$'' (Lemma 9, p.593; proof p.605)

\im\label{0411201710} Inequality like (\ref{0311221427}), for deterministic $\sff$, without $\Phi_T$, 
under ``$[A1]$'' on p.587 and ``$[A3^\flat]$'' 
(Lemma 8. p.592, proof p.596; $\Leftarrow$ Lemma 7 + Lemma 10 + Lemma 11)
\im\label{0411201711} Inequality like (\ref{0311221427}), for deterministic $\sff$, without $\Phi_T$, 
under ``$[A1']$'' on p.563 and ``$[A3^\flat]$'' (Proposition 1, p.593; $\Leftarrow$ Lemma 8)

\im\label{0411201713} Inequality like (\ref{0311221427}), for deterministic $\sff$, without $\Phi_T$, 
under ``$[A1']$'' and ``$[A3]$'' on p.564 (Theorem 1, p.565; proof p.593 $\Leftarrow$ Proposition 1 + Lemma 9)
\end{enumerate}

We will only sketch a route to 
Theorem \ref{0311221323} 
since the proof is almost the same as 
Theorem 1 of \cite{yoshida2004partial}, that is proved in Section 6 of \cite{yoshida2004partial} pp.585-621. 
Condition $[P1]$ of this paper is ``$[A1']$'', and it implies ``$[A1^\sharp]$'' in the present situation. 
Condition $[P2]$ is nothing but ``$[A2]$'' of \cite{yoshida2004partial}. 
The above Steps \ref{0411201701}-\ref{0411201708} run as they are.

On the other hand, a careful treatment of $\calc$-measurable functions $\sff$ is necessary in Step \ref{0411201709}. 
Moreover, for convenience of applications in this paper, we have adopted Condition $[P3]$, which is Condition ``$[A3^\natural]$'' 
introduced in Remark 15 of \cite{yoshida2004partial}. 
This requires some modifications in Step \ref{0411201712}. 
Since the resulting inequalities are slightly different, we will fix these issues and rebuild a proof as follows. 
}
\halflineskip

\noindent (II) 
We fill Step \ref{0411201709}. 
On p.599 of \cite{yoshida2004partial}, 
the truncation functional $\psi_T$ is defined as the indicator function of the set 
\beas 
\Omega_0(T)\cap\big\{s_TI_\sfd\leq\text{Var}_{{\fred\calc}}[T^{-1/2}Z_T]\leq u_TI_\sfd\big\}, 
\eeas
where the event $\Omega_0(T)$ is specified on p.586. 
In place of Inequality (34) (p.605) in the proof of its Lemma 11 (p.601) of \cite{yoshida2004partial}, we now estimate 
\beas
\bigg\|\psi_T{\fblue\Phi_T}\bigg(E_\calc\big[\sff(T^{-1/2}Z'_T)\big]-\Psi_{T,p,\calc}[\sff]\bigg)\bigg\|_1
\eeas
as follows. 
\begin{en-text}
First, 
\beas &&
\bigg\|(1-\Phi_T)\psi_T\bigg(E_\calc\big[\sff(T^{-1/2}Z'_T)\big]-\Psi_{T,p,\calc}[\sff]\bigg)\bigg\|_1
\nn\\&\simleq&
\big\|1-\Phi_T\big\|_1T^{\gamma\beta}
+ 
\big\|1-\Phi_T\big\|_{q'}\big\|\Psi_{T,\sfp,\calc}[1+|\cdot|^{\sfp_0}]\big\|_q
\nn\\&=&
O(T^{-L})
\eeas
as $T\to\infty$ for every $L>0$, 
due to (\ref{0311221825}) and (\ref{0211221828}), 
where $q$ is an index appearing in (\ref{0311221825}) and $q'=q/(q-1)$. 
\end{en-text
\begin{en-text}
\beas
\bigg\|\psi_T\bigg(E_\calc\big[\sff(T^{-1/2}Z'_T)\big]-\Psi_{T,p,\calc}[\sff]\bigg)\bigg\|_1
\eeas
as follows. First, 
\beas &&
\bigg\|(1-\Phi_T)\psi_T\bigg(E_\calc\big[\sff(T^{-1/2}Z'_T)\big]-\Psi_{T,p,\calc}[\sff]\bigg)\bigg\|_1
\nn\\&\simleq&
\big\|1-\Phi_T\big\|_1T^{\gamma\beta}
+ 
\big\|1-\Phi_T\big\|_{q'}\big\|\Psi_{T,\sfp,\calc}[1+|\cdot|^{\sfp_0}]\big\|_q
\nn\\&=&
O(T^{-L})
\eeas
as $T\to\infty$ for every $L>0$, 
due to (\ref{0311221825}) and (\ref{0211221828}), 
where $q$ is an index appearing in (\ref{0311221825}) and $q'=q/(q-1)$. 
\end{en-text}
Based on Sweeting's smoothing inequality applied 
(inside of the $L^1$-norm below) 
to 
the $\calc$-measurable function $\sff$ and 
the conditional probability measures 
$P_\calc^{T^{-1/2}Z'_T}$ and $\Psi_{T,\sfp,\calc}$, 
we obtain an inequality corresponding to (34) of Yoshida \cite{yoshida2004partial}: 
\bea\label{03230119}&&
\bigg\|\Phi_T\psi_T\bigg(E_\calc\big[\sff(T^{-1/2}Z'_T)\big]-\Psi_{T,p,\calc}[\sff]\bigg)\bigg\|_1
\nn\\&\leq&
CM\bigg\{
\bigg\|\Phi_T\psi_T\int_{\bbR^\sfd}h(|x|)
\big|\calk_{T^{-K_1}}\ast\big(P_\calc^{T^{-1/2}Z'_T}-\Psi_{T,\sfp,\calc}\big)\big|(dx)\bigg\|_1
\nn\\&&\hspace{30pt}
+\bigg\|\big(E_\calc^{T^{-1/2}Z'_T}+{\fblue\Phi_T}|\Psi_{T,\sfp,\calc}|\big)[h(|\cdot|)]\bigg\|_1
\int_{x:|x|\geq aT^\zeta}h(|x|)\calk(dx)
\nn\\&&\hspace{30pt}
+\bigg\|\big(E_\calc^{T^{-1/2}Z'_T}+{\fblue\Phi_T}|\Psi_{T,\sfp,\calc}|\big)[h(|\cdot|)]\bigg\|_1
\delta(\alpha)^{\lfloor a^{-1/2}T^\zeta\rfloor}
\bigg\}
\nn\\&&
+C'\bigg\|\psi_T\Phi_T\int_{\bbR^\sfd}
\sup_{x:|x|\leq a T^{-(K+\zeta)}\lfloor(2a)^{-1}T^\zeta\rfloor}
\>\omega_\sff(x+y,2aT^{-K_1})\Psi_{T,\sfp,\calc}^+(dy)\bigg\|_1.
\eea
Here 
$\calk$ is a probability measure on $\bbR^\sfd$ having the Fourier transform with a compact support, and $\calk_\ep$ is defined by 
$\calk_\ep(A)=\calk(\ep^{-1}A)$for $A\in\bbB_\sfd$ and $\ep>0$. 
$K_1$ and $K$ are arbitrary positive constants such that $K_1>K$, and 
$\zeta=(K_1-K)/2$. 
{\fred A positive constnat $\alpha>1/2$ and $\delta(\alpha)=\alpha^{-1}(1-\alpha)\in(0,1)$. }
$C$ is a constant depending on $\sfp,\sfd,\alpha$, and $C'=(2\alpha-1)^{-1}$, 
and $a$ is a positive number. 
The function $h(x)=1+x^{\sfp_0}$. 
For the first three terms in $\{...\}$ on the right-hand side of (\ref{03230119}), 
the same estimates are valid as in 
Yoshida \cite{yoshida2004partial}, p.605, {\fblue except that here we use (\ref{0311221825}) having localization by $\Phi_T$.}
The last term is estimated with 
\beas 
\bigg\|\psi_T\Phi_T
\omega\big(\sff;T^{-K},\Psi_{T,\sfp,\calc}^+\big)\bigg\|_1
\eeas
for large $T$, 
{\fblue where $\omega(\sff,\ep,\nu)=\int\omega_\sff(x,\ep)\nu(dx)$ for $\ep>0$ and a measure $\nu$}. 
Thus, we obtained the following lemma, that is a counterpart to Lemma 11 of 
Yoshida \cite{yoshida2004partial}, p.601. 
\begin{lemma}\label{0311230140}
Assume Conditions  (\ref{0311221825}) is fulfilled. 
Then, for given constants $K_1$ and $K$ satisfying $K_1>K>0$, 
there exist positive constants $C_\sfd$ and $\delta''$ such that 
\bea\label{0311221522}&&
\bigg\|\psi_T{\fblue\Phi_T}\bigg(E_\calc\big[\sff(T^{-1/2}Z_T)\big]-\Psi_{T,p,\calc}[\sff]\bigg)\bigg\|_1
\nn\\&\leq&
C_\sfd M\sum_{\alpha:|\alpha|\leq\sfd+1+\sfp_0}
\int\big\|\psi_T\partial_u^\alpha\big[\big(
H_T(u,\calc)e^{\tti u\cdot e_T}-\wh{\Psi}_{T,\sfp,\calc}(u)\big)
\wh{\calk}(T^{-K_1}u)\big]\big\|_1du
\nn\\&&
+C_\sfd\bigg\|\psi_T\Phi_T
\omega\big(\sff;T^{-K},\Psi_{T,\sfp,\calc}^+\big)\bigg\|_1
+\ol{o}_{\check{\cale}(M,{\fblue\sfp_0})}\big(T^{-(\sfp-2+\delta'')/2}\big)
\eea
as $T\to\infty$, that is, this inequality is valid for $\sff\in\check{\cale}(M,\gamma)$, 
having the error term of order $o\big(T^{-(\sfp-2+\delta'')/2}\big)$ 
uniformly in $\sff\in\check{\cale}(M,{\fblue\sfp_0})$. 
\end{lemma}
{\fblue 
\halflineskip\noindent
(III) We are now in Step \ref{0411201712}. 
Lemma 9 of \cite{yoshida2004partial} (p.593) shows that ``$[A3]$'' (p.564) implies 
\bd
\im[{\bf $[A3^\flat]$}] (p.591 of \cite{yoshida2004partial}). 
There exist positive constants $\eta_1,\eta_2,\eta_3,B$ ($\eta_1+\eta_2<1$, $\eta_3<1$), and 
truncation functionals $\Psi_j:(\Omega,\calf)\to\big([0,1],\bbB([0,1])\big)$ 
such that 
\bd
\im[(i)] $\ds \sup_{u:|u|\geq B}\big|E_{\wh{\calc}(j)}\big[\Psi_j\exp(\tti u\cdot Z_{I(j)})\big]\big|\leq\eta_1$ a.s. for every $j$. 
\im[(ii)] For the functionals 
\beas 
p_j\big(\wh{\calc}(j)\big) 
&:=& 
E_{\wh{\calc}(j)}\big[(1-\Psi_j)+2\big(1-\varphi(T^{-\beta}Z_{I(j)})\big)\big], 
\eeas
it holds that 
\beas 
P\bigg[\#\big\{j;\>p_j\big(\wh{\calc}(j)\big)\geq\eta_2\big\}\geq\eta_3n'(T)\bigg] &=& o(T^{-M_1})
\eeas
as $T\to\infty$. 
\ed
\ed
Here $M_1$ is a positive constant. 

We are now assuming $[P3]$. 
The parameters $\eta_i$ ($i=2,3$) in $[P3]$ will be denoted by $\eta_i^\natural$  ($i=2,3$) respectively. 
So, we have 
\bea\label{0411210548}
P\bigg[\sum_j {\fred P}_\calc\big[\wt{\Phi}(j)\leq 1-\eta_2^\natural\big]>\eta_3^\natural n'(T)\bigg] 
&=& 
o(T^{-L})
\eea
as $T\to\infty$. 
Choose positive constants $\eta_i$ ($i=2,3$) and $\ep$ as $\eta_2^\natural<\eta_2$ and $\eta_1+\eta_2<1$, and 
$\eta_3^\natural<\eta_3^\natural+\ep<\eta_3<1$. 
The proof of Lemma 9 of \cite{yoshida2004partial} (p.593), which we will follow, is on pp.605-608 of \cite{yoshida2004partial}. 
We define the truncation functional $\Psi_j$ is by 
\beas 
\Psi_j 
&=& 
\psi_j1_{\big\{\sup_{u:|u|\geq B}\big|E_{\wh{\calc}(j)}\big[\Psi_j\exp(\tti u\cdot Z_{I(j)})\big]\big|\leq\eta_1\big\}}, 
\eeas
where $\psi_j$ is given in $[P3]$. Then (i) of $[A3^\flat]$ is obvious. 
As was done in Step (a) (pp.605-606) and Step (c) (p.608) of the proof of Lemma 9 of \cite{yoshida2004partial}, 
for $[A3^\flat]$ (ii), it suffices to show 
\bea\label{0411210529}
P\big[\Omega_0(T)\cap\{W_2\leq\ep\}\big] &=& O(T^{-L})
\eea
as $T\to\infty$, for every $L>0$, where 
$W_2=n'(T)^{-1/2}\sum_j\big\{\eta_3-E_\calc[\zeta_j]\big\}$ 
for $\zeta_j=1_{\big\{\wh{p}_j(\calc(j))\geq\eta_2\big\}}$. 

We modify Step (b) (pp.607-608) as follows. 
The conditional expectation $E_\calc[\zeta_j]$ admits the estimate
$E_\calc[\zeta_j]\leq\Phi_1(j)+\Phi_2(j)$ with 
\beas 
\Phi_1(j)
&=& 
P_\calc\big[E_{\wh{\calc}(j)}[1-\Psi_j]\geq\eta_2^\natural\big]
\yeq
P_\calc\big[\wt{\Phi}(j)\leq1-\eta_2^\natural\big]
\eeas
and 
\beas 
\Phi_2(j)
&=&
P_\calc\bigg[
E_{\wh{\calc}(j)}\big[2\big(1-\varphi(T^{-\beta}Z_{I(j)})\big)\big]\geq\eta_2-\eta_2^\natural\bigg].
\eeas
As in Step (b), $\Phi_2(j)=cT^{-c'-1}$ on $\Omega_0(T)$ with some positive constants $c$ and $c'$, thanks to the choice of $\Omega_0(T)$. 

Now we have 
\beas &&
P\bigg[\Omega_0(T)\cap\bigg\{\sum_jE_\calc[\zeta_j]>(\eta_3-\ep)n'(T)\bigg\}\bigg]
\nn\\&\leq&
P\bigg[\Omega_0(T)\cap\bigg\{\sum_j\Phi_1(j)+\sum_j\Phi_2(j)>(\eta_3-\ep)n'(T)\bigg\}\bigg]
\nn\\&\leq&
P\bigg[\Omega_0(T)\cap\bigg\{\sum_jP_\calc\big[\wt{\Phi}(j)\leq1-\eta_2^\natural\big]+cT^{-c'}>(\eta_3-\ep)n'(T)\bigg\}\bigg]
\nn\\&\leq&
P\bigg[\Omega_0(T)\cap\bigg\{\sum_jP_\calc\big[\wt{\Phi}(j)\leq1-\eta_2^\natural\big]>\eta_3^\natural n'(T)\bigg\}\bigg]
\nn\\&&
+P\bigg[\Omega_0(T)\cap\bigg\{cT^{-c'}>(\eta_3-\eta_3^\natural-\ep)n'(T)\bigg\}\bigg]
\nn\\&=&
O(T^{-L})\quad(T\to\infty)
\eeas
for every $L>0$, due to (\ref{0411210548}). 
Therefore, we obtain (\ref{0411210529}) by the estimate 
\beas 
P\big[\Omega_0(T)\cap\{W_2\leq\ep\}\big]
&\leq&
P\bigg[\Omega_0(T)\cap\bigg\{\sum_jE_\calc[\zeta_j]\geq(\eta_3-\ep)n'(T)\bigg\}\bigg]
\yeq O(T^{-L})
\eeas
as $T\to\infty$. 
}

\halflineskip
\noindent
(IV) 
Proof of Theorem \ref{0311221323}. 
Following Yoshida \cite{yoshida2004partial}, 
we estimate 
the integral on the right-hand side of (\ref{0311221522}) 
by dividing it into 
two integrals over $\{u;\>|u|\geq l(T)\}$ (Step \ref{0411201708}) 
and 
$\{u;\>|u|\leq l(T)\}$ (Step \ref{0411201707}), 
$l(T)=\lfloor T^{\ep_4}\rfloor$ defined 
with a sufficiently small positive number $\ep_4$ in Lemma 2 (p.587), 
which ensures $[A1]$ (p.587) under $[P1]$ i.e. $[A1']$ (p.563). 
In Step \ref{0411201710}, 
the resulting estimates were combined in Lemma 8 (p.592) to give, in Step \ref{0411201711}, 
Proposition 1 (p.593), and Theorem 1 (p.565) in Step \ref{0411201713}, 
that is rephrased in the present situation as Theorem \ref{0311221323}. 

{\fblue 
We remark that on the line 4, p.599 of \cite{yoshida2004partial}, which is in the proof of Lemma 8 of \cite{yoshida2004partial}, 
Condition (1) on p.565 is used to detach $\psi_T$ from $\Psi_{T,p,\calc}$ 
on the left-hand side of the inequality in Lemma 11 of \cite{yoshida2004partial}, p.601. 
In the present situation, since the definition of $\Delta_T^\Phi(\sff)$ 
keeps the factor $\Phi_T$ as (\ref{0311221401}), 
we can detach $\psi_T$ from $\Phi_T\Psi_{T,p,\calc}[\sff]$ on the left-hand side of (\ref{0311221522}) when using Lemma \ref{0311221522} 
by paying 
\bea\label{0412050305}
\|(1-\psi_T)\Phi_T\|_{q'}\big\|\Phi_T\Psi_{T,p,\calc}[1+|\cdot|^{\sfp_0}]\big\|_{q}
\eea
for it, where $q'=q/(q-1)$. Here we used $\Phi_T^2=\Phi_T$.  
We need (\ref{0311221825}) to bound the second factor on the right-hand side of (\ref{0412050305}). 
Moreover, 
\beas
\|(1-\psi_T)\Phi_T\|_{q'}
&\leq&
\big\|(1-1_{\Omega_0(T)})\Phi_T\big\|_{q'}+\big\|(1-1_{\{{\rm{Var}}_\calc[T^{-1/2}Z_T]\leq u_TI_\sfd\}})\Phi_T\big\|_{q'}
\nn\\&&
+\big\|(1-1_{\{{\rm{Var}}_\calc[T^{-1/2}Z_T]\geq s_TI_\sfd\}})\Phi_T\big\|_{q'}
\nn\\&\leq&
P\big[\Omega_0(T)^c]^{1/q'}+P\big[{\rm{Var}}_\calc[T^{-1/2}Z_T]>u_TI_\sfd]^{1/q'}
\nn\\&&
+P\big[\{{\rm{Var}}_\calc[T^{-1/2}Z_T]< s_TI_\sfd\}\cap\{\Phi_T=1\}]^{1/q'}. 
\eeas
We have $P\big[\Omega_0(T)^c]^{1/q'}=O(T^{-(\sfp-2+\delta'')})$ for some $\delta''$ (Lemma 2 of \cite{yoshida2004partial}, p.587), 
and also $P\big[{\rm{Var}}_\calc[T^{-1/2}Z_T]>u_TI_\sfd]=O(T^{-L})$ for any $L>0$.

}
For the proof of Theorem \ref{0311221323}, finally, 
what is different in our situation is estimation of the second term 
on the right-hand side of (\ref{0311221522}); 
it was originally on p.600 of \cite{yoshida2004partial}. 
As mentioned there, 
the measure 
$\Psi_{T,\sfp,\calc}^+$ has an expression 
\bea\label{0311231431} 
\phi\big(x;0,\text{Var}_\calc[T^{-1/2}Z_T]\big)
\bigg(\sum_{{\bf n}\in\bbZ_+^\sfd:|{\bf n}|\leq 3(\sfp-2)}d_{\bf n}(\calc)x^{\bf n}\bigg)^+
\eea
and the coefficients $d_{\bf n}(\calc)$ involve 
the conditional cumulants of $T^{-1/2}Z_T$ of degree at most $\sfp$, and 
$\big(\det\text{Var}_\calc[T^{-1/2}Z_T]\big)^{-k}$ 
($k=0,1,...,3(\sfp-2)$). 
Therefore 
\bea\label{0311231421}
\big\|\psi_T|d_{\bf n}(\calc)|\big\|_2
&\simleq&
s_T^{-3(\sfp-2)\sfd}
\eea
since the factors other than the determinants are in $L^\inftym$. 
On the other hand, 
\bea\label{0311231427} &&
\bigg\|\psi_T\Phi_T|d_{\bf n}(\calc)|
\int_{\bbR^\sfd}\omega_\sff(x;T^{-K})|x|^{|{\bf n}|}\phi\big(x;0,\text{Var}_\calc[T^{-1/2}Z_T]\big)
dx\bigg\|_1
\nn\\&\leq&
\big\|\psi_T|d_{\bf n}(\calc)|\big\|_2
\bigg\|\psi_T\Phi_T
\int_{\bbR^\sfd}\omega_\sff(x;T^{-K})|x|^{|{\bf n}|}\phi\big(x;0,\text{Var}_\calc[T^{-1/2}Z_T]\big)
dx\bigg\|_2
\nn\\&\leq&
u_T^{\sfd/2}s_T^{-\sfd/2}
\big\|\psi_T|d_{\bf n}(\calc)|\big\|_2
\bigg\|\psi_T\Phi_T
\int_{\bbR^\sfd}\omega_\sff(x;T^{-K})|x|^{|{\bf n}|}
\phi\big(x;0,u_TI_\sfd\big)
dx\bigg\|_2
\eea
since 
\beas 
\phi\big(x;0,\text{Var}_\calc[T^{-1/2}Z_T]\big)
&\leq&
(\det s_TI_\sfd)^{-1/2}(\det u_TI_\sfd)^{1/2}\phi\big(x;0,u_TI_\sfd\big)
\eeas
when $\psi_T=1$. 
Moreover, 
\bea\label{0311231429}&&
\bigg\|\psi_T\Phi_T
\int_{\bbR^\sfd}\omega_\sff(x;T^{-K})|x|^{|{\bf n}|}
\phi\big(x;0,u_TI_\sfd\big)
dx\bigg\|_2
\nn\\&\leq&
\bigg\|\psi_T\Phi_T
\bigg(\int_{\bbR^\sfd}\omega_\sff(x;T^{-K})^2\phi\big(x;0,u_TI_\sfd\big)dx\bigg)^{1/2}
\bigg(\int_{\bbR^\sfd}|x|^{2|{\bf n}|}\phi\big(x;0,u_TI_\sfd\big)dx\bigg)^{1/2}
\bigg\|_2
\nn\\&\leq&
\bigg\|\Phi_T
\bigg(\int_{\bbR^\sfd}\omega_\sff(x;T^{-K})^2\phi\big(x;0,u_TI_\sfd\big)dx\bigg)^{1/2}\bigg\|_2
\bigg(\int_{\bbR^\sfd}|x|^{2|{\bf n}|}\phi\big(x;0,u_TI_\sfd\big)dx\bigg)^{1/2}
\nn\\&\simleq&
u_T^{|{\bf n}|/2}
\omega_2^\Phi\big(\sff;T^{-K},N(0,u_TI_\sfd)\big)
\eea
From (\ref{0311231431}), (\ref{0311231427}) and (\ref {0311231429}), 
we obtain \beas&&
\bigg\|\psi_T\Phi_T
\omega\big(\sff;T^{-K},\Psi_{T,\sfp,\calc}^+\big)\bigg\|_1
\nn\\&\leq&
\sum_{{\bf n}\in\bbZ_+^\sfd:|{\bf n}|\leq3(\sfp-2)}
\bigg\|\psi_T\Phi_T|d_{\bf n}(\calc)|
\int_{\bbR^\sfd}\omega_\sff(x;T^{-K})|x|^{|{\bf n}|}\phi\big(x;0,\text{Var}_\calc[T^{-1/2}Z_T]\big)
dx\bigg\|_1
\nn\\&\simleq&
u_T^{\gamma(1)}s_T^{-\gamma(2)}\omega_2^\Phi\big(\sff;T^{-K},N(0,u_TI_\sfd)\big). 
\eeas
This completes the sketch of the proof of Theorem \ref{0311221323}. 

\subsection{Proof of Theorem \ref{0404131140}}\label{0411110631}
For $\sff\in\check{\cale}(M,\gamma)$, let 
\beas 
\sff^*(z)&=&\Phi_T\sff(G_Tz). 
\eeas 
Then $\sff^*\in\check{\cale}(M',\gamma)$ for some positive constant $M'$ 
by the assumption (\ref{0404131039}). 
We apply Theorem \ref{0311221323} to the $\calc$-measurable random function $\sff^*$ instead of $\sff$. 

From (\ref{0404131247}) and (\ref{0404131255}), 
\bea\label{0404131256}
\chi_{T,r,\calc}^*(v)
&=& 
\chi_{T,r,\calc}(G_T\>\!^{\star} v).
\eea
By the definition of ${P}_{T,r,\calc}$, 
\beas 
\exp\bigg(\sum_{r=2}^\infty\ep^{r-2}(r!)^{-1}\chi_{T,r,\calc}^*(v)\bigg)
&=& 
\exp\big(2^{-1}\chi_{T,2,\calc}^*(v)\big)\sum_{r=0}^\infty\ep^rT^{-r/2}
{P}_{T,r,\calc}\big(v;\chi_{T,\cdot,\calc}^*\big),
\eeas
and we obtain 
\bea\label{0404140436}
\exp\bigg(\sum_{r=2}^\infty\ep^{r-2}(r!)^{-1}\chi_{T,r,\calc}(u)\bigg)
&=& 
\exp\big(2^{-1}\chi_{T,2,\calc}(u)\big)\sum_{r=0}^\infty\ep^rT^{-r/2}
{P}_{T,r,\calc}\big((G_T^\star)^{-1}u;\chi_{T,\cdot,\calc}^*\big),
\nn\\&&
\eea
by plugging $(G_T^\star)^{-1}u$ into $v$ and by using (\ref{0404131256}). 
Comparing (\ref{0404181443}) and (\ref{0404140436}), we see 
\beas
P_{T,r,\calc}\big(u;\chi_{T,\cdot,\calc}\big)
&=&
{P}_{T,r,\calc}\big((G_T^\star)^{-1}u;\chi_{T,\cdot,\calc}^*\big), 
\eeas
or equivalently, 
\bea\label{0404140438}
P_{T,r,\calc}\big(G_T^\star v;\chi_{T,\cdot,\calc}\big)
&=&
{P}_{T,r,\calc}\big(v;\chi_{T,\cdot,\calc}^*\big).  
\eea

On the event 
\beas &&
\big\{\text{Var}_\calc[T^{-1/2}Z_T]\text{ is non-degenerate }\big\}\cap\{\Phi_T=1\}
\nn\\&=&
\big\{\text{Var}_\calc[Z_T^*]\text{ is non-degenerate }\big\}\cap\{\Phi_T=1\}, 
\eeas
we have 
\beas 
\Psi_{T,\sfp,\calc}[\sff^*] 
&=& 
\int \Phi_Tf(G_Tz)\Psi_{T,\sfp,\calc}\sff(dz)
\nn\\&=&
\int \Phi_Tf(G_Tz)\frac{d\calf^{-1}\big[\wh{\Psi}_{T,\sfp,\calc}\big]}{dz}(z)dz
\nn\\&=&
\int \Phi_Tf(G_Tz)\sum_{r=0}^{\sfp-2}T^{-r/2}{P}_{T,r,\calc}\big(-\tti\partial_z;\chi_{T,\cdot,\calc}\big)
\phi\big(z;0,\text{Var}_\calc[T^{-1/2}Z_T]\big)dz
\nn\\&=&
\int \Phi_Tf(G_Tz)\sum_{r=0}^{\sfp-2}T^{-r/2}{P}_{T,r,\calc}\big(G_T^\star(-\tti \partial_y);\chi_{T,\cdot,\calc}\big)
\phi\big(G_T^{-1}y;0,\text{Var}_\calc[T^{-1/2}Z_T]\big)dz
\eeas
for $y=G_Tz$, 
since
\beas 
\frac{\partial}{\partial z}
 &=& G_T^\star\frac{\partial}{\partial y}. 
\eeas
Moreover, 
\beas 
\Psi_{T,\sfp,\calc}[\sff^*] 
&=& 
\int \Phi_Tf(G_Tz)\sum_{r=0}^{\sfp-2}T^{-r/2}{P}_{T,r,\calc}\big((-\tti \partial_y);\chi_{T,\cdot,\calc}^*\big)
\phi\big(y;0,\text{Var}_\calc[Z_T^*]\big)|\det G_T|dz
\eeas
for $y=G_Tz$, 
if (\ref{0404140438}) is used to rewrite the random symbols 
${P}_{T,r,\calc}\big(G_T^\star(-\tti \partial_y);\chi_{T,\cdot,\calc}\big)$. 
Therefore, 
\beas 
\Psi_{T,\sfp,\calc}[\sff^*] 
&=& 
\int \Phi_Tf(y)\sum_{r=0}^{\sfp-2}T^{-r/2}{P}_{T,r,\calc}\big((-\tti \partial_y);\chi_{T,\cdot,\calc}^*\big)
\phi\big(y;0,\text{Var}_\calc[Z_T^*]\big)dy
\nn\\&&
\hspace{100pt}(\text{by change of variables from }z\text{ to }y=G_Tz)
\nn\\&=&
\Phi_T \Psi_{T,\sfp,\calc}^*[f]
\eeas

{\fblue 
Now we apply Theorem \ref{0311221323} to $\sff^*$ to prove the desired inequality 
with $a_0^{-1}u_T$ for a given $u_T$, where $a_0$ is a positive constant, and a constant $K^*>K$ for given $K$. 
More precisely, for some $M^*>0$ depending on also $a_0$, we obtain
\beas\label{0411220620}&&
\sup_{\sff\in\check{\cale}(M,\sfp_0)}\big\|\Phi_TE_\calc[\sff(Z_T^*)]-\Phi_T\Psi_{T,\sfp,\calc}^*[\sff]\big\|_1
\nn\\&=&
\sup_{\sff\in\check{\cale}(M,\sfp_0)}\big\|\Phi_T\big(E_\calc[\sff^*(T^{-1/2}Z_T)]-\Psi_{T,\sfp,\calc}[\sff^*]\big)\big\|_1
\qquad(\Phi_T^2=\Phi_T)
\nn\\&\leq& 
M^*\bigg\{\bigg(P\bigg[\text{Var}_\calc\big[T^{-1/2}Z_T\big]<s_TI_\sfd,\>\Phi_T=1\bigg]\bigg)^\theta
\nn\\&&\hspace{30pt}
+u_T^{\gamma(1)}s_T^{-\gamma(2)}\sup_{\sff\in\check{\cale}(M,{\fblue\sfp_0})}
\omega^\Phi_2\big(\sff^*;T^{-K^*},\phi(x;0,a_0^{-1}u_TI_\sfd)\big)
\bigg\}
+o(T^{-(\sfp-2+\delta^*)/2})
\eeas
as $T\to\infty$, 
if $s_T\geq T^{-c'}$ for large $T$. 
We notice that
\beas 
\omega^\Phi_2\big(\sff^*;T^{-K^*},\phi(x;0,a_0^{-1}u_TI_\sfd)\big)
&\leq&
a_2\omega^\Phi_2\big(\sff;a_1T^{-K^*},\phi(x;0,u_TI_\sfd)\big)
\nn\\&\leq&
a_2\omega^\Phi_2\big(\sff;T^{-K},\phi(x;0,u_TI_\sfd)\big)
\eeas
for large $T$, suitably choosing $a_0$, 
where $a_i$ ($i=0,1,2$) are positive constants associated with the bound (\ref{0404131039}) and independent of $\sff$; 
use $a_0a_0^{-1}u_T=u_T$ inside $\phi$ when changing variable $z$ to $y=G_Tz$ in the integral in the modulus of continuity in measure. 
%
%
\begin{en-text}
We remark that the contribution of $(1-\Phi_T)\sff(Z_T^*)$ to the expectation is of $o(T^{-(\sfp-2+\delta^*)/2})$. 
It is also the case for $(1-\Phi_T)\Psi_{T,\sfp,\calc}^*[\sff]$. 
Indeed, they are of $O(T^{-L})$ for arbitrary positive constant $L$ by the H\"older inequality
with $\|1-\Phi_T\|_p= O(T^{-L})$ for any $p>1$ and $L>0$, since $E_\calc[|T^{-1/2}Z_T|^{\sfp_0}]$ is 
estimated by the $\calc$-conditional cumulants of $T^{-1/2}Z_T$, 
then $E_\calc[\sff(Z_T^*)]$ ($T>1$) 
is uniformly bounded 
in $L^{q_*}$ for some $q_*>1$ 
(use the estimate of the $\calc$-conditional cumulants of $T^{-1/2}Z_T$ on pp.597-598 of \cite{yoshida2004partial}, with the first property in (\ref{0404131039})), 
and $\Psi_{T,\sfp,\calc}^*[\sff]$ is assumed to admit the estimate (\ref{0411231312}). 
\end{en-text}
}
\qed

\onelineskip
\noindent
{\bf Acknowledgements} \vspace{2mm}\\
The authors thank Professor Kengo Kamatani for his valuable comments on numerical computations, and Dr. Ruohan Zhan for her helpful comments from a practitioner point of view. 

\bibliographystyle{spmpsci}      
\bibliography{bibtex-201211120-20210212-20210907+-20221107+}

\end{document}

%% file: nakamacro030101+.tex
\def\infm{{\infty\text{--}}}
\def\inftym{\infm}
\def\koko{{\coloroy{koko}}}
\def\bd{\begin{description}}
\def\ed{\end{description}}

\def\D2{\bbD_{2,\infty-}}

\def\A{{\bf A}}
\def\B{{\bf B}}

\def\D{{\bf D}}
\def\E{{\bf E}}

\def\M{{\bf M}}

\def\W{{\bf W}}

\def\calb{{\cal B}}
\def\calc{{\cal C}}
\def\cald{{\cal D}}
\def\cale{{\cal E}}
\def\calf{{\cal F}}
\def\calg{{\cal G}}
\def\calh{{\cal H}}
\def\cali{{\cal I}}
\def\calj{{\cal J}}
\def\calk{{\cal K}}
\def\call{{\cal L}}
\def\calm{{\cal M}}
\def\caln{{\cal N}}

\def\calv{{\cal V}}
\def\calw{{\cal W}}

\def\ds{\displaystyle}
\def\yeq{\>=\>}
\def\yleq{\>\leq\>}
\def\ygeq{\>\geq\>}

\def\sff{{\sf f}}
\def\sfk{{\sf k}}
\def\sfm{{\sf m}}
\def\sfg{{\sf g}}
\def\sfd{{\sf d}}
\def\sfp{{\sf p}}
\def\sfr{{\sf r}}

\def\simleq{\ \raisebox{-.7ex}{$\stackrel{{\textstyle <}}{\sim}$}\ }

\def\ep{\epsilon}
\def\half{\frac{1}{2}}

\def\Iku{\Rightarrow}

\def\up{\uparrow}
\def\down{\downarrow}

\def\y{\vspace*{3mm}\\}
\def\halflineskip{\vspace*{3mm}}
\def\nn{\nonumber}
\def\be{\begin{equation}}
\def\ee{\end{equation}}
\def\bea{\begin{eqnarray}}
\def\eea{\end{eqnarray}}
\def\beas{\begin{eqnarray*}}
\def\eeas{\end{eqnarray*}}
\def\bi{\begin{itemize}}
\def\ei{\end{itemize}}
\def\im{\item}
\def\bd{\begin{description}}
\def\ed{\end{description}}
\def\l{\left}
\def\r{\right}

\def\dotc{\stackrel{\circ}{C}}

\newcommand{\bbB}{{\mathbb B}}

\newcommand{\bbD}{{\mathbb D}}

\newcommand{\bbI}{{\mathbb I}}
\newcommand{\bbJ}{{\mathbb J}}
\newcommand{\bbK}{{\mathbb K}}

\newcommand{\bbM}{{\mathbb M}}
\newcommand{\bbN}{{\mathbb N}}

\newcommand{\bbR}{{\mathbb R}}
\newcommand{\bbS}{{\mathbb S}}
\newcommand{\bbT}{{\mathbb T}}

\newcommand{\bbV}{{\mathbb V}}

\newcommand{\bbY}{{\mathbb Y}}
\newcommand{\bbZ}{{\mathbb Z}}

\def\tti{{\tt i}}
\def\onelineskip{\halflineskip\halflineskip}
\newcommand{\sfa}{{\sf a}}

\def\sfr{{\sf r}}
\def\sfp{{\sf p}}
\def\sfd{{\sf d}}
\def\sfm{{\sf m}}